\documentclass[aps,pra,floatfix,showpacs,preprint,onecolumn]{revtex4-2}
\usepackage{graphicx}
\usepackage{amsmath, amsfonts, amssymb, bm}

\usepackage{amstext}
\usepackage{mathrsfs}
\usepackage{hyperref}
\usepackage{cleveref}

\usepackage{color}

\begin{document}
\title{WKB electron wave functions in a tightly focused laser beam}
\author{A.~Di Piazza}
\email{dipiazza@mpi-hd.mpg.de}
\affiliation{Max Planck Institute for Nuclear Physics, Saupfercheckweg 1, D-69117 Heidelberg, Germany}

\begin{abstract}
Available laser technology is opening the possibility of testing QED experimentally in the so-called strong-field regime. This calls for developing theoretical tools to investigate strong-field QED processes in electromagnetic fields of complex spacetime structure. Here, we propose a scheme to compute electron wave functions in tightly focused laser beams by taking into account exactly the complex spacetime structure of the fields. The scheme is solely based on the validity of the Wentzel-Kramers-Brillouin (WKB) approximation and the resulting wave functions, unlike previously proposed ones [Phys. Rev. Lett. \textbf{113}, 040402 (2014)], do not rely on approximations on the classical electron trajectory. Moreover, a consistent procedure is indicated to take into account higher-order quantum effects within the WKB approach depending on higher-and-higher powers of the Planck constant. In the case of a plane-wave background field the found wave functions exactly reduce to the Volkov states, which are then written in a new and fully quasiclassical form. Finally, by using the leading-order WKB wave functions to compute the probabilities of nonlinear Compton scattering and nonlinear Breit-Wheeler pair production, it is explicitly shown that, if additionally the energies of the charges are sufficiently large that the latter are not significantly deflected by the field, the corresponding Baier's formulas are exactly reproduced for an otherwise arbitrary classical electron/positron trajectory.
\end{abstract}

\maketitle

\section{Introduction}

Shortly after the invention of the laser theoreticians started investigating how QED processes can be affected or even primed by coherent light \cite{Reiss_1962,Nikishov_1964,Goldman_1964,Brown_1964}. The theoretical framework employed in these pioneering works was the so-called Furry picture \cite{Furry_1951,Landau_b_4_1982}, where the electromagnetic field of the laser is treated as a given, classical background field and the electron-positron spinor field is quantized in the presence of that background field. In this way the effects of the laser field could have been included self-consistently in the calculations. A fundamental requirement to work within the Furry picture is the possibility of solving the Dirac equation analytically in the presence of the background electromagnetic field. In the laser case this is clearly an impossible task if the corresponding field features both spatial and temporal focusing due to the complexity of the spacetime structure of the field. However, if the spatial focusing of the laser field can be ignored, i.e., if the laser field can be approximated as a plane wave, the Dirac equation admits an exact, analytical solution, and the corresponding electron states are known as Volkov states \cite{Volkov_1935,Landau_b_4_1982}. 

The computation of the probabilities of the two basic QED processes in a laser field, the emission of a single photon by an electron (nonlinear Compton scattering) and the electron-positron pair production by a photon (nonlinear Breit-Wheeler pair production), by employing the Volkov states to describe electrons and positrons \cite{Reiss_1962,Nikishov_1964,Goldman_1964,Brown_1964}, indicated that the total probabilities of these processes were controlled by the two Lorentz- and gauge-invariant parameters $\xi=|e|F_0/m\omega_0$ and $\chi=\sqrt{-(F_{0,\mu\nu}q^{\nu})^2}/mF_{cr}$ (see also the reviews \cite{Mitter_1975,Ritus_1985,Ehlotzky_2009,Reiss_2009,Di_Piazza_2012,Dunne_2014} and the monographs \cite{Landau_b_4_1982,Greiner_b_1985,Fradkin_b_1991,Baier_b_1998}). Here, we have introduced the electron mass $m$, the electron charge $e<0$, and the critical field of QED $F_{cr}=m^2/\hbar|e|$, and units with $\epsilon_0=c=1$ together with the metric tensor $\eta^{\mu\nu}=\text{diag}(+1,-1,-1,-1)$ are employed throughout. Moreover, the background laser field is characterized by the amplitude $F_0^{\mu\nu}=(\bm{E}_0,\bm{B}_0)$ of the electromagnetic field tensor, with $|\bm{E}_0|,|\bm{B}_0|\sim F_0$, and by the typical angular frequency $\omega_0$. Finally, the four-vector $q^{\mu}$ indicates the four-momentum of the incoming particle ($q^2=m^2$ for electrons/positrons and $q^2=0$ for photons). The so-called strong-field QED regime is characterized by both parameters $\xi$ and $\chi$ being of the order of or larger than unity. The parameter $\xi$, known as classical nonlinearity parameter, is a classical parameter and in general controls the importance of relativistic effects and of nonlinear effects in the laser-field amplitude. The parameter $\chi$, known as quantum nonlinearity parameter, instead controls pure quantum effects like the importance of photon recoil in nonlinear Compton scattering and, in the case of electrons and positrons, it corresponds to the amplitude of the laser field in the rest frame of the particle in units of $F_{cr}$. The first experimental results on the two mentioned basic strong-field QED processes have been obtained by employing an optical laser beam of intensity of the order of $10^{18}\text{-}10^{19}\;\text{W/cm$^2$}$ ($\xi\sim 0.1\text{-}1$) but colliding with an electron beam of energy of about $45\;\text{GeV}$ ($\chi\sim 0.1\text{-}1$) in the well-known E-144 SLAC experimental campaign \cite{Bula_1996,Burke_1997}.

The rapid advancement of laser technology is opening the possibility of investigating strong-field QED phenomena in the highly nonlinear regime where $\xi\gg 1$ whereas $\chi\gtrsim 1$. Laser intensities exceeding $10^{22}\;\text{W/cm$^2$}$ have already achieved experimentally \cite{Yoon_2019} and several multipetawatt facilities are under construction or planned \cite{APOLLON_10P,ELI,CoReLS,Bromage_2019,XCELS}, which can overcome the present record by one-two orders of magnitude. In addition, ultrarelativistic electron beams can nowadays be produced not only in conventional accelerators but also via laser wakefield acceleration \cite{Leemans_2014}. Indeed, experiments about radiation reaction at the edge between the classical and the quantum regime have been already carried out by employing laser-accelerated electron beams \cite{Cole_2018,Poder_2018} (see Refs. \cite{Wistisen_2018,Wistisen_2019} for correspondingly recent experimental results on quantum radiation reaction in crystalline fields). Moreover, two experimental campaigns at DESY \cite{Abramowicz_2019} and at SLAC \cite{Meuren_2020} are in preparation, which aim at accurate experimental results on strong-field QED by employing multiterawatt-class lasers and high-quality electron beams normally used for producing x-ray radiation.

On the theory side numerous studies have been published on nonlinear Compton scattering \cite{Narozhny_2000,Ivanov_2004,Boca_2009,Harvey_2009,Mackenroth_2010,Boca_2011,Mackenroth_2011,
Seipt_2011,Seipt_2011b,Dinu_2012,Krajewska_2012,Dinu_2013,Seipt_2013,Krajewska_2014,
Wistisen_2014,Harvey_2015,Seipt_2016,Seipt_2016b,Angioi_2016,Harvey_2016b,Angioi_2018,
Di_Piazza_2018,Alexandrov_2019,Di_Piazza_2019,Ilderton_2019_b,Seipt_2020,King_2020_b} and nonlinear Breit-Wheeler pair production \cite{Narozhny_2000,Roshchupkin_2001,Reiss_2009,
Heinzl_2010b,Mueller_2011b,Titov_2012,Nousch_2012,Krajewska_2013b,Jansen_2013,
Augustin_2014,Meuren_2015,Meuren_2016,Di_Piazza_2019,King_2020,Seipt_2020} (see also the reviews \cite{Mitter_1975,Ritus_1985,Ehlotzky_2009,Di_Piazza_2012,Roshchupkin_2012}). Recently, also higher-order processes like nonlinear double Compton scattering \cite{Loetstedt_2009,Seipt_2012,Mackenroth_2013,King_2015,Dinu_2019} and trident pair production \cite{Hu_2010,Ilderton_2011,King_2013,Dinu_2018,Mackenroth_2018,Dinu_2020,Torgrimsson_2020}
have been investigated in the presence of a plane wave by employing the Volkov states and the corresponding Volkov propagator.

The availability of an exact solution of the Dirac equation in a plane-wave field has certainly provided an enormous insight into processes in intense laser fields. Moreover, the early experiments reported in Refs. \cite{Bula_1996,Burke_1997} employed picosecond optical laser pulses focused on an area of the order of $60\;\text{$\mu$m$^2$}$. This explains why the experimental results could have been reproduced by starting from the probability of the corresponding processes in a plane wave. The more recent experiments reported in Refs.~\cite{Cole_2018, Poder_2018} are carried out employing femtosecond pulses focused down to an area of few square wavelengths, and future experiments aiming at even higher intensities will possibly employ shorter and more tightly focused laser pulses. There are already theoretical tools which enable one to study processes in the presence of laser fields of complex spacetime structure, for which no exact analytical solution of the Dirac equation is available. We mention here the so-called locally constant field approximation (LCFA), which allows one to write the probabilities of a strong-field QED process in the presence of an arbitrary background laser field as an average over the corresponding result in a constant crossed field \cite{Reiss_1962,Ritus_1985,Baier_b_1998,Di_Piazza_2012}. However, generally speaking, the LCFA applies only for laser pulses characterized by $\xi\gg 1$ (see the studies \cite{Baier_1989,Khokonov_2002,Di_Piazza_2007,Wistisen_2015,Harvey_2015,Dinu_2016,
Di_Piazza_2018,Blackburn_2018,Alexandrov_2019,Di_Piazza_2019, Ilderton_2019_b,Podszus_2019,Ilderton_2019,Raicher_2021} for investigations about the limitations of the LCFA), whereas some of the upcoming experimental campaigns \cite{Abramowicz_2019,Meuren_2020} may also aim at least initially at investigating more moderate intensity regimes, not to mention the fact that the condition $\xi\gg 1$ cannot be fulfilled for all electron-laser interaction points. Another widely used tool is represented by Baier's formulas of nonlinear Compton scattering and nonlinear Breit-Wheeler pair production, which express the corresponding probabilities as integrals over the classical trajectories of the charged particles \cite{Baier_1967,Baier_1968,Baier_1969,Baier_b_1998} (see also the book \cite{Akhiezer_b_1996}). Baier's formulas are based on the quasiclassical operator technique, allow one to obtain results at the leading order in the quasiclassical, ultrarelativistic limit, when the energies of the charges are sufficiently large that the latter are only barely deflected by the background field, and they are especially useful for numerical calculations. However, the employed operator technique does not provide a general prescription on how to calculate neither the amplitude of a generic QED process nor higher-order corrections. 

Apart from these general approaches, we also mention that effects of the laser spatial focusing in nonlinear Compton and Thomson scattering (the latter process corresponding to the classical emission of radiation, where QED effects like recoil can be neglected) have been investigated numerically in Ref. \cite{Li_2015} and in Ref. \cite{Harvey_2016}, respectively. Also, analytical expressions of scalar Wentzel-Kramers-Brillouin (WKB) wave functions have been found in Ref. \cite{Heinzl_2016} for a specific class of background fields depending on the spacetime coordinates via the quantity $(fx)$ like a plane wave but with $f^{\mu}$ not being necessarily lightlike as in a plane wave. The classical and quantum dynamics of a scalar particle in the background field of two counterpropagating plane waves has been studied in Ref. \cite{King_2016}. Moreover, in Ref. \cite{Heinzl_2017} Poincar\'e symmetry and superintegrability were exploited to construct the exact solution of the Dirac equation for an external field approximating the transverse structure of a radially polarized laser field close to the focus (see also Ref. \cite{Bagrov_b_2014}). Finally, we also mention Refs. \cite{Orzalesi_1974,Orzalesi_1975}, where the author develops an eikonal perturbation theory especially suitable for application to hadron dynamics.

In Refs.~\cite{Di_Piazza_2014,Di_Piazza_2015,Di_Piazza_2016,Di_Piazza_2017_b} approximated expressions of the electron states and of the electron propagator in a tightly focused laser beam have been found by starting from the WKB approximation and applied to investigate nonlinear Compton scattering and nonlinear Breit-Wheeler pair production. The findings in Refs.~\cite{Di_Piazza_2014,Di_Piazza_2015,Di_Piazza_2016,Di_Piazza_2017_b} are based on the physical assumption that the energies of the charges are so large that they are only barely deflected by the focused laser field. Mathematically this is the case in the customarily considered (almost) counterpropagating setup for an incoming electron energy $\varepsilon$ much larger than $\max(m,m\xi)$, an approximation scheme which is particularly useful in the ultrarelativistic regime for moderate laser intensities but also at those high intensities to be reached in the near future. It is worth stressing, in fact, that the quantity $\eta=\max(m,m\xi)/\varepsilon$ is automatically much smaller than unity for present and upcoming experimental conditions, provided one aims at investigating the strong-field regime of QED. Indeed, for an ultrarelativistic electron initially counterpropagating with respect to the laser beam, in order to enter the strong-field QED regime (say at $\chi>1$), by assuming the laser to be a Ti:sapphire laser ($\omega_0=1.55\;\text{eV}$) and to have a soon feasible intensity of $I_0\sim 10^{23}\;\text{W/cm$^2$}$  \cite{APOLLON_10P,ELI} (corresponding to $\xi=150$), it is necessary that $\varepsilon\gtrsim 500\;\text{MeV}$ such that it is $\varepsilon/m\approx 10^3$. Now, in Ref. \cite{Di_Piazza_2014} the explicit classical trajectory of an electron in a generic background field was found at the next-to-leading order in the parameter $\eta$ having in mind the case of a tightly focused laser beam. This, in turn, allowed us to determine the classical action of the electron in the field analytically under the same approximation and to write the electron states explicitly in terms of the background electromagnetic field in a form similar to that of Volkov states \cite{Di_Piazza_2015} (these states reduce to the corresponding ones presented in Refs. \cite{Blankenbecler_1987,Akhiezer_1993} in the case of a background time-independent scalar potential). Those states are particularly useful to carry out analytical calculations because the electron trajectory is explicitly expressed in terms of the background field.

In the present paper, we continue the investigation of nonlinear Compton scattering and nonlinear Breit-Wheeler pair production in the presence of a tightly focused laser beam by elaborating on the method presented in Refs.~\cite{Di_Piazza_2014,Di_Piazza_2015,Di_Piazza_2016,Di_Piazza_2017_b}. First, we explicitly find the electron states at the leading order in the WKB approximation but without approximating the classical electron trajectory as in Refs.~\cite{Di_Piazza_2014,Di_Piazza_2015,Di_Piazza_2016,Di_Piazza_2017_b}. In this way, the validity of the presented states is not limited by particular features of the classical electron trajectory but only by quantum conditions, which are derived below. Moreover, we show that for a background plane wave the obtained wave functions exactly reduce to the Volkov states. As a by-product, then, we derive a new form of the Volkov states, where their quasiclassical structure is manifest also in the spinor structure. In addition, we use the found electron and positron states to write the amplitudes of nonlinear Compton scattering and nonlinear Breit-Wheeler pair production, which are then more accurate than those obtained before but which are more suitable for a numerical computation based on the charged particles classical trajectory. By still making no approximations about the classical trajectory but by computing the probabilities of nonlinear Compton scattering and nonlinear Breit-Wheeler pair production up to leading order in $\eta$ \footnote{This does not contradict the above statement about the expansion of the trajectory with respect to $\eta$ in Refs.~\cite{Di_Piazza_2014,Di_Piazza_2015,Di_Piazza_2016,Di_Piazza_2017_b}. As mentioned in Ref. \cite{Di_Piazza_2014} there can be situations, where the external field features particular symmetries, like in a crystal, where the approximated solution of the trajectory presented there is not valid even though the electron energy is the largest dynamical energy scale in the problem.}, we exactly reproduce Baier's formulas based on the particles classical trajectory in the given electromagnetic field (this was verified in Refs. \cite{Di_Piazza_2016,Di_Piazza_2017_b} only for the approximated trajectory found in Ref. \cite{Di_Piazza_2014}). This already casts Baier's method into a self-consistent approach, which can be in principle extended to higher-order processes and whose validity conditions are under control. Moreover, the initial amplitudes obtained here beyond the leading-order expansion in $\eta$ are more general than Baier's formulas as only effects proportional to $\hbar$ are neglected (limitations of the Baier's formulas as compared with the WKB approach were already noticed in Ref. \cite{Raicher_2019} in the case of radiation by an electron in a rotating electric field). Finally, a prescription is provided to compute higher-order quantum corrections of the found electron states within the WKB approach, which may be useful if, for example, more accurate results are required in a specific problem. 

\section{Notation}
In the present section we introduce the notation employed in the paper. For the sake of definiteness, we assume that the laser field main propagation direction corresponds to the negative $z$ axis of the coordinate system. Thus, it is convenient to introduce the light-cone coordinates 
\begin{align}
T=\frac{t+z}{2}, && \bm{x}_{\perp}=(x,y), && \phi=t-z
\end{align}
for a spacetime point with coordinates $x^{\mu}=(t,\bm{x})=(t,x,y,z)$. The light-cone coordinates of an arbitrary four-vector $v^{\mu}=(v_0,\bm{v})$ are defined as $v_+=(v_0+v_z)/2$, $\bm{v}_{\perp}=(v_x,v_y)$, and $v_-=v_0-v_z$. The same definition is extended to the Dirac gamma matrices $\gamma^{\mu}=(\gamma^0,\bm{\gamma})$: $\gamma_+=(\gamma^0+\gamma^3)/2$, $\bm{\gamma}_{\perp}=(\gamma^1,\gamma^2)$, and $\gamma_-=\gamma^0-\gamma^3$. Moreover, the derivatives 
\begin{align}
\frac{\partial}{\partial T}=\frac{\partial}{\partial t}+\frac{\partial}{\partial z}, && \frac{\partial}{\partial \phi}=\frac{1}{2}\left(\frac{\partial}{\partial t}-\frac{\partial}{\partial z}\right)
\end{align}
with respect to the light-cone coordinates $T$ and $\phi$ can be derived from the relations
\begin{align}
\frac{\partial}{\partial t}=\frac{1}{2}\frac{\partial}{\partial T}+\frac{\partial}{\partial \phi},&& \frac{\partial}{\partial z}=\frac{1}{2}\frac{\partial}{\partial T}-\frac{\partial}{\partial \phi},
\end{align}
whereas the derivatives with respect to the transverse coordinates form the two-dimensional vector $\bm{\nabla}_{\perp}=\partial/\partial \bm{x}_{\perp}$. Also, the scalar product between two four-vectors $u^{\mu}$ and $v^{\mu}$ can be written as 
\begin{equation}
(uv)=u_+v_-+u_-v_+-\bm{u}_{\perp}\cdot\bm{v}_{\perp}
\end{equation}
and the four-divergence of a vector field $G^{\mu}(x)$ as
\begin{equation}
\partial_{\mu} G^{\mu}=\frac{\partial G_+}{\partial T}+\frac{\partial G_-}{\partial \phi}+\bm{\nabla}_{\perp}\cdot\bm{G}_{\perp}.
\end{equation}
It is convenient to introduce the four-dimensional quantities: $n^{\mu}=(1,\bm{n})$ and $\tilde{n}^{\mu}=(1,-\bm{n})/2$, with $\bm{n}=(0,0,1)$ being the unit vector along the $z$ direction, and $a_j^{\mu}=(0,\bm{a}_j)$, where $j=1,2$, with $\bm{a}_1$ and $\bm{a}_2$ being the unit vectors along the $x$ and the $y$ direction, respectively. It is clear that the four-dimensional quantities $n^{\mu}$, $\tilde{n}^{\mu}$, and $a^{\mu}_j$ fulfill the completeness relation: $\eta^{\mu\nu}=n^{\mu}\tilde{n}^{\nu}+\tilde{n}^{\mu}n^{\nu}-a_1^{\mu}a_1^{\nu}-a_2^{\mu}a_2^{\nu}$ (note that $(n\tilde{n})=1$ and $(a_1a_1)=(a_2a_2)=-1$, whereas all other possible scalar products among $n^{\mu}$, $\tilde{n}^{\mu}$, and $a^{\mu}_j$ vanish). By using the quantities $n^{\mu}$, $\tilde{n}^{\mu}$, and $a^{\mu}_j$ one can express the light-cone coordinates as $T=(\tilde{n}x)$, $\bm{x}_{\perp}=-((xa_1),(xa_2))$, and $\phi=(nx)$. Analogously, the light-cone coordinates of an arbitrary four-vector $v^{\mu}=(v^0,\bm{v})$ are given by $v_+=(\tilde{n}v)$, $\bm{v}_{\perp}=-((va_1),(va_2))$, and $v_-=(nv)$, whereas, by using the ``hat'' notation for the contraction of a four-vector and the four Dirac gamma matrices, one obtains $\gamma_+=\hat{\tilde{n}}$, $\bm{\gamma}_{\perp}=-(\hat{a}_1,\hat{a}_2)$, and $\gamma_-=\hat{n}$. Concerning the derivatives with respect to the coordinates, the following relations hold: $\partial/\partial T=\partial_T=(n\partial)$, $\bm{\nabla}_{\perp}=((a_1\partial),(a_2\partial))$, and $\partial/\partial\phi=\partial_{\phi}=(\tilde{n}\partial)$. 

Having in mind physical situations in which the incoming particle initially (almost) counterpropagates with respect to the laser field, it is natural to use/interpret the light-cone variable $T$ as the light-cone time and the other three coordinates $\bm{x}_{\text{lc}}=(\bm{x}_{\perp},\phi)$ as the light-cone spatial coordinates. Correspondingly, in the case of an on-shell electron four-momentum $p^{\mu}=(\varepsilon,\bm{p})$, with $\varepsilon=\sqrt{m^2+\bm{p}^2}$, we consider $\bm{p}_{\text{lc}}=(\bm{p}_{\perp},p_+)$ as the three spatial light-cone components of the four-momentum and $p_-=(m^2+\bm{p}_{\perp}^2)/2p_+$ as the remaining time light-cone component. The components of the three-dimensional quantity $\bm{x}_{\text{lc}}$ ($\bm{p}_{\text{lc}}$) are indicated by means of Latin indexes running from $1$ to $3$, with $x_{\text{lc},1}=x$, $x_{\text{lc},2}=y$, and $x_{\text{lc},3}=\phi$ ($p_{\text{lc},1}=p_x$, $x_{\text{lc},2}=p_y$, and $p_{\text{lc},3}=p_+$). Analogous considerations hold for a generic photon four-momentum.

Concerning the background electromagnetic field, we assume it to be described by the four-vector potential $A^{\mu}(x)=(V(x),\bm{A}(x))$ satisfying the Lorenz-gauge condition $\partial_{\mu}A^{\mu}(x)=0$ together with the additional constraint $A_-(x)=0$, i.e., $A_z(x)=V(x)$ and the asymptotic conditions $\lim_{T\to\pm\infty}A^{\mu}(T,\bm{x}_{\text{lc}})=0$. For the sake of convenience, a fixed value $T_0$ of the light-cone time $T$ is chosen below to assign asymptotic conditions of the electron classical trajectory. We assume that the absolute value of $T_0$ is sufficiently large that we can ignore the field there, i.e., $A^{\mu}(T_0,\bm{x}_{\text{lc}})=0$.

Note that the considerations based on the incoming particle being initially almost counterpropagating with respect to the laser field can be easily reformulated if this is not the case, by appropriately adapting the light-cone coordinates in such a way that they are (almost) aligned with respect to the initial velocity of the particle.

\section{Classical action, Hamilton-Jacobi equation, and the van Vleck determinant}
As it is well known \cite{Landau_b_3_1977}, in the WKB method the classical action plays a fundamental role. Therefore, it is convenient here to report some known results about the classical action, which have been appropriately adapted to the use of light-cone coordinates.

The action is originally defined as the generator of the canonical transformation from the coordinates and conjugated momenta of a dynamical system at a generic time to the corresponding quantities at the initial time, which also explains why knowing the action corresponds to solving the equations of motion of the system \cite{Goldstein_b_2002}. In particular, the action is a generator of type two, i.e., it depends on the old coordinated (those at the generic time) and on the new constant momenta (those at the initial time). In this respect, we recall that in the relativistic domain the classical action $S(x)$ of an electron in the presence of an electromagnetic field described by the four-potential $A^{\mu}(x)$ is computed as a \emph{complete} solution of the relativistic Hamilton-Jacobi equation \cite{Landau_b_2_1975} 
\begin{equation}
\label{HJ}
(\partial_{\mu}S+eA_{\mu})(\partial^{\mu}S+eA^{\mu})-m^2=0,
\end{equation}
which in light-cone coordinates and in the Lorenz gauge with $A_-(x)=0$ reads
\begin{equation}
\label{HJ_LC}
\partial_T S=\frac{m^2+(\bm{\nabla}_{\perp}S-e\bm{A}_{\perp})^2}{2(\partial_{\phi}S+eV)}.
\end{equation}
A complete solution of a first-order differential equation is a solution containing as many independent integration constants as the number of variables and the meaning of the word ``independent'' in this context will be clarified below \cite{Goldstein_b_2002}. Since the Hamilton-Jacobi equation only contains derivatives of the action, an integration constant is always an additive constant, which we assume to be determined by the initial condition on the action. Indeed, we are interested in the solution $S_p(x;T_0)$ of the Hamilton-Jacobi equation, which fulfills the initial condition $S_p(T_0,\bm{x}_{\text{lc}};T_0)=-(p_+\phi+p_-T_0-\bm{p}_{\perp}\cdot\bm{x}_{\perp})$ [recall that $A^{\mu}(T_0,\bm{x}_{\text{lc}})=0$]. As we have mentioned already, since the four-momentum is assumed to be on-shell and $p^0=\varepsilon=+\sqrt{m^2+\bm{p}^2}$, the three remaining independent integration constants are chosen to be the components $\bm{p}_{\text{lc}}=(\bm{p}_{\perp},p_+)$ of the four-momentum, with $p_+>0$ and $p_-=(m^2+\bm{p}^2_{\perp})/2p_+$. The independence of these constants means that the (absolute value of the) determinant 
\begin{equation}
\label{vV_det_1}
\left|\det\left(\frac{\partial^2 S_p}{\partial p_{\text{lc},i}\partial x_{\text{lc},j}}\right)\right|
\end{equation}
never vanishes and we will show below that this is indeed the case here. Once the initial condition is given, it is clear, for example, how to integrate numerically the Hamilton-Jacobi equation (\ref{HJ_LC}). It is worth stressing here that, by rewriting the Hamilton-Jacobi equation [see Eq. (\ref{HJ})] as
\begin{equation}
(\partial_tS+eV)^2-(\partial_zS-eV)^2=m^2+(\bm{\nabla}_{\perp}S-e\bm{A}_{\perp})^2,
\end{equation}
one sees that the quantity $P_{e,0}(x)=-\partial_tS(x)-eV(x)$ is either always strictly larger than $|P_{e,z}(x)|$, with $P_{e,z}(x)=\partial_zS(x)-eA_z(x)=\partial_zS(x)-eV(x)$ (this notation will be clear below), and then it is positive or always strictly smaller than the quantity $-|P_{e,z}(x)|$ and then it is negative. Correspondingly, the two quantities $P_{e,-}(x)=P_{e,0}(x)-P_{e,z}(x)=-[\partial_tS(x)+eV(x)]-[\partial_zS(x)-eV(x)]=-\partial_TS$ and $P_{e,+}(x)=[P_{e,0}(x)+P_{e,z}(x)]/2=\{-[\partial_tS(x)+eV(x)]+[\partial_zS(x)-eV(x)]\}/2=-\partial_{\phi}S(x)-eV(x)$ are always either both positive or both negative [see also Eq. (\ref{HJ_LC})]. As we will see below in the determination of the electron states, the first (second) case corresponds to positive-energy (negative-energy) electron states. These cases are characterized by different initial conditions on the action and the above one corresponds to the case of positive-energy states, which also corresponds to the standard choice in classical electrodynamics. 

Once the action $S_p(x;T_0)$ is known, the corresponding classical trajectory can be determined. In fact, by using the theory of canonical transformations, the independent components $\bm{P}_{e,\text{lc}}(x;T_0)=(\bm{P}_{e,\perp}(x;T_0),P_{e,+}(x;T_0))$ of the kinetic momentum of the electron at the light-cone time $T$ are given by $\bm{P}_{e,\perp}(x;T_0)=\bm{\nabla}_{\perp}S_p(x;T_0)-e\bm{A}_{\perp}(x)$ and $P_{e,+}(x;T_0)=-\partial_{\phi}S_p(x;T_0)-eA_+(x)$, which can be written in a manifestly covariant form as $P_e^{\mu}(x;T_0)=(\mathcal{E}_e(x;T_0),\bm{P}_e(x;T_0))=-\partial^{\mu}S_p(x;T_0)-eA^{\mu}(x)$ \cite{Landau_b_2_1975,Goldstein_b_2002}. Also, the initial electron coordinates $\bm{x}_{0,\text{lc}}=(\bm{x}_{0,\perp},\phi_0)$ at $T=T_0$ are given as functions of $x^{\mu}$ and $\bm{p}_{\text{lc}}$ via the relations $\bm{x}_{0,\perp}-(\bm{p}_{\perp}/p_+)T_0=\bm{\nabla}_{\bm{p}_{\perp}}S_p(x;T_0)$ and $-\phi_0+(p_-/p_+)T_0=\partial_{p_+}S_p(x;T_0)$, with $p_-=(m^2+\bm{p}_{\perp}^2)/2p_+$ \cite{Goldstein_b_2002}. Note that the initial condition above corresponds to the electron asymptotically moving along the straight line $\bm{x}_{\perp}=\bm{x}_{\perp}(T;T_0)=\bm{x}_{0,\perp}+(\bm{p}_{\perp}/p_+)(T-T_0)$ and $\phi=\phi(T;T_0)=\phi_0+(p_-/p_+)(T-T_0)$. Now, the determinant in Eq. (\ref{vV_det_1}) can be written as
\begin{equation}
\label{vV_det_2}
\left|\det\left(\frac{\partial x_{\text{lc},0,i}}{\partial x_{\text{lc},j}}\right)\right|.
\end{equation}
This expression shows that the requirement that this determinant never vanishes allows one to invert the relations $\bm{x}_{0,\perp}-(\bm{p}_{\perp}/p_+)T_0=\bm{\nabla}_{\bm{p}_{\perp}}S_p(x;T_0)$ and $-\phi_0+(p_-/p_+)T_0=\partial_{p_+}S_p(x;T_0)$ and to obtain the trajectory $\bm{x}_{\text{lc}}=\bm{x}_{\text{lc}}(T;T_0,\bm{x}_{0,\text{lc}},\bm{p}_{\text{lc}})$. Then, by replacing these quantities in the relations $\bm{P}_{e,\perp}(x;T_0)=\bm{\nabla}_{\perp}S_p(x;T_0)-e\bm{A}_{\perp}(x)$ and $P_{e,+}(x;T_0)=-\partial_{\phi}S_p(x;T_0)-eA_+(x)$, one also obtains the independent light-cone components of the kinetic four-momentum as functions of $T$, $T_0$, $\bm{x}_{0,\text{lc}}$, and $\bm{p}_{\text{lc}}$.

It is useful to report here another alternative form of the determinant in Eqs. (\ref{vV_det_1}) and (\ref{vV_det_2}). In fact, from the general relation $P_e^{\mu}(x;T_0)=-\partial^{\mu}S_p(x;T_0)-eA^{\mu}(x)$ between the kinetic four-momentum of the electron at the light-cone time $T$ and the spacetime derivatives of the action, one can also rewrite the determinant in Eq. (\ref{vV_det_1}) as
\begin{equation}
\label{vV_det_3}
\left|\det\left(\frac{\partial P_{e,\text{lc},j}}{\partial p_{\text{lc},i}}\right)\right|,
\end{equation}
showing that if it is different from zero the initial light-cone kinetic momenta of the electron can also be expressed in terms of the corresponding quantities at the light-cone time $T$.

A customary way of solving the Hamilton-Jacobi equation, which transparently relates this equation with the Newtonian equations of motion, is the method of characteristics \cite{Evans_b_2010}. In general, the basic idea is to imagine that the spacetime hypersurface $S=S_p(x;T_0)$ can be constructed as the union of an infinite number of characteristic curves, in such a way that instead of solving a partial derivatives differential equation, one solves the ordinary differential equations fulfilled by the characteristic curves. Indeed, the latter are parametrized by the proper time $\tau$ as $x^{\mu}=x^{\mu}(\tau;T_0)$ and satisfy the equations
\begin{equation}
\label{Eq_mot_x}
m\frac{dx^{\mu}}{d\tau}=\Pi_e^{\mu},
\end{equation}
where $\Pi_e^{\mu}(\tau;T_0)=P_e^{\mu}(x(\tau;T_0);T_0)$. Together with the definition $P_e^{\mu}(x;T_0)=-\partial^{\mu}S_p(x;T_0)-eA^{\mu}(x)$ and with the fact that the action $S_p(x;T_0)$ satisfies the Hamilton-Jacobi equation, Eq. (\ref{Eq_mot_x}) implies that
\begin{equation}
\label{Eq_mot_p}
m\frac{d^2x^{\mu}}{d\tau^2}=\frac{d\Pi_e^{\mu}}{d\tau}=(\partial^{\nu}P_e^{\mu})\frac{dx_{\nu}}{d\tau}=(\partial^{\nu}P_e^{\mu}-\partial^{\mu}P_e^{\nu})\frac{dx_{\nu}}{d\tau}=eF^{\mu\nu}\frac{dx_{\nu}}{d\tau},
\end{equation}
which is nothing but the Lorentz equation in the external electromagnetic field $F^{\mu\nu}(x)=\partial^{\mu}A^{\nu}(x)-\partial^{\nu}A^{\mu}(x)$. In order to solve the system of equations (\ref{Eq_mot_x})-(\ref{Eq_mot_p}), we have to provide initial conditions. By assuming to fix the initial conditions at $\tau=0$, we have $x^{\mu}(0;T_0)=(T_0,\bm{x}_{0,\text{lc}})$ and $\Pi_e^{\mu}(0;T_0)=-\partial^{\mu}S_p(x_0;T_0)=p^{\mu}$. Note that, although the initial conditions $\bm{x}_{0,\text{lc}}$ on the positions are arbitrarily chosen (within the manifold $T=T_0$), the ones on the four-momentum components $\bm{\Pi}_{e,\text{lc}}(\tau;T_0)$ have to be compatible with the original Hamilton-Jacobi equation, i.e., $\Pi^2_e(0;T_0)=(\partial S_p(T_0,\bm{x}_{0,\text{lc}};T_0))^2=p^2=m^2$ and with the initial conditions on the action, which is automatically the case here because $S_p(T_0,\bm{x}_{\text{lc}};T_0)=-(p_+\phi+p_-T_0-\bm{p}_{\perp}\cdot\bm{x}_{\perp})$. These initial conditions are said to be admissible and the corresponding Cauchy problem is well posed \cite{Evans_b_2010}. Note that the initial on-shell condition on the four-momentum and the Lorentz equation guarantee that the Hamilton-Jacobi equation $\Pi_e^2(\tau;T_0)=m^2$ is always fulfilled along any characteristic curve. 

The action along an arbitrary characteristic curve can be constructed by introducing the function $\Sigma_p(\tau;T_0)=S_p(x(\tau;T_0);T_0)$, which fulfills the equation
\begin{equation}
\label{Eq_mot_Sigma}
\frac{d\Sigma_p}{d\tau}=(\partial_{\mu}S_p)\frac{dx^{\mu}}{d\tau}=-m-e\frac{(\Pi_eA)}{m}.
\end{equation}
Now, the procedure is to solve the equations of motion (\ref{Eq_mot_x})-(\ref{Eq_mot_p}) and Eq. (\ref{Eq_mot_Sigma}) for the action by fixing generic initial conditions $\bm{x}_{0,\text{lc}}$, $\bm{p}_{\text{lc}}$, and $\Sigma_p(0;T_0)=S_p(x(0;T_0);T_0)=-(p_+\phi_0+p_-T_0-\bm{p}_{\perp}\cdot\bm{x}_{0,\perp})$. In this way, one obtains the functions $x^{\mu}=x^{\mu}(\tau;T_0,\bm{x}_{0,\text{lc}},\bm{p}_{\text{lc}})$, $\Pi_e^{\mu}(\tau;T_0,\bm{x}_{0,\text{lc}},\bm{p}_{\text{lc}})$, and $\Sigma_p(\tau;T_0,\bm{x}_{0,\text{lc}},\bm{p}_{\text{lc}})$, where, for the sake of clarity, we have explicitly indicated here the dependence on the initial light-cone coordinates and momenta. Now, as we have already discussed and as we will prove below, the four equations $x^{\mu}=x^{\mu}(\tau;T_0,\bm{x}_{0,\text{lc}},\bm{p}_{\text{lc}})$ can be inverted to obtain the functions $\tau=\tau(x;T_0,\bm{p}_{\text{lc}})$ and $\bm{x}_{0,\text{lc}}=\bm{x}_{0,\text{lc}}(x;T_0,\bm{p}_{\text{lc}})$ \cite{Evans_b_2010}. Then, the action is finally obtained as $S_p(x;T_0)=\Sigma_p(\tau(x;T_0,\bm{p}_{\text{lc}});T_0,\bm{x}_{0,\text{lc}}(x;T_0,\bm{p}_{\text{lc}}),\bm{p}_{\text{lc}})$. As an illustrative example, we report in the Appendix \ref{App_A} the derivation of the classical action of an electron in a plane wave by means of the method of characteristics.

Note that, since $dT/d\tau=\Pi_{e,+}/m>0$, an alternative possibility is to parametrize the trajectory by using directly the light-cone time $T$ and to obtain the trajectories as functions $\bm{x}_{\text{lc}}=\bm{x}_{\text{lc}}(T;T_0,\bm{x}_{0,\text{lc}},\bm{p}_{\text{lc}})$. As we have already observed, these relations can be inverted and one can write $\bm{x}_{0,\text{lc}}=\bm{x}_{0,\text{lc}}(x;T_0,\bm{p}_{\text{lc}})$. In this case, the action is represented as $S_p(x;T_0)=\Sigma_p(T;T_0,\bm{x}_{0,\text{lc}}(x;T_0,\bm{p}_{\text{lc}}),\bm{p}_{\text{lc}})$, where, for the sake of notational simplicity, we have used the same symbol $\Sigma_p$ for the function of the proper time and of the light-cone time.

\subsection{The van Vleck determinant}
\label{vV}
The last step of the procedure described above to compute the action relies on the possibility of expressing the proper time $\tau$ and the initial light-cone coordinates $\bm{x}_{0,\text{lc}}$ at the light-cone time $T_0$ as functions of the coordinates $x^{\mu}$ at the generic light-cone time $T$ (the initial light-cone components of the momentum are always $\bm{p}_{\text{lc}}$). In order to prove this possibility mathematically it is convenient to introduce the so-called van Vleck determinant \cite{van_Vleck_1928} (see also Refs. \cite{Schiller_1962_a,Schiller_1962_b}). Again, due to the use of light-cone coordinates, we introduce here a slightly different definition of the van Vleck determinant than the one discussed in the literature \cite{van_Vleck_1928,Schiller_1962_a,Schiller_1962_b}. Anticipating that below we need the van Vleck determinant both for an electron and for a positron, we limit to the first case, we indicate the van Vleck determinant for the electron as $D_e(x;T_0)$, and we define it as [see also Eqs. (\ref{vV_det_1}) and (\ref{vV_det_2})]
\begin{equation}
D_e(x;T_0)=\left|\det\left(\frac{\partial^2 S_p}{\partial p_{\text{lc},i}\partial x_{\text{lc},j}}\right)\right|=\left|\det\left(\frac{\partial x_{\text{lc},0,i}}{\partial x_{\text{lc},j}}\right)\right|.
\end{equation}
Recalling the discussion around Eq. (\ref{vV_det_1}), we conclude that if $D_e(x;T_0)\neq 0$, then the action $S_p(x;T_0)$ represents a complete solution of the Hamilton-Jacobi equation, which allows one to construct the solution of the Lorentz equation of motion. Alternatively, one can solve the Lorentz equation of motion and build the action. 

Being obtained from the classical action $S_p(x;T_0)$, the van Vleck determinant $D_e(x;T_0)$ can also be computed by means of the method of characteristics. By focusing on the trajectory $x^{\mu}=x^{\mu}(\tau;T_0)$ corresponding to the initial conditions $x^{\mu}(0;T_0)=x_0^{\mu}$ on the position, we consider a four-dimensional infinitesimally thin tube $\Omega$ along (and containing) the trajectory between two proper times $\tau_1$ and $\tau_2>\tau_1$ \cite{Rubinow_1963}. By applying the Gauss theorem to the quantity $\partial_{\mu} P_e^{\mu}(x;T_0)$, we obtain
\begin{equation}
\label{Eq_vV}
\int_{\Omega}d^4x\,\partial_{\mu}P_e^{\mu}(x;T_0)=\int_{\Sigma(\Omega)}d\Sigma_{\mu}P_e^{\mu}(x;T_0),
\end{equation}
where $\Sigma(\Omega)$ is the three-dimensional surface enclosing the four-dimensional tube $\Omega$, with $d\Sigma^{\mu}$ pointing outwards. Now, the integral over the infinitesimal four-volume on the left-hand side of this equation can be written as
\begin{equation}
\int_{\Omega}d^4x\,\partial_{\mu}P_e^{\mu}(x;T_0)=\int_{\tau_1}^{\tau_2}d\tau\,dV_r(\tau)\partial_{\mu}P_e^{\mu}(x(\tau;T_0);T_0)=\int_{\tau_1}^{\tau_2}d\tau\,dV_r(\tau)\partial_{\mu}\Pi_e^{\mu}(\tau;T_0),
\end{equation}
where, using the fact that the infinitesimal four-volume $d^4x$ is a Lorentz scalar quantity, we have written it as the product of the proper time $d\tau$ times the three-dimensional volume $dV_r(\tau)$ in the instantaneous rest frame of the electron at $\tau$. Now, the side part of the three-dimensional surface $\Sigma(\Omega)$ is perpendicular to the four-momentum by definition. By exploiting the fact that the quantity $d\Sigma_{\mu}P_e^{\mu}(x;T_0)$ is a Lorentz scalar quantity and by recalling that the surface orientation is outward $\Omega$, we can write the right-hand side of Eq. (\ref{Eq_vV}) as 
\begin{equation}
\int_{\Sigma(\Omega)}d\Sigma_{\mu}P_e^{\mu}(x;T_0)=m[dV_r(\tau_2)-dV_r(\tau_1)].
\end{equation}
By taking now $\tau_2$ to be larger than $\tau_1$ by an infinitesimal amount $d\tau$, we obtain the Lorentz-invariant equation
\begin{equation}
\label{Eq_vV_diff}
m\frac{d\, dV_r(\tau)}{d\tau}=dV_r(\tau)\partial_{\mu}\Pi_e^{\mu}(\tau;T_0),
\end{equation}
where, we recall, the spacetime point $x$ corresponds to the four-position of the electron at the proper time $\tau$. In order to express the infinitesimal volume $dV_r(\tau)$ in the instantaneous rest frame of the electron in terms of the corresponding infinitesimal volume $d^3x_{\text{lc}}(\tau;T_0)$ in the laboratory frame, we notice that $d\tau dV_r(\tau)=dTd^3x_{\text{lc}}(\tau;T_0)$ and we obtain $dV_r(\tau)=d^3x_{\text{lc}}(\tau;T_0)\Pi_{e,+}(\tau;T_0)/m$. Thus, by integrating Eq. (\ref{Eq_vV_diff}) between the initial proper time $0$ and a generic proper time $\tau$, we obtain
\begin{equation}
\label{vV_det}
\frac{d^3x_{\text{lc}}(0;T_0)}{d^3x_{\text{lc}}(\tau;T_0)}=\frac{\Pi_{e,+}(\tau;T_0)}{p_+}e^{-\frac{1}{m}\int_0^{\tau}d\tau'(\partial \Pi_e)}.
\end{equation}
Finally, by recalling that 
\begin{equation}
\frac{d^3x_{\text{lc}}(0;T_0)}{d^3x_{\text{lc}}(\tau;T_0)}=\left\vert\det\left(\frac{\partial x_{\text{lc},0,i}}{\partial x_{\text{lc},j}}\right)\right\vert,
\end{equation}
we conclude that the van Vleck determinant $\Delta_e(\tau;T_0)=D_e(x(\tau;T_0);T_0)$ along that trajectory, can be expressed as
\begin{equation}
\label{vV_tau}
\Delta_e(\tau;T_0)=\frac{\Pi_{e,+}(\tau;T_0)}{p_+}e^{-\frac{1}{m}\int_0^{\tau}d\tau'(\partial \Pi_e)},
\end{equation}
which shows that it indeed never vanishes.

Another interesting relation can be obtained by noticing that the van Vleck determinant satisfies the equation
\begin{equation}
\begin{split}
\frac{d}{d\tau}\left[\frac{\Delta_e(\tau;T_0)}{\Pi_{e,+}(\tau;T_0)}\right]&=\partial_{\mu}\left[\frac{D_e(x(\tau;T_0);T_0)}{P_{e,+}(x(\tau;T_0);T_0)}\right]\frac{P_e^{\mu}(x(\tau;T_0);T_0)}{m}=-\frac{(\partial \Pi_e)}{m}\frac{\Delta_e(\tau;T_0)}{\Pi_{e,+}(\tau;T_0)}\\
&=-\frac{\partial_{\mu}P_e^{\mu}(x(\tau;T_0);T_0)}{m}\frac{D_e(x(\tau;T_0);T_0)}{P_{e,+}(x(\tau;T_0);T_0)},
\end{split}
\end{equation}
where in the second equality we have used Eq. (\ref{vV_tau}). This equation, in fact, allows one to construct a conserved quantity as it implies that the four-current
\begin{equation}
\label{J_tau}
\mathcal{J}_e^{\mu}(x;T_0)=D_e(x;T_0)\frac{p_+}{P_{e,+}(x;T_0)}\frac{P_e^{\mu}(x;T_0)}{m},
\end{equation}
computed at a generic spacetime point $x$ is divergenceless, i.e., $(\partial\mathcal{J}_e(x;T_0))=0$.

We conclude by observing, as in the previous paragraph, that all the considerations done in terms of the proper time $\tau$ can be also carried out directly in terms of the light-cone time $T$, which is what we will ultimately use below. For example, the expressions of the van Vleck determinant at the spacetime point $x$ and of the four-current in Eq. (\ref{J_tau}) become
\begin{align}
\label{vV_T}
D_e(x;T_0)&=\frac{P_{e,+}(x;T_0)}{p_+}e^{-\int_{T_0}^T\frac{dT'}{P_{e,+}}(\partial P_e)},\\
\label{J_T}
\mathcal{J}_e^{\mu}(x;T_0)&=\frac{P_e^{\mu}(x;T_0)}{m}e^{-\int_{T_0}^T\frac{dT'}{P_{e,+}}(\partial P_e)}.
\end{align}

\section{Electron states}
\label{ES}

In the present section we derive the electron states both with positive and negative energies in the presence of an arbitrary spacetime shaped laser beam within the WKB approach. 

The starting point is the Dirac equation
\begin{equation}
\label{Dirac}
[\gamma^{\mu}(i\hbar\partial_{\mu}-eA_{\mu})-m]\Psi=0,
\end{equation}
for the Dirac spinor field $\Psi(x)$. Based on the general argument that the de Broglie wavelength of an ultrarelativistic particle is small, we apply the WKB method \cite{Landau_b_3_1977} and we look for a solution of the Dirac equation of the form \cite{Pauli_1932,Rubinow_1963,Di_Piazza_2014}
\begin{equation}
\label{Psi}
\Psi(x)=e^{\frac{i}{\hbar}S(x)}\Theta(x).
\end{equation}
Thus, the spinor $\Theta(x)$ has to fulfill the equation
\begin{equation}
\label{Dirac_Theta}
[\gamma^{\mu}(\partial_{\mu}S+eA_{\mu})+m]\Theta=i\hbar\gamma^{\mu}\partial_{\mu}\Theta.
\end{equation}
So far, the method is the same as that described in Refs. \cite{Pauli_1932,Rubinow_1963,Di_Piazza_2014}. Following those references one can first neglect the term proportional to $\hbar$ and obtains the equation
\begin{equation}
\label{Dirac_Theta_0}
[\gamma^{\mu}(\partial_{\mu}S+eA_{\mu})+m]\Theta^{(0)}=0
\end{equation}
for the zeroth-order spinor $\Theta^{(0)}(x)$. This equation admits a nontrivial solution only if $\det[\gamma^{\mu}(\partial_{\mu}S+eA_{\mu})+m]=0$, which implies that the quantity $S(x)$ has to satisfy the Hamilton-Jacobi equation (\ref{HJ}) and that it can be identified with the classical action \cite{Landau_b_2_1975}. However, at this point it is inconvenient to proceed and account for higher orders in $\hbar$ because the matrix on the left-hand side of Eq. (\ref{Dirac_Theta}) cannot be inverted. Thus, starting from Eq. (\ref{Dirac_Theta}), we follow a different procedure.

By recalling the general idea of conveniently transforming the Dirac equation into a second-order differential equation \cite{Landau_b_4_1982}, we look for a solution of the form
\begin{equation}
\label{Theta}
\Theta(x)=\frac{1}{2m}\{\gamma^{\mu}[-\partial_{\mu}S(x)-eA_{\mu}(x)+i\hbar\partial_{\mu}]+m\}\Phi(x),
\end{equation}
such that the spinor $\Phi(x)$ fulfills the equation
\begin{equation}
\label{Dirac_Phi_i}
\left\{(\partial S+eA)^2-m^2-i\hbar\left[2(\partial_{\mu}S+eA_{\mu})\partial^{\mu}+\partial_{\mu}(\partial^{\mu}S+eA^{\mu})-\frac{ie}{2}\sigma^{\mu\nu}F_{\mu\nu}\right]-\hbar^2\square\right\}\Phi=0,
\end{equation}
where $\sigma^{\mu\nu}=(i/2)[\gamma^{\mu},\gamma^{\nu}]$ and $\square=\partial_{\mu}\partial^{\mu}$. This equation is suitable for a perturbative expansion in $\hbar$ and at the leading order in $\hbar$, i.e., by setting $\hbar$ equal to zero, one recovers the condition that $S(x)$ corresponds to the classical action. 

So far the method applies to both positive- and negative-energy states. Below we consider separately these two cases and we start from the positive-energy states.

\subsection{Positive-energy states}
The distinction between positive- and negative-energy states takes place already at the leading order in $\hbar$. In fact, the electron states with positive energies are identified by choosing $S(x)$ as the classical action $S_p(x;T_0)$, discussed in the previous section, which fulfills the initial condition $S_p(T_0,\bm{x}_{\text{lc}};T_0)=-(p_+\phi+p_-T_0-\bm{p}_{\perp}\cdot\bm{x}_{\perp})$ at the fixed light-cone time $T_0$ for a given on-shell four-momentum $p^{\mu}=(\varepsilon,\bm{p})$, with positive energy $\varepsilon=\sqrt{m^2+\bm{p}^2}$, or, in light-cone components, $p_-=(m^2+\bm{p}_{\perp}^2)/2p_+$, with $p_{\pm}>0$. 

Once we have identified the function $S(x)$ with the classical action $S_p(x;T_0)$, Eq. (\ref{Dirac_Phi_i}) is solved by an arbitrary spinor $\Phi(x)$ for $\hbar=0$. This is expected because Eq. (\ref{Dirac_Theta}) for $\hbar=0$ only implies that $[\hat{P}_e(x;T_0)-m]\Theta^{(0)}(x)=0$ [see also Eq. (\ref{Dirac_Theta_0}) and recall that $P_e^{\mu}(x;T_0)=-\partial^{\mu}S_p(x;T_0)-eA^{\mu}(x)$], which is fulfilled for an arbitrary choice of $\Phi(x)$ [see Eq. (\ref{Theta}) for $\hbar=0$]. Thus, we proceed further and we rewrite Eq. (\ref{Dirac_Phi_i}) taking into account that the function $S(x)$ has been chosen as the classical action $S_p(x;T_0)$. By indicating the corresponding spinor as $\Phi_p(x;T_0)$, the term independent of $\hbar$ in Eq. (\ref{Dirac_Phi_i}) identically vanishes and that equation becomes
\begin{equation}
\label{Dirac_Phi}
\left(2P_{e,\mu}\partial^{\mu}+(\partial P_{e})+\frac{ie}{2}\sigma^{\mu\nu}F_{\mu\nu}+i\hbar\square\right)\Phi_p=0.
\end{equation}
At the leading order in $\hbar$, i.e., by imagining to expand the spinor $\Phi_p(x;T_0)$ in powers of $\hbar$ starting from $\Phi_p^{(0)}(x;T_0)$ and by setting $\hbar=0$ in Eq. (\ref{Dirac_Phi}), we obtain an equation for $\Phi_p^{(0)}(x;T_0)$:
\begin{equation}
\label{Dirac_Phi_0}
\left(2P_{e,\mu}\partial^{\mu}+(\partial P_e)+\frac{ie}{2}\sigma^{\mu\nu}F_{\mu\nu}\right)\Phi_p^{(0)}=0.
\end{equation}
In order to solve this equation, we notice that the differential operator acting on $\Phi_p^{(0)}(x;T_0)$ commutes with the matrix $\hat{P}_e(x;T_0)$:
\begin{equation}
\begin{split}
\left[2P_{e,\mu}\partial^{\mu}+(\partial P_e)+\frac{ie}{2}\sigma^{\mu\nu}F_{\mu\nu},\hat{P}_e\right]&=2P_{e,\mu}\gamma_{\nu}\partial^{\mu}P_e^{\nu}-i[\sigma^{\mu\nu}\partial_{\mu}P_{e,\nu},\hat{P}_e]\\
&=2P_{e,\mu}\gamma_{\nu}\partial^{\mu}P_e^{\nu}+[(\eta^{\mu\nu}-\gamma^{\nu}\gamma^{\mu})\partial_{\mu}P_{e,\nu},\hat{P}_e]\\
&=2P_{e,\mu}\gamma_{\nu}\partial^{\mu}P_e^{\nu}-2\gamma^{\nu}P_e^{\mu}\partial_{\mu}P_{e,\nu}+2P_e^{\nu}\gamma^{\mu}\partial_{\mu}P_{e,\nu}=0,
\end{split}
\end{equation}
where in the last step we have used the fact that $P^2_e(x;T_0)=m^2$. In this way, we can choose 
$\Phi_p^{(0)}(x;T_0)$ to be an eigenstate of the matrix $\hat{P}_e(x;T_0)$ and we require that $\hat{P}_e(x;T_0)\Phi_p^{(0)}(x;T_0)=m\Phi_p^{(0)}(x;T_0)$. It is known from the free theory that the solution of the equation $\hat{P}_e(x;T_0)\Phi_p^{(0)}(x;T_0)=m\Phi_p^{(0)}(x;T_0)$ can be written as \cite{Landau_b_4_1982}
\begin{equation}
\label{Phi_0}
\begin{split}
\Phi^{(0)}_p(x;T_0)&=\frac{1}{\sqrt{2p_+\mathcal{V}_0}}\begin{pmatrix}
\sqrt{\mathcal{E}_e(x;T_0)+m}w_p(x;T_0)\\
\frac{\bm{P}_e(x;T_0)\cdot\bm{\sigma}}{\sqrt{\mathcal{E}_e(x;T_0)+m}}w_p(x;T_0)
\end{pmatrix}\\
&=\frac{\hat{P}_e(x;T_0)+m}{\sqrt{2p_+\mathcal{V}_0[\mathcal{E}_e(x;T_0)+m]}}\begin{pmatrix}
w_p(x;T_0)\\
0
\end{pmatrix},
\end{split}
\end{equation}
where $\bm{\sigma}$ are the Pauli matrices, $w_p(x;T_0)$ is an arbitrary two-dimensional spinor, and where the light-cone normalization volume $\mathcal{V}_0$ along the perpendicular and the $\phi$ directions has been introduced.

Now, it can be shown that, if $\Phi_p^{(0)}(x;T_0)$ satisfies the equation $\hat{P}_e(x;T_0)\Phi_p^{(0)}(x;T_0)=m\Phi_p^{(0)}(x;T_0)$, then Eq. (\ref{Dirac_Phi_0}) can be written as $[\hat{P}_e(x;T_0)+m]\gamma^{\mu}\partial_{\mu}\Phi_p^{(0)}=0$ [which coincides with Eq. (9) of the Supplemental Material of Ref. \cite{Di_Piazza_2014}]. Moreover, if $\Phi_p^{(0)}(x;T_0)$ has the form in Eq. (\ref{Phi_0}), then Eq. (\ref{Dirac_Phi_0}) becomes
\begin{equation}
\frac{1}{\sqrt{2p_+\mathcal{V}_0}}\begin{pmatrix}
\sqrt{\mathcal{E}_e+m}\left[2P_e^{\mu}\partial_{\mu}+(\partial_{\mu}P_e^{\mu})-ie\bm{\sigma}\cdot\left(\bm{B}-\frac{\bm{P}_e\times\bm{E}}{\mathcal{E}_e+m}\right)\right]w_p\\
\frac{\bm{P}_e\cdot\bm{\sigma}}{\sqrt{\mathcal{E}_e+m}}\left[2P_e^{\mu}\partial_{\mu}+(\partial_{\mu}P_e^{\mu})-ie\bm{\sigma}\cdot\left(\bm{B}-\frac{\bm{P}_e\times\bm{E}}{\mathcal{E}_e+m}\right)\right]w_p
\end{pmatrix}=0,
\end{equation}
which is satisfied if the two-dimensional spinor $w_p(x;T_0)$ satisfies the equation
\begin{equation}
\label{Eq_w}
P_e^{\mu}\partial_{\mu}w_p=-\frac{1}{2}(\partial_{\mu}P_e^{\mu})w_p+\frac{ie}{2}\bm{\sigma}\cdot\left(\bm{B}-\frac{\bm{P}_e\times\bm{E}}{\mathcal{E}_e+m}\right)w_p,
\end{equation}
where $\bm{E}(x)$ and $\bm{B}(x)$ are the background electric and magnetic field, respectively. This equation can be solved by means of the method of characteristics by introducing the electron's proper time $\tau$ according to the equation $dx^{\mu}/d\tau=\Pi_e^{\mu}/m$ as explained in the previous section (see also Ref. \cite{Fock_1937} for an alternative method to solve the Dirac equation based on the introduction of the particle proper time). By setting
\begin{equation}
\label{w_rho}
w_p(x;T_0)=e^{-\frac{1}{2m}\int_0^{\tau}d\tau'(\partial P_e)}r_p(x;T_0)=e^{-\frac{1}{2}\int_{T_0}^T\frac{dT'}{P_{e,+}}(\partial P_e)}r_p(x;T_0),
\end{equation}
we conclude that the two-component spinor $r_p(x;T_0)$ satisfies the equation [see Eq. (\ref{Eq_w})]
\begin{equation}
\label{Eq_rho}
P_e^{\mu}\partial_{\mu}r_p=\frac{ie}{2}\bm{\sigma}\cdot\left(\bm{B}-\frac{\bm{P}_e\times\bm{E}}{\mathcal{E}_e+m}\right)r_p.
\end{equation}
By using again the method of characteristics, one first defines the function $\rho_{p,s}(\tau;T_0)=r_{p,s}(x(\tau;T_0);T_0)$ along a generic trajectory and then, by determining the actual trajectories $x^{\mu}=x^{\mu}(\tau;T_0,\bm{x}_{0,\text{lc}},\bm{p}_{\text{lc}})$ depending on the initial conditions as $x^{\mu}(0;T_0,\bm{x}_{0,\text{lc}},\bm{p}_{\text{lc}})=x_0^{\mu}$, one finds the solution $\rho_{p,s}(\tau;T_0)=\rho_{p,s}(\tau;T_0,\bm{x}_{0,\text{lc}},\bm{p}_{\text{lc}})$ of Eq. (\ref{Eq_rho}) along the generic trajectory and then obtains $r_{p,s}(x;T_0)=\rho_{p,s}(\tau(x;T_0,\bm{p}_{\text{lc}});T_0,\bm{x}_{0,\text{lc}}(x;T_0,\bm{p}_{\text{lc}}),\bm{p}_{\text{lc}})$. In this way, one can easily show from this equation that the two-component spinor $r_p(x;T_0)$ has constant norm $r^{\dag}_p(x;T_0)r_p(x;T_0)$, which can then be set equal to unity. Moreover, the average vector $\bm{s}_p(x;T_0)=r_p^{\dag}(x;T_0)\bm{\sigma}r_p(x;T_0)$ satisfies the equation
\begin{equation}
\label{Eq_zeta}
P_e^{\mu}\partial_{\mu}\bm{s}_p=e\bm{s}_p\times\left(\bm{B}-\frac{\bm{P}_e\times\bm{E}}{\mathcal{E}_e+m}\right),
\end{equation}
which is equivalent to the Bargmann-Michel-Telegdi (BMT) equation $mds_p^{\mu}/d\tau=eF^{\mu\nu}s_{p,\nu}$ for the electron polarization four-vector \cite{Landau_b_4_1982}
\begin{equation}
s_p^{\mu}(x;T_0)=\left(\frac{\bm{s}_p(x;T_0)\cdot\bm{P}_e(x;T_0)}{m},\bm{s}_p(x;T_0)+\frac{\bm{s}_p(x;T_0)\cdot\bm{P}_e(x;T_0)}{m[\mathcal{E}_e(x;T_0)+m]}\bm{P}_e(x;T_0)\right),
\end{equation}
computed along the electron trajectory. In order to identify the two-component spinor $r_p(x;T_0)$ uniquely, initial conditions have to be assigned. By choosing for the sake of definiteness the direction of the momentum $\bm{p}$ as spin-quantization direction at $T_0$, we are led to introduce the discrete spin quantum number $s=\pm 1$, depending on the two possible orientations of the vector $\bm{s}_p(T_0,\bm{x}_{\text{lc}};T_0)$, and we correspondingly indicate the spinor $\Phi_p(x;T_0)$ as $\Phi_{p,s}(x;T_0)$ and analogously $\Phi^{(0)}_p(x;T_0)$ as $\Phi^{(0)}_{p,s}(x;T_0)$ (as for the initial four-momentum $p^{\mu}$, all characteristic curves are characterized by the same spin quantum number $s$). In this way, we conclude that ignoring the term linear in $\hbar$ in Eq. (\ref{Dirac_Phi}) amounts to determine the dynamics of the electron spin or, more precisely, to the electron magnetic moment associated with its spin according to the classical BMT equation (with the electron gyromagnetic equal to two). Indeed, we have seen that the resulting equation (\ref{Dirac_Phi_0}) turns out to be independent of $\hbar$. Consequently the spinor $\Phi^{(0)}_{p,s}(x;T_0)$ does not depend on $\hbar$ either and this is why we continue to indicate it with the upper index $(0)$, i.e.,
\begin{equation}
\label{Phi_0_1}
\Phi^{(0)}_{p,s}(x;T_0)=\frac{e^{-\frac{1}{2}\int_{T_0}^T\frac{d\tilde{T}}{P_{e,+}}(\partial P_e)}}{\sqrt{2p_+\mathcal{V}_0}}\begin{pmatrix}
\sqrt{\mathcal{E}_e(x;T_0)+m}r_{p,s}(x;T_0)\\
\frac{\bm{P}_e(x;T_0)\cdot\bm{\sigma}}{\sqrt{\mathcal{E}_e(x;T_0)+m}}r_{p,s}(x;T_0)
\end{pmatrix}.
\end{equation}
As we have already mentioned, Eq. (\ref{Dirac_Phi_0}) corresponds to the next-to-leading order approximation of the original Eq. (\ref{Dirac_Phi}). This explains why the exponential function in Eq. (\ref{Phi_0_1}) has to be interpreted as the leading-order (imaginary) quantum correction to the phase of the electron wave function, whose WKB leading order is $S_p(x;T_0)/\hbar$, i.e., it is $O(\hbar^{-1})$ [see Eq. (\ref{Psi})].

At this point, the higher-order corrections in $\hbar$ of the solution of Eq. (\ref{Dirac_Phi}) can be written in a formal way by introducing the proper-time evolution operator $\mathcal{U}_e(\tau,\tau')$, which solves the equation
\begin{equation}
2m\frac{d\mathcal{U}_e}{d\tau}=-\left((\partial P_{e})+\frac{ie}{2}\sigma^{\mu\nu}F_{\mu\nu}\right)\mathcal{U}_e,
\end{equation}
with the initial condition $\mathcal{U}_e(\tau,\tau)=1$. In general, one may need the operator $\mathcal{U}_e(\tau,\tau')$ for arbitrary proper times $\tau$ and $\tau'$. Therefore, by employing both the time ordering operator $\mathcal{T}_>$ and the time antiordering operator $\mathcal{T}_<$, the operator $\mathcal{U}_e(\tau,\tau')$ can be written as
\begin{equation}
\mathcal{U}_e(\tau,\tau')=e^{-\frac{1}{2m}\int_{\tau'}^{\tau}d\tilde{\tau}(\partial P_e)}\left[\theta(\tau-\tau')\mathcal{T}_>\left(e^{-\frac{ie}{4m}\int_{\tau'}^{\tau}d\tilde{\tau}\sigma^{\mu\nu}F_{\mu\nu}}\right)+\theta(\tau'-\tau)\mathcal{T}_<\left(e^{-\frac{ie}{4m}\int_{\tau'}^{\tau}d\tilde{\tau}\sigma^{\mu\nu}F_{\mu\nu}}\right)\right],
\end{equation}
where $\theta(\cdot)$ is the step function. By using this expression, one can formally write the exact solution of Eq. (\ref{Dirac_Phi}) in the form
\begin{equation}
\label{Phi}
\Phi_{p,s}(x;T_0)=\Phi^{(0)}_{p,s}(x;T_0)-\frac{i\hbar}{2}\int_{T_0}^T\frac{dT'}{P_{e,+}}\mathcal{U}_e(T,T')\square\Phi_{p,s},
\end{equation}
where
\begin{equation}
\begin{split}
\mathcal{U}_e(T,T')&=e^{-\frac{1}{2}\int_{T'}^T\frac{d\tilde{T}}{P_{e,+}}(\partial P_e)}\left[\theta(T-T')\mathcal{T}_>\left(e^{-\frac{ie}{4}\int_{T'}^T\frac{d\tilde{T}}{P_{e,+}}\sigma^{\mu\nu}F_{\mu\nu}}\right)\right.\\
&\quad+\left.\theta(T'-T)\mathcal{T}_<\left(e^{-\frac{ie}{4}\int_{T'}^T\frac{d\tilde{T}}{P_{e,+}}\sigma^{\mu\nu}F_{\mu\nu}}\right)\right],
\end{split}
\end{equation}
which is particularly suitable for an expansion in $\hbar$ [with an abuse of notation we have replaced in $\mathcal{U}_e(\tau,\tau')$ the dependence on $\tau$ and $\tau'$ with the dependence on $T$ and $T'$]. Once the spinor $\Phi_{p,s}(x;T_0)$ is obtained, one can use the definitions in Eqs. (\ref{Theta}) and (\ref{Psi}) to obtain the corresponding positive-energy solution of the Dirac equation, which can be denoted as $U_{p,s}(x;T_0)$. In general, in order to apply the Furry picture for computing transition probabilities one needs so-called in- and out-states, which reduce to free states at asymptotic early and late times, respectively. By using the upper index $(\text{in})$ [$(\text{out})$] to the denote the quantities which reduce to the corresponding free ones in the asymptotic past [future], we have that $U^{(\text{in})}_{p,s}(x)=\lim_{T_0\to-\infty}U_{p,s}(x;T_0)$ [$U^{(\text{out})}_{p,s}(x)=\lim_{T_0\to\infty}U_{p,s}(x;T_0)$].

Now, we focus onto the zeroth-order solution, in the sense explained above, and notice from Eq. (\ref{Theta}) that at the leading order in $\hbar$ it is $\Theta_{p,s}^{(0)}(x;T_0)=\Phi_{p,s}^{(0)}(x;T_0)$, because in the preexponential spinor corrections proportional to $\hbar$ can be neglected. In this way, we conclude that the positive-energy electron state $U^{(0)}_{p,s}(x;T_0)$ including terms up to $O(\hbar^0)$ is given by
\begin{equation}
\label{U_T_0}
U^{(0)}_{p,s}(x;T_0)=e^{\frac{i}{\hbar}S_p(x;T_0)}\frac{e^{-\frac{1}{2}\int_{T_0}^T\frac{d\tilde{T}}{P_{e,+}}(\partial P_e)}}{\sqrt{2p_+\mathcal{V}_0}}\begin{pmatrix}
\sqrt{\mathcal{E}_e(x;T_0)+m}r_{p,s}(x;T_0)\\
\frac{\bm{P}_e(x;T_0)\cdot\bm{\sigma}}{\sqrt{\mathcal{E}_e(x;T_0)+m}}r_{p,s}(x;T_0)
\end{pmatrix}.
\end{equation}
We first discuss the normalization of the state $U^{(0)}_{p,s}(x;T_0)$ by computing the four-current $J^{(0)\,\mu}_{e;p,s}(x;T_0)=\bar{U}^{(0)}_{p,s}(x;T_0)\gamma^{\mu}U^{(0)}_{p,s}(x;T_0)$. By recalling the properties of the free states \cite{Landau_b_4_1982}, one easily obtains
\begin{equation}
\label{J_e_0}
J^{(0)\,\mu}_{e;p,s}(x;T_0)=\frac{P^{\mu}_e(x;T_0)}{p_+\mathcal{V}_0}e^{-\int_{T_0}^T\frac{d\tilde{T}}{P_{e,+}}(\partial P_e)}=\frac{P^{\mu}_e(x;T_0)}{P_{e,+}(x;T_0)}\frac{D_e(x;T_0)}{\mathcal{V}_0},
\end{equation}
where in the second equality we have used the expression of the van Vleck determinant in Eq. (\ref{vV_T}). This four-current has the desired properties that it corresponds to the standard normalization of one particle in the light-cone volume $\mathcal{V}_0$ at the initial light-cone time $T_0$ and that, due to Eq. (\ref{J_T}), is conserved: $(\partial J^{(0)}_{e;p,s})=0$. Indeed, this shows the importance of including the corrections proportional to $\hbar^0$ in the quasiclassical wave functions, which was the original motivation of the work \cite{van_Vleck_1928} by van Vleck. Also, the expression in Eq. (\ref{J_e_0}) of the electron four-current suggests to interpret the light-cone volume $\mathcal{V}_0$ as [see also the discussions below Eqs. (\ref{Q_C}) and (\ref{Q_BW})]
\begin{equation}
\label{V}
\mathcal{V}_0=\int_{\mathcal{V}_0} d^3x_{0,\text{lc}}=\int_{\mathcal{V}_T} d^3x_{\text{lc}}\,D_e(x;T_0)=\int_{\mathcal{V}_T} d^3x_{\text{lc}}\,\frac{P_{e,+}(x;T_0)}{p_+}e^{-\int_{T_0}^T\frac{d\tilde{T}}{P_{e,+}}(\partial P_e)},
\end{equation}
where $\mathcal{V}_T$ is the transformed light-cone volume of $\mathcal{V}_0$, such that the wave function (\ref{U_T_0}) can be cast in the form
\begin{equation}
U^{(0)}_{p,s}(x;T_0)=\sqrt{\frac{D_e(x;T_0)}{\mathcal{V}_0}}\frac{e^{\frac{i}{\hbar}S_p(x;T_0)}}{\sqrt{2P_{e,+}(x;T_0)}}\begin{pmatrix}
\sqrt{\mathcal{E}_e(x;T_0)+m}r_{p,s}(x;T_0)\\
\frac{\bm{P}_e(x;T_0)\cdot\bm{\sigma}}{\sqrt{\mathcal{E}_e(x;T_0)+m}}r_{p,s}(x;T_0)
\end{pmatrix},
\end{equation}
which precisely corresponds to the nonrelativistic form derived by van Vleck in Ref. \cite{van_Vleck_1928}. Finally, we note that the same normalization as above is customarily employed also in the case of a plane wave (see below). 

The state in Eq. (\ref{U_T_0}) can be written in an alternative and physically suggestive form by recalling that all the quantities are assumed to be computed by means of the method of characteristics, which only relies on the classical electron trajectory obtained by solving the Lorentz equation [and on the solutions of the equations for the action $S_p(x;T_0)$ and for the two-dimensional spinor $r_{p,s}(x;T_0)$]. By recalling the expression of the van Vleck determinant and the identity (\ref{vV_det}), we obtain
\begin{equation}
\begin{split}
U^{(0)}_{p,s}(x;T_0)&=\int_{\mathcal{V}_0} \frac{d^3 x_{0,\text{lc}}}{\Delta_e(T;T_0)}\delta^3(\bm{x}_{\text{lc}}-\bm{x}_{\text{lc}}(T;T_0,\bm{x}_{0,\text{lc}},\bm{p}_{\text{lc}}))\\
&\quad\times e^{\frac{i}{\hbar}\Sigma_p(T;T_0)}\frac{e^{-\frac{1}{2}\int_{T_0}^T\frac{d\tilde{T}}{\Pi_{e,+}}(\partial \Pi_e)}}{\sqrt{2p_+\mathcal{V}_0}}\begin{pmatrix}
\sqrt{\Pi_{e,0}(T;T_0)+m}\rho_{p,s}(T;T_0)\\
\frac{\bm{\Pi}_e(T;T_0)\cdot\bm{\sigma}}{\sqrt{\Pi_{e,0}(T;T_0)+m}}\rho_{p,s}(T;T_0)
\end{pmatrix}\\
&=\int_{\mathcal{V}_0} \frac{d^3 x_{0,\text{lc}}}{\sqrt{\Delta_e(T;T_0)}}\delta^3(\bm{x}_{\text{lc}}-\bm{x}_{\text{lc}}(T;T_0,\bm{x}_{0,\text{lc}},\bm{p}_{\text{lc}}))\\
&\quad\times\frac{e^{\frac{i}{\hbar}\Sigma_p(T;T_0)}}{\sqrt{2\Pi_{e,+}(T;T_0)\mathcal{V}_0}}\begin{pmatrix}
\sqrt{\Pi_{e,0}(T;T_0)+m}\rho_{p,s}(T;T_0)\\
\frac{\bm{\Pi}_e(T;T_0)\cdot\bm{\sigma}}{\sqrt{\Pi_{e,0}(T;T_0)+m}}\rho_{p,s}(T;T_0)
\end{pmatrix},
\end{split}
\end{equation}
where the electron trajectory $\bm{x}_{\text{lc}}=\bm{x}_{\text{lc}}(T;T_0,\bm{x}_{0,\text{lc}},\bm{p}_{\text{lc}})$ is the one corresponding to the initial condition $\bm{x}_{\text{lc}}(0)=\bm{x}_{0,\text{lc}}$ on the light-cone position, parametrized with respect to the light-cone $T$ and where, as in the case of the evolution operator $\mathcal{U}_e(T,T')$, we used the same symbol for the functions along the trajectory for either the proper time or the light-cone time being used to parametrize it. In this way, the state $U^{(0)}_{p,s}(x;T_0)$ with the given on-shell four-momentum $p^{\mu}$ and spin quantum number $s$ is represented as an infinite linear combination of ``freelike'' states with the free action, four-momentum, and spin four-vector replaced with the corresponding quantities evaluated along all possible classical electron trajectories in the external field corresponding to arbitrary initial positions, each contribution being weighted via the inverse square root of the van Vleck determinant.

Starting from Eq. (\ref{U_T_0}), the positive-energy electron in-states and out-states including terms up to $O(\hbar^0)$ are given by
\begin{align}
\label{U_in}
U^{(0,\text{in})}_{p,s}(x)&=e^{\frac{i}{\hbar}S_p^{(\text{in})}(x)}\frac{e^{-\int_{-\infty}^Td\tilde{T}(\partial P^{(\text{in})}_e)/2P^{(\text{in})}_{e,+}}}{\sqrt{2p_+\mathcal{V}_0}}u^{(0,\text{in})}_{p,s}(x),\\
\label{U_out}
U^{(0,\text{out})}_{p,s}(x)&=e^{\frac{i}{\hbar}S_p^{(\text{out})}(x)}\frac{e^{-\int_{\infty}^Td\tilde{T}(\partial P^{(\text{out})}_e)/2P^{(\text{out})}_{e,+}}}{\sqrt{2p_+\mathcal{V}_0}}u^{(0,\text{out})}_{p,s}(x),
\end{align}
where
\begin{align}
\label{u_in}
u^{(0,\text{in})}_{p,s}(x)&=\begin{pmatrix}
\sqrt{\mathcal{E}_e^{(\text{in})}(x)+m}r^{(\text{in})}_{p,s}(x)\\
\frac{\bm{P}^{(\text{in})}_e(x)\cdot\bm{\sigma}}{\sqrt{\mathcal{E}_e^{(\text{in})}(x)+m}}r^{(\text{in})}_{p,s}(x)
\end{pmatrix},\\
\label{u_out}
u^{(0,\text{out})}_{p,s}(x)&=\begin{pmatrix}
\sqrt{\mathcal{E}_e^{(\text{out})}(x)+m}r^{(\text{out})}_{p,s}(x)\\
\frac{\bm{P}^{(\text{out})}_e(x)\cdot\bm{\sigma}}{\sqrt{\mathcal{E}_e^{(\text{out})}(x)+m}}r^{(\text{out})}_{p,s}(x)
\end{pmatrix}.
\end{align}
As we have discussed in Ref. \cite{Di_Piazza_2015}, by approximating the electron trajectory and the spin evolution up to the leading order in the parameter $\eta$ for the electron trajectory, it is possible to put the above states in a form, which is reminiscent of the Volkov states, for computing the probabilities of strong-field QED processes. 

In the next two paragraphs we discuss a new expression of the Volkov states and the conditions of validity of approximations used to obtain the states in Eqs. (\ref{U_in})-(\ref{U_out}).

\subsubsection{A fully quasiclassical form of the Volkov states}
It is often stated that Volkov states have the unique feature of having a quasiclassical structure although they are an exact solution of the Dirac equation \cite{Ritus_1985,Di_Piazza_2012}. It is certainly true that Volkov states feature the typical exponential of the classical action divided by $\hbar$, which is typical, as we have also seen above, of quasiclassical wave functions. However, it is not correspondingly evident that the spinor structure of the Volkov states, in the form in which they have been written (see, e.g., \cite{Landau_b_4_1982}), is also quasiclassical, as it does not resemble, for example, Eq. (\ref{U_T_0}). Here, we show that our method allows one to write the Volkov states in a new, closed form, whose spinor structure is also manifestly quasiclassical. This also explicitly implies that the present method provides the exact Volkov wave functions in the case of a background plane wave.

According to the above notation, we assume that the background field only depends on the coordinate $T$, i.e., it is of the form $A^{\mu}(T)$. By assuming the same structure of the states until Eq. (\ref{Dirac_Phi}), we notice that one can choose the spinor $\Phi_p(x;T_0)$ to depend only on $T$ in this case. Indeed, from the results in the Appendix \ref{App_A}, one obtains that the four-momentum $P_e^{\mu}(x;T_0)$ only depends on the coordinate $T$ and it formally coincides with the solution $\Pi_e^{\mu}(T;T_0)$ of the Lorentz equation reported in the Appendix \ref{App_A}. For this reason, we indicate the four-momentum $P_e^{\mu}(x;T_0)$ as $P_e^{\mu}(T;T_0)$. As a result, the last term proportional to $\hbar$ in Eq. (\ref{Dirac_Phi}) identically vanishes and Eq. (\ref{Dirac_Phi_0}) is exact in this case. Moreover, since the plus component of the four-momentum is a constant of motion, the term $(\partial P_e)$ also vanishes (see the Appendix \ref{App_A}). By writing again the spinor $\Phi_p(T;T_0)$ as in Eq. (\ref{Phi_0}) but with all quantities only depending only on $T$, we obtain
\begin{equation}
\label{Phi_0_PW}
\Phi_p(T;T_0)=\frac{1}{\sqrt{2p_+\mathcal{V}_0}}\begin{pmatrix}
\sqrt{\mathcal{E}_e(T;T_0)+m}r_p(T;T_0)\\
\frac{\bm{P}_e(T;T_0)\cdot\bm{\sigma}}{\sqrt{\mathcal{E}_e(T;T_0)+m}}r_p(T;T_0)
\end{pmatrix}=\frac{\hat{P}_e(T;T_0)+m}{\sqrt{2p_+\mathcal{V}_0[\mathcal{E}_e(T;T_0)+m]}}\begin{pmatrix}
r_p(T;T_0)\\
0
\end{pmatrix},
\end{equation}
then the two-dimensional spinor $r_p(T;T_0)$ has to fulfill the equation [see also the discussion below Eq. (\ref{Eq_rho})]
\begin{equation}
\label{Eq_rho_PW}
p_+\frac{dr_p}{dT}=\frac{ie}{2}\bm{\sigma}\cdot\left(\bm{B}-\frac{\bm{P}_e\times\bm{E}}{\mathcal{E}_e+m}\right)r_p.
\end{equation}
Since in the present case of a background plane wave and within the chosen gauge, it is $\bm{B}(T)=-\bm{n}\times \bm{E}(T)$ (recall that the propagation direction of the wave is $-\bm{n}$), with $\bm{E}(T)=-(1/2)d\bm{A}(T)/dT$, Eq. (\ref{Eq_rho_PW}) can be solved analytically and in closed form. By introducing the spin quantum number $s$ as above and fixing the initial condition as $r_{p,s}(T_0;T_0)=r_{0;p,s}$, the solution reads
\begin{equation}
\label{r_PW}
\begin{split}
r_{p,s}(T;T_0)&=\sqrt{\frac{\mathcal{E}_e(T;T_0)+m}{\varepsilon+m}}\left\{1+\frac{e}{4p_+}\bm{\sigma}\cdot\left[\bm{n}+\frac{\bm{P}_e(T;T_0)}{\mathcal{E}_e(T;T_0)+m}\right]\bm{\sigma}\cdot\bm{A}(T)\right\}r_{0;p,s}\\
&=\sqrt{\frac{\mathcal{E}_e(T;T_0)+m}{\varepsilon+m}}\left\{1+\frac{e}{4p_+}\frac{\bm{P}_e(T;T_0)\cdot\bm{A}(T)}{\mathcal{E}_e(T;T_0)+m}\right.\\
&\left.\quad+\frac{ie}{4p_+}\bm{\sigma}\cdot\left[\bm{n}\times\bm{A}(T)+\frac{\bm{P}_e(T;T_0)\times\bm{A}(T)}{\mathcal{E}_e(T;T_0)+m}\right]\right\} r_{0;p,s},
\end{split}
\end{equation}
as it can be easily checked by substituting this expression in Eq. (\ref{Eq_rho_PW}).

The resulting wave function in Eqs. (\ref{Phi_0_PW}) and (\ref{r_PW}) has to be compared with the traditional form of the corresponding Volkov wave function $U_{V;p,s}(x;T_0)$, which in our notation can be written as $U_{V;p,s}(x;T_0)=e^{\frac{i}{\hbar}S_p(x;T_0)}\Phi_{V;p,s}(T;T_0)$, where $S_p(x;T_0)$ is given by Eq. (\ref{S_PW}) and where \cite{Landau_b_4_1982}
\begin{equation}
\label{Phi_V_PW}
\begin{split}
\Phi_{V;p,s}(T;T_0)&=\frac{1}{\sqrt{2p_+\mathcal{V}_0}}\left[1+e\frac{\hat{\tilde{n}}\hat{A}(T)}{2p_+}\right]\begin{pmatrix}
\sqrt{\varepsilon+m}r_{0;p,s}\\
\frac{\bm{p}\cdot\bm{\sigma}}{\sqrt{\varepsilon+m}}r_{0;p,s}
\end{pmatrix}\\
&=\frac{1}{\sqrt{2p_+\mathcal{V}_0}}\left[1+e\frac{\hat{\tilde{n}}\hat{A}(T)}{2p_+}\right]\frac{\hat{p}+m}{\sqrt{\varepsilon+m}}\begin{pmatrix}
r_{0;p,s}\\
0
\end{pmatrix}\\
&=\frac{\hat{P}_e(T;T_0)+m}{\sqrt{2p_+\mathcal{V}_0(\varepsilon+m)}}\left[1+e\frac{\hat{\tilde{n}}\hat{A}(T)}{2p_+}\right]\begin{pmatrix}
r_{0;p,s}\\
0
\end{pmatrix}.
\end{split}
\end{equation}
Indeed, after some algebra and, in particular, by using the third equality in Eq. (\ref{Phi_V_PW}), it can be shown that $\Phi_{p,s}(T;T_0)=\Phi_{V;p,s}(T;T_0)$, where
\begin{equation}
\Phi_{p,s}(T;T_0)=\frac{1}{\sqrt{2p_+\mathcal{V}_0}}\begin{pmatrix}
\sqrt{\mathcal{E}_e(T;T_0)+m}r_{p,s}(T;T_0)\\
\frac{\bm{P}_e(T;T_0)\cdot\bm{\sigma}}{\sqrt{\mathcal{E}_e(T;T_0)+m}}r_{p,s}(T;T_0)
\end{pmatrix},
\end{equation}
with $r_{p,s}(T;T_0)$ given by Eq. (\ref{r_PW}). Finally, by replacing the spinor $\Phi_{p,s}(T;T_0)$ into Eq. (\ref{Theta}), one sees that the term proportional to $\hbar$ identically vanishes because $\hat{\tilde{n}}^2=0$. Thus, we obtain that $\Theta_{p,s}(T;T_0)=\Phi_{p,s}(T;T_0)$ and then that the positive-energy Volkov state $U_{V;p,s}(x;T_0)$ can be written in the fully quasiclassical form
\begin{equation}
\label{U_V}
\begin{split}
U_{V;p,s}(x;T_0)&=\frac{e^{\frac{i}{\hbar}S_p(x;T_0)}}{\sqrt{2p_+\mathcal{V}_0}}\begin{pmatrix}
\sqrt{\mathcal{E}_e(T;T_0)+m}r_{p,s}(T;T_0)\\
\frac{\bm{P}_e(T;T_0)\cdot\bm{\sigma}}{\sqrt{\mathcal{E}_e(T;T_0)+m}}r_{p,s}(T;T_0)
\end{pmatrix},
\end{split}
\end{equation}
with $r_{p,s}(T;T_0)$ given by Eq. (\ref{r_PW}).  Note that the corresponding four-current $J^{\mu}_{V;p,s}(x;T_0)=\bar{U}_{V;p,s}(x;T_0)\gamma^{\mu}U_{V;p,s}(x;T_0)=P^{\mu}_e(T;T_0)/p_+\mathcal{V}_0$ is automatically conserved because the plus component of the four-momentum is a constant of motion in the case of the plane wave $A^{\mu}(T)$.

\subsubsection{Conditions of validity of the WKB approach}

Concerning the conditions of validity of the approximations used to obtain the states in Eqs. (\ref{U_in})-(\ref{U_out}), we stress that the only employed approximation is that quantum corrections proportional to $\hbar$ in Eqs. (\ref{Theta}) and (\ref{Phi}) have been neglected. Due to the fact that in general the WKB expansion is asymptotic  (see, e.g. \cite{Taya_2021}) and due to the complex, multicomponent structure of the electron states, a complete and quantitative analysis of the conditions under which higher-order corrections in $\hbar$ can be neglected must also rely on numerical analyses. However, general conditions can be derived, which are based on the following arguments and considerations starting from Eq. (\ref{Dirac_Phi}). As we have seen, if the background field were a plane wave, i.e., in the present context, if it depends only on $T$, then one can directly seek for a solution of Eq. (\ref{Dirac_Phi}) depending only on $T$ and the last term containing $i\hbar\square$ would vanish (as we have noticed, the quantity $(\partial P_e)$ vanishes in this case because the plus component of the four-momentum is a constant of motion). Thus, one concludes that the corrections brought about by the term containing $i\hbar\square$ arise due to the spatial focusing of the background field or, in general, to its spacetime features beyond the plane wave. If we consider the typical example of a Gaussian beam with electromagnetic field amplitude $F_0$, central angular frequency $\omega_0$ (central wavelength $\lambda_0=2\pi/\omega_0$), and spatial focusing radius $\sigma_0$ (Rayleigh length $l_R=\pi\sigma_0^2/\lambda_0$), we can estimate $|\partial_T|\sim \omega_0$, $|\partial_{\phi}|\sim 1/l_R=2/\omega_0\sigma_0^2$, and $|\bm{\nabla}_{\perp}|\sim 1/\sigma_0$, such that $|\square|\sim 1/\sigma_0^2$. In order to ascertain the conditions of validity of the approximations used, it is first sufficient to compare the quantity $\hbar/\sigma_0^2$ with two of the first three terms in Eq. (\ref{Dirac_Phi}). By choosing the first one and the third one, we obtain the conditions $\hbar/\sigma_0^2\ll \omega_0p_+$ and $\hbar/\sigma_0^2\ll|e|F_0$, which, apart from numerical factors which can be ignored at the present level of accuracy, can be written as $\lambda^2_C/\sigma_0^2\ll \chi/\xi$ and $\lambda_0\lambda_C/\sigma_0^2\ll \xi$, where $\lambda_C=\hbar/m\approx 3.9\times 10^{-11}\;\text{cm}$ is the Compton wavelength. 

In addition to these conditions, we recall that in obtaining the states in Eqs. (\ref{U_in})-(\ref{U_out}), we have also neglected the term proportional to $\hbar$ in Eq. (\ref{Theta}). If we kept that term, we would obtain in general that $\Theta^{(0)}(x)=(1+i\hbar\gamma^{\mu}\partial_{\mu}/2m)\Phi^{(0)}(x)$. At this point, one would conclude that, since the largest derivative is the one with respect to $T$ and it is of the order of $\omega_0$, the condition here is $\hbar\omega_0\ll m$. However, this condition is certainly too restrictive because we already know, for example, that in the plane-wave case the WKB solution is exact independently of the frequency of the field. Indeed, as we have seen in the case of a plane wave, this contribution does actually vanish identically due to the matrix structure. Thus, based on this, we again expect physically that the condition depends on the fact that the background field is spatially focused and then that it is of the form $\hbar/\sigma_0\ll m$, i.e., $\lambda_C/\sigma_0\ll 1$ (the corresponding condition on the Rayleigh length is less restrictive). 

In all cases, as we concluded in Ref. \cite{Di_Piazza_2014}, these conditions are less restrictive than the one $\eta\ll 1$, which then would allow one to obtain explicit analytical expressions of the classical action/trajectory and then of the electron states in terms of the background electromagnetic field. In conclusion, the states in Eqs. (\ref{U_in})-(\ref{U_out}) have more general validity than those finally obtained in Refs. \cite{Di_Piazza_2014,Di_Piazza_2015} but require numerical methods to be efficiently employed.

\subsection{Negative-energy states}
\label{NES}
The derivation of the negative-energy states proceeds analogously to the positive-energy states. The main difference is in the choice of the initial condition for the classical action. Now, in fact, we indicate the action as $S_{-p}(x;T_0)$ as we determine it by imposing the initial condition $S_{-p}(T_0,\bm{x}_{\text{lc}};T_0)=p_+\phi+p_-T_0-\bm{p}_{\perp}\cdot\bm{x}_{\perp}$ again for the on-shell four-momentum $p^{\mu}$ [$p_-=(m^2+\bm{p}_{\perp}^2)/2p_+$, with $p_{\pm}>0$]. In the present case, the kinetic four-momentum is defined as $P_p^{\mu}(x;T_0)=(\mathcal{E}_p(x;T_0),\bm{P}_p(x;T_0))=\partial^{\mu}S_{-p}(x;T_0)+eA^{\mu}(x)$, with the index $p$ standing for ``positron'' (as in the previous paragraph the index ``e'' stood for ``electron''). In fact, as we will see below, $P_p^{\mu}(x;T_0)$ represents the kinetic four-momentum of a positron in the background field and one can already see that, recalling the Hamilton-Jacobi equation (\ref{HJ}), $P^2_p(x;T_0)=m^2$. As before, the initial conditions are such that the positron asymptotically moves along the straight line $\bm{x}_{\perp}(T)=\bm{x}_{0,\perp}+(\bm{p}_{\perp}/p_+)(T-T_0)$ and $\phi(T)=\phi_0+(p_-/p_+)(T-T_0)$, with the general definitions $\bm{x}_{0,\perp}-(\bm{p}_{\perp}/p_+)T_0=-\bm{\nabla}_{\bm{p}_{\perp}}S_{-p}(x;T_0)$ and $-\phi_0+(p_-/p_+)T_0=-\partial_{p_+}S_{-p}(x;T_0)$.

After using the same definitions for the spinors $\Theta_{-p}(x;T_0)$ and $\Phi_{-p}(x;T_0)$ as in Eqs. (\ref{Psi}) and (\ref{Theta}), respectively, the original Dirac equation is equivalent to the equation
\begin{equation}
\label{Dirac_Phi_p}
\left(2P_{p,\mu}\partial^{\mu}+(\partial P_p)-\frac{ie}{2}\sigma^{\mu\nu}F_{\mu\nu}-i\hbar\square\right)\Phi_{-p}=0
\end{equation}
for the spinor $\Phi_{-p}(x;T_0)$. In order to determine the zeroth-order spinor $\Phi_{-p}^{(0)}(x;T_0)$, we set $\hbar=0$ in Eq. (\ref{Dirac_Phi_p}) and we obtain
\begin{equation}
\label{Dirac_Phi_0_p}
\left(2P_{p,\mu}\partial^{\mu}+(\partial P_p)-\frac{ie}{2}\sigma^{\mu\nu}F_{\mu\nu}\right)\Phi_{-p}^{(0)}=0.
\end{equation}
This equation corresponds to Eq. (\ref{Dirac_Phi_0}) for the spinor $\Phi^{(0)}_p(x;T_0)$ but with the replacements $P^{\mu}_e(x;T_0)\to P^{\mu}_p(x;T_0)$ and $e\to -e$. Analogously as in the positive-energy case, the operator acting on $\Phi_{-p}^{(0)}(x;T_0)$ commutes with the matrix $\hat{P}_p(x;T_0)$ and we can choose the spinor $\Phi_{-p}^{(0)}(x;T_0)$ to satisfy the eigenvalue equation $\hat{P}_p(x;T_0)\Phi_{-p}^{(0)}(x;T_0)=-m\Phi_{-p}^{(0)}(x;T_0)$. Thus, according to the free theory \cite{Landau_b_4_1982}, the spinor $\Phi_{-p}^{(0)}(x;T_0)$ can be written as
\begin{equation}
\label{Phi_0_p}
\Phi^{(0)}_{-p}(x;T_0)=\frac{1}{\sqrt{2p_+\mathcal{V}_0}}\begin{pmatrix}
\frac{\bm{P}_p(x;T_0)\cdot\bm{\sigma}}{\sqrt{\mathcal{E}_p(x;T_0)+m}}w_{-p}(x;T_0)\\
\sqrt{\mathcal{E}_p(x;T_0)+m}w_{-p}(x;T_0)
\end{pmatrix},
\end{equation}
where $w_{-p}(x;T_0)$ is an arbitrary two-dimensional spinor. Analogously as for the positive-energy states, by replacing the expression in Eq. (\ref{Phi_0_p}) of the zeroth-order spinor $\Phi_{-p}^{(0)}(x;T_0)$ in Eq. (\ref{Dirac_Phi_0_p}), we obtain that it is satisfied if the two-dimensional spinor $w_{-p}(x;T_0)$ satisfies the equation
\begin{equation}
\label{Eq_w_p}
P_p^{\mu}\partial_{\mu}w_{-p}=-\frac{1}{2}(\partial_{\mu}P_p^{\mu})w_{-p}-\frac{ie}{2}\bm{\sigma}\cdot\left(\bm{B}-\frac{\bm{P}_p\times\bm{E}}{\mathcal{E}_p+m}\right)w_{-p},
\end{equation}
which, again, corresponds to Eq. (\ref{Eq_w}) but with $P^{\mu}_e(x;T_0)\to P^{\mu}_p(x;T_0)$ and $e\to -e$. Also this equation can be solved by applying the method of characteristics. In this case, we introduce the positron proper time $\tau$ according to the equation
\begin{equation}
\label{Eq_mot_x_p}
m\frac{dx^{\mu}}{d\tau}=\Pi_p^{\mu},
\end{equation}
where $\Pi_p^{\mu}(\tau;T_0)=P^{\mu}_p(x(\tau;T_0);T_0)$, with $x^{\mu}=x^{\mu}(\tau;T_0)$ being a generic positron trajectory. This equation implies that
\begin{equation}
\label{Eq_mot_p_p}
m\frac{d^2x^{\mu}}{d\tau^2}=\frac{d\Pi_p^{\mu}}{d\tau}=-eF^{\mu\nu}\frac{dx_{\nu}}{d\tau},
\end{equation}
which is the Lorentz equation for a positron in the external field. Analogously as before, one can also introduce the action function $\Sigma_{-p}(\tau;T_0)=S_{-p}(x(\tau;T_0);T_0)$ computed along the positron trajectory at hand and the procedure is now to solve the equations of motion (\ref{Eq_mot_x_p}) and (\ref{Eq_mot_p_p}) and the equation
\begin{equation}
\frac{d\Sigma_{-p}}{d\tau}=(\partial_{\mu}S_{-p})\frac{dx^{\mu}}{d\tau}=m-e\frac{(\Pi_pA)}{m}
\end{equation}
for the action $\Sigma_{-p}(\tau;T_0)$ computed along the positron trajectory for generic initial conditions $\bm{x}_{0,\text{lc}}$, $\bm{p}_{\text{lc}}$, and $\Sigma_{-p}(0;T_0)=S_{-p}(x(0;T_0);T_0)=p_+\phi_0+p_-T_0-\bm{p}_{\perp}\cdot\bm{x}_{0,\perp}$. In this way, one obtains the functions $x^{\mu}=x^{\mu}(\tau;T_0,\bm{x}_{0,\text{lc}},\bm{p}_{\text{lc}})$, $\Pi_p^{\mu}(\tau;T_0,\bm{x}_{0,\text{lc}},\bm{p}_{\text{lc}})$, and $\Sigma_{-p}(\tau;T_0,\bm{x}_{0,\text{lc}},\bm{p}_{\text{lc}})$. The four equations $x^{\mu}=x^{\mu}(\tau;T_0,\bm{x}_{0,\text{lc}},\bm{p}_{\text{lc}})$ can be inverted to obtain the functions $\tau=\tau(x;T_0,\bm{p}_{\text{lc}})$ and $\bm{x}_{0,\text{lc}}=\bm{x}_{0,\text{lc}}(x;T_0,\bm{p}_{\text{lc}})$ \cite{Evans_b_2010}. Indeed, one can also in this case show that the (positron) van Vleck determinant is given by
\begin{equation}
D_p(x;T_0)=\frac{P_{p,+}(x;T_0)}{p_+}e^{-\frac{1}{2}\int_{T_0}^T\frac{d\tilde{T}}{P_{p,+}}(\partial P_p)}
\end{equation}
and it never vanishes. Finally, the action $S_{-p}(x;T_0)$ is obtained in the usual way as $S_{-p}(x;T_0)=\Sigma_{-p}(\tau(x;T_0,\bm{p}_{\text{lc}});T_0,\bm{x}_{0,\text{lc}}(x;T_0,\bm{p}_{\text{lc}}),\bm{p}_{\text{lc}})$. 

Going back to Eq. (\ref{Eq_w_p}) and by setting
\begin{equation}
\label{w_rho_p}
w_{-p}(x;T_0)=e^{-\frac{1}{2m}\int_0^{\tau}d\tau'(\partial P_p)}r_{-p}(x;T_0)=e^{-\frac{1}{2}\int_{T_0}^T\frac{dT'}{P_{p,+}}(\partial P_p)}r_{-p}(x;T_0),
\end{equation}
we conclude that the two-component spinor $r_{-p}(x;T_0)$ satisfies the equation
\begin{equation}
\label{Eq_rho_p}
P_p^{\mu}\partial_{\mu}r_{-p}=-\frac{ie}{2}\bm{\sigma}\cdot\left(\bm{B}-\frac{\bm{P}_p\times\bm{E}}{\mathcal{E}_p+m}\right)r_{-p}.
\end{equation}
As before, this equation implies that the two-component spinor $r_{-p}(x;T_0)$ has constant norm $r^{\dag}_{-p}(x;T_0)r_{-p}(x;T_0)$, which can be set equal to unity. Moreover, the average vector $\bm{s}_{-p}(x;T_0)=r_{-p}^{\dag}(x;T_0)\bm{\sigma}r_{-p}(x;T_0)$ satisfies the equation
\begin{equation}
\label{Eq_zeta_p}
P_p^{\mu}\partial_{\mu}\bm{s}_{-p}=-e\bm{s}_{-p}\times\left(\bm{B}-\frac{\bm{P}_p\times\bm{E}}{\mathcal{E}_p+m}\right),
\end{equation}
which is equivalent to the BMT equation $mds_{-p}^{\mu}/d\tau=-eF^{\mu\nu}s_{-p,\nu}$ along the positron trajectory for the positron polarization four-vector \cite{Landau_b_4_1982}
\begin{equation}
s_{-p}^{\mu}(x;T_0)=\left(\frac{\bm{s}_{-p}(x;T_0)\cdot\bm{P}_p(x;T_0)}{m},\bm{s}_{-p}(x;T_0)+\frac{\bm{s}_{-p}(x;T_0)\cdot\bm{P}_p(x;T_0)}{m[\mathcal{E}_p(x;T_0)+m]}\bm{P}_p(x;T_0)\right).
\end{equation}
Also in this case we choose the direction of the momentum $\bm{p}$ as spin-quantization direction at $T_0$ and we introduce the discrete spin quantum number $s=\pm 1$, depending on the two possible orientations of the vector $\bm{s}_{-p}(T_0,\bm{x}_{\text{lc}};T_0)$. Correspondingly we indicate $\Phi_{-p}(x;T_0)$ rather as $\Phi_{-p,-s}(x;T_0)$, following the notation in Ref. \cite{Landau_b_4_1982} and analogously $\Phi^{(0)}_{-p}(x;T_0)$ rather as $\Phi^{(0)}_{-p,-s}(x;T_0)$.

Concerning higher-order corrections in $\hbar$, we observe that the proper-time evolution operator $\mathcal{U}_p(\tau,\tau')$ solves here the equation [see Eq. (\ref{Dirac_Phi_p})]
\begin{equation}
2m\frac{d\mathcal{U}_p}{d\tau}=-\left((\partial P_p)-\frac{ie}{2}\sigma^{\mu\nu}F_{\mu\nu}\right)\mathcal{U}_p,
\end{equation}
with the initial condition $\mathcal{U}_p(\tau,\tau)=1$, and then
\begin{equation}
\mathcal{U}_p(\tau,\tau')=e^{-\frac{1}{2m}\int_{\tau'}^{\tau}d\tilde{\tau}(\partial P_p)}\left[\theta(\tau-\tau')\mathcal{T}_>\left(e^{\frac{ie}{4m}\int_{\tau'}^{\tau}d\tilde{\tau}\sigma^{\mu\nu}F_{\mu\nu}}\right)+\theta(\tau'-\tau)\mathcal{T}_<\left(e^{\frac{ie}{4m}\int_{\tau'}^{\tau}d\tilde{\tau}\sigma^{\mu\nu}F_{\mu\nu}}\right)\right].
\end{equation}
In this way, we can write the equation for the spinor $\Phi_{-p,-s}(x;T_0)$ as the integral equation
\begin{equation}
\label{Phi_p_p}
\Phi_{-p,-s}(x;T_0)=\Phi^{(0)}_{-p,-s}(x;T_0)+\frac{i\hbar}{2}\int_{T_0}^T\frac{dT'}{P_{p,+}}\mathcal{U}_p(T,T')\square\Phi_{-p,-s},
\end{equation}
where
\begin{equation}
\begin{split}
\mathcal{U}_p(T,T')&=e^{-\frac{1}{2}\int_{T'}^T\frac{d\tilde{T}}{P_{p,+}}(\partial P_p)}\left[\theta(T-T')\mathcal{T}_>\left(e^{\frac{ie}{4}\int_{T'}^T\frac{d\tilde{T}}{P_{p,+}}\sigma^{\mu\nu}F_{\mu\nu}}\right)\right.\\
&\quad\left.+\theta(T'-T)\mathcal{T}_<\left(e^{\frac{ie}{4}\int_{T'}^T\frac{d\tilde{T}}{P_{p,+}}\sigma^{\mu\nu}F_{\mu\nu}}\right)\right].
\end{split}
\end{equation}
Also in the present case, once the spinor $\Phi_{-p,-s}(x;T_0)$ is obtained, one can use the definitions in Eqs. (\ref{Theta}) and (\ref{Psi}) to obtain the corresponding negative-energy solution of the Dirac equation, which we indicate as $V_{p,s}(x;T_0)$. As before, the in- and out-states are defined as $V^{(\text{in})}_{p,s}(x)=\lim_{T_0\to-\infty}V_{p,s}(x;T_0)$ and $V^{(\text{out})}_{p,s}(x)=\lim_{T_0\to\infty}V_{p,s}(x;T_0)$, respectively.

Finally, as in the previous paragraph, we focus on the zeroth-order solution and we observe that at this order it is $\Theta_{-p,-s}^{(0)}(x;T_0)=\Phi_{-p,-s}^{(0)}(x;T_0)$ [see also Eq. (\ref{Theta})]. Thus, the leading-order state $V^{(0)}_{p,s}(x;T_0)$, which includes terms up to $O(\hbar^0)$, reads
\begin{equation}
\label{V_T_0}
V^{(0)}_{p,s}(x;T_0)=e^{\frac{i}{\hbar}S_{-p}(x;T_0)}\frac{e^{-\frac{1}{2}\int_{T_0}^T\frac{d\tilde{T}}{P_{p,+}}(\partial P_p)}}{\sqrt{2p_+\mathcal{V}_0}}\begin{pmatrix}
\frac{\bm{P}_p(x;T_0)\cdot\bm{\sigma}}{\sqrt{\mathcal{E}_p(x;T_0)+m}}r_{-p,-s}(x;T_0)\\
\sqrt{\mathcal{E}_p(x;T_0)+m}r_{-p,-s}(x;T_0)
\end{pmatrix}.
\end{equation}
The corresponding four-current $J^{(0)\,\mu}_{p;p,s}(x;T_0)=\bar{V}^{(0)}_{p,s}(x;T_0)\gamma^{\mu}V^{(0)}_{p,s}(x;T_0)$ is given by
\begin{equation}
J^{(0)\,\mu}_{p;p,s}(x;T_0)=\frac{P^{\mu}_p(x;T_0)}{p_+\mathcal{V}_0}e^{-\int_{T_0}^T\frac{d\tilde{T}}{P_{p,+}}(\partial P_p)}.
\end{equation}
As in the positive-energy case, the four-current $J^{(0)\,\mu}_{p;p,s}(x;T_0)$ has the desired properties that it corresponds to the standard normalization of one particle in the light-cone volume $\mathcal{V}_0$ at the initial time $T_0$ and that it is conserved: $(\partial J^{(0)}_{p;p,s})=0$.

It is useful to mention that the state $V^{(0)}_{p,s}(x;T_0)$ can be equivalently written as
\begin{equation}
\begin{split}
V^{(0)}_{p,s}(x;T_0)&=\int_{\mathcal{V}_0} \frac{d^3 x_{0,\text{lc}}}{\sqrt{\Delta_p(T;T_0)}}\delta^3(\bm{x}_{\text{lc}}-\bm{x}_{\text{lc}}(T;T_0,\bm{x}_{0,\text{lc}},\bm{p}_{\text{lc}}))\\
&\quad\times\frac{e^{\frac{i}{\hbar}\Sigma_{-p}(T;T_0)}}{\sqrt{2\Pi_{p,+}(T;T_0)\mathcal{V}_0}}\begin{pmatrix}
\frac{\bm{\Pi}_p(T;T_0)\cdot\bm{\sigma}}{\sqrt{\Pi_{p,0}(T;T_0)+m}}\rho_{-p,-s}(T;T_0)\\
\sqrt{\Pi_{p,0}(T;T_0)+m}\rho_{-p,-s}(T;T_0)
\end{pmatrix},
\end{split}
\end{equation}
where, analogously to the positive-energy case, the functions $\rho_{-p,-s}(T;T_0)$ and $\Delta_p(T;T_0)$ are the functions $\rho_{-p,-s}(\tau;T_0)=r_{-p,-s}(x(\tau;T_0);T_0)$ and $\Delta_p(\tau;T_0)=D_p(x(\tau;T_0);T_0)$ computed along the positron trajectory $\bm{x}_{\text{lc}}=\bm{x}_{\text{lc}}(T;T_0,\bm{x}_{0,\text{lc}},\bm{p}_{\text{lc}})$ corresponding to the initial condition $\bm{x}_{\text{lc}}(T_0)=\bm{x}_{0,\text{lc}}$ on the light-cone position and parametrized with respect to the light-cone time $T$. 

Then, one obtains the negative-energy electron in-states and out-states, which include terms up to $O(\hbar^0)$, as
\begin{align}
\label{V_in}
V^{(0,\text{in})}_{p,s}(x)&=e^{\frac{i}{\hbar}S_{-p}^{(\text{in})}(x)}\frac{e^{-\frac{1}{2}\int_{-\infty}^Td\tilde{T}(\partial P^{(\text{in})}_p)/P^{(\text{in})}_{p,+}}}{\sqrt{2p_+\mathcal{V}_0}}v^{(0,\text{in})}_{p,s}(x),\\
\label{V_out}
V^{(0,\text{out})}_{p,s}(x)&=e^{\frac{i}{\hbar}S_{-p}^{(\text{out})}(x)}\frac{e^{-\frac{1}{2}\int_{\infty}^Td\tilde{T}(\partial P^{(\text{out})}_p)/P^{(\text{out})}_{p,+}}}{\sqrt{2p_+\mathcal{V}_0}}v^{(0,\text{out})}_{p,s}(x),
\end{align}
where
\begin{align}
\label{v_in}
v^{(0,\text{in})}_{p,s}(x)&=\begin{pmatrix}
\frac{\bm{P}^{(\text{in})}_p(x)\cdot\bm{\sigma}}{\sqrt{\mathcal{E}_p^{(\text{in})}(x)+m}}r^{(\text{in})}_{-p,-s}(x)\\
\sqrt{\mathcal{E}_p^{(\text{in})}(x)+m}r^{(\text{in})}_{-p,-s}(x)
\end{pmatrix},\\
\label{v_out}
v^{(0,\text{out})}_{p,s}(x)&=\begin{pmatrix}
\frac{\bm{P}^{(\text{out})}_p(x)\cdot\bm{\sigma}}{\sqrt{\mathcal{E}_p^{(\text{out})}(x)+m}}r^{(\text{out})}_{-p,-s}(x)\\
\sqrt{\mathcal{E}_p^{(\text{out})}(x)+m}r^{(\text{out})}_{-p,-s}(x)
\end{pmatrix}.
\end{align}

Finally, for the sake of completeness, we also report the negative-energy Volkov state $V_{V;p,s}(x;T_0)$ in the full quasiclassical form
\begin{equation}
\begin{split}
V_{V;p,s}(x;T_0)&=\frac{e^{\frac{i}{\hbar}S_{-p}(x;T_0)}}{\sqrt{2p_+\mathcal{V}_0}}\begin{pmatrix}
\frac{\bm{P}_p(T;T_0)\cdot\bm{\sigma}}{\sqrt{\mathcal{E}_p(T;T_0)+m}}r_{-p,-s}(T;T_0)\\
\sqrt{\mathcal{E}_p(T;T_0)+m}r_{-p,-s}(T;T_0)
\end{pmatrix},
\end{split}
\end{equation}
where the action $S_{-p}(x;T_0)$ is given by Eq. (\ref{S_PW}) with the substitution $p^{\mu}\to -p^{\mu}$, where the four-momentum $P_p^{\mu}(T;T_0)=(\mathcal{E}_p(T;T_0),\bm{P}_p(T;T_0))$ is given by Eq. (\ref{P_e_PW}) with the replacement $e\to -e$, and where [see Eq. (\ref{r_PW})] 
\begin{equation}
\begin{split}
r_{-p,-s}(T;T_0)&=\sqrt{\frac{\mathcal{E}_p(T;T_0)+m}{\varepsilon+m}}\left\{1-\frac{e}{4p_+}\bm{\sigma}\cdot\left[\bm{n}+\frac{\bm{P}_p(T;T_0)}{\mathcal{E}_p(T;T_0)+m}\right]\bm{\sigma}\cdot\bm{A}(T)\right\}r_{0;-p,-s}\\
&=\sqrt{\frac{\mathcal{E}_p(T;T_0)+m}{\varepsilon+m}}\left\{1-\frac{e}{4p_+}\frac{\bm{P}_p(T;T_0)\cdot\bm{A}(T)}{\mathcal{E}_p(T;T_0)+m}\right.\\
&\left.\quad-\frac{ie}{4p_+}\bm{\sigma}\cdot\left[\bm{n}\times\bm{A}(T)+\frac{\bm{P}_p(T;T_0)\times\bm{A}(T)}{\mathcal{E}_p(T;T_0)+m}\right]\right\} r_{0;-p,-s},
\end{split}
\end{equation}
with $r_{0;-p,-s}=r_{-p,-s}(T_0;T_0)$.

\section{Nonlinear single Compton scattering}

In this section we use the positive-energy electron states in Eqs. (\ref{U_in})-(\ref{u_out}), to compute the emission spectrum of nonlinear Compton scattering and, for the sake of notational simplicity, we remove the upper index $(0)$ as there is no possibility of confusion. For the same reason, the constant $\hbar$ will be set equal to unity.

We indicate as $p^{\mu}=(\varepsilon,\bm{p})$ ($p^{\prime\,\mu}=(\varepsilon',\bm{p}')$), with $\varepsilon=\sqrt{m^2+\bm{p}^2}$ ($\varepsilon'=\sqrt{m^2+\bm{p}^{\prime\,2}}$), and $s$ ($s'$) the four-momentum and the spin quantum number of the incoming (outgoing) electron, respectively,  and as $k^{\mu}=(\omega,\bm{k})$, with $\omega=|\bm{k}|$, and $l$ the four-momentum and the polarization index of the emitted photon, respectively. The $S$-matrix transition element $S^{(e^-\to e^-\gamma)}$ is given by
\begin{equation}
S^{(e^-\to e^-\gamma)}=-ie\int dT\int_{\mathcal{V}_0} d^3x_{\text{lc}}\,\bar{U}^{(\text{out})}_{p',s'}(x)\frac{\hat{e}^*_{k,l}e^{i(kx)}}{\sqrt{2k_+\mathcal{V}_0}}U^{(\text{in})}_{p,s}(x),
\end{equation}
where $e^{\mu,*}_{k,l}$ indicates the polarization four-vector of the emitted photon. The differential emission probability with respect to the photon light-cone three-momentum $\bm{k}_{\text{lc}}$ and averaged (summed) over the initial (final) discrete quantum numbers, is given by
\begin{equation}
\label{dP_NCS}
\begin{split}
&\frac{dP^{(e^-\to e^-\gamma)}}{d^3k_{\text{lc}}}=\lim_{\mathcal{V}_0\to\infty}\frac{e^2}{2}\mathcal{V}_0^2\sum_{s,s',l}\int \frac{d^3p'_{\text{lc}}}{(2\pi)^6}\int dT dT'\int_{\mathcal{V}_0} d^3x_{\text{lc}}d^3x'_{\text{lc}}\bar{U}^{(\text{out})}_{p',s'}(x)\frac{\hat{e}^*_{k,l}e^{i(kx)}}{\sqrt{2k_+\mathcal{V}_0}}U^{(\text{in})}_{p,s}(x)\\
&\quad\times\bar{U}^{(\text{in})}_{p,s}(x')\frac{\hat{e}_{k,l}e^{-i(kx')}}{\sqrt{2k_+\mathcal{V}_0}}U^{(\text{out})}_{p',s'}(x')\\
&\quad=\lim_{\mathcal{V}_0\to\infty}\frac{e^2}{2}\sum_{s,s',l}\int \frac{d^3p'_{\text{lc}}}{(2\pi)^6\mathcal{V}_0}\int dT dT'\int_{\mathcal{V}_0} \frac{d^3x_{\text{lc}}d^3x'_{\text{lc}}}{8p'_+k_+p_+}\bar{u}^{(\text{out})}_{p',s'}(x)\hat{e}^*_{k,l}u^{(\text{in})}_{p,s}(x)\bar{u}^{(\text{in})}_{p,s}(x')\hat{e}_{k,l}u^{(\text{out})}_{p',s'}(x')\\
&\qquad\times e^{-\int_{\infty}^Td\tilde{T}(\partial P^{\prime(\text{out})}_e)/2P^{\prime(\text{out})}_{e,+}-\int_{-\infty}^Td\tilde{T}(\partial P^{(\text{in})}_e)/2P^{(\text{in})}_{e,+}-\int_{\infty}^{T'}d\tilde{T}(\partial P^{\prime(\text{out})}_e)/2P^{\prime(\text{out})}_{e,+}-\int_{-\infty}^{T'}d\tilde{T}(\partial P^{(\text{in})}_e)/2P^{(\text{in})}_{e,+}}\\
&\qquad\times e^{i[-S_{p'}^{(\text{out})}(x)+(kx)+S_p^{(\text{in})}(x)+S_{p'}^{(\text{out})}(x')-(kx')-S_p^{(\text{in})}(x')]}.
\end{split}
\end{equation}
Now, we will systematically approximate this expression by assuming that the electron is ultrarelativistic, it is initially (almost) counterpropagating with respect to the laser field, and its energy is the largest dynamical energy in the problem, i.e., the classical light-cone components of the electron four-momenta satisfy the hierarchy $\Pi_{e,+}(T;T_0)\gg\max(m,|\bm{\Pi}_{e,\perp}(T;T_0)|)\gg\Pi_{e,-}(T;T_0)$ (see the Appendix \ref{App_B}). By referring to the initial electron energy and to the laser classical nonlinearity parameter $\xi$, we can formulate the above conditions in a more transparent form requiring that $\varepsilon\gg \max(m,m\xi)$ and we will keep only leading-order terms in the ratio $\eta=\max(m,m\xi)/\varepsilon$. Otherwise we do not make additional assumptions on the electron trajectory. Analogous conditions and approximations are assumed for the final electron energy. In the Appendix \ref{App_B}, we report some considerations on this regime of interaction, based on the general structure of the equations of motion and not on approximating the solution of the equations of motion, i.e., the electron trajectory. One can see, for example, that at the leading order in $\eta$ one can ignore the real exponential functions corresponding to the van Vleck determinants in Eq. (\ref{dP_NCS}) as [see Eq. (\ref{vV_T})]
\begin{equation}
\begin{split}
\Delta_e(T;T_0)&=\frac{\Pi_{e,+}(T;T_0)}{p_+}e^{-\int_{T_0}^T\frac{dT'}{\Pi_{e,+}}(\partial \Pi_e)}=\frac{\Pi_{e,+}(T;T_0)}{p_+}e^{-\int_{T_0}^T\frac{dT'}{\Pi_{e,+}}(\partial_{T'}\Pi_{e,+}+\partial_{\phi}\Pi_{e,-}+\bm{\nabla}_{\perp}\cdot\bm{\Pi}_{e,\perp})}\\
&=e^{-\int_{T_0}^T\frac{dT'}{\Pi_{e,+}}(\partial_{\phi}\Pi_{e,-}+\bm{\nabla}_{\perp}\cdot\bm{\Pi}_{e,\perp})}=1+O(\eta).
\end{split}
\end{equation}

Recalling the concept of formation length \cite{Ter-Mikaelian_b_1972,Baier_b_1998,Baier_2005}, it is convenient to pass to the average and relative spacetime variables $x_+^{\mu}=(x^{\mu}+x^{\prime\mu})/2$ and $x_-^{\mu}=x^{\mu}-x^{\prime\mu}$, respectively (see also Refs. \cite{Di_Piazza_2017,Di_Piazza_2021}). Indeed, under the above conditions, one can easily ascertain that the formation lengths in the variables $\bm{x}_{\text{lc}}$ can be neglected at the leading order in $\eta$ \cite{Di_Piazza_2017,Di_Piazza_2021} (see also the Appendix \ref{App_B}). Thus, one can set $\bm{x}_{-,\text{lc}}=\bm{0}$ everywhere in Eq. (\ref{dP_NCS}) except in the actions, where a first-order expansion on those variables has to be carried out (the reason will be clear below):
\begin{equation}
\begin{split}
&\frac{dP^{(e^-\to e^-\gamma)}}{d^3k_{\text{lc}}}\approx\lim_{\mathcal{V}_0\to\infty}\frac{e^2}{2}\sum_{s,s',l}\int \frac{d^3p'_{\text{lc}}}{(2\pi)^6\mathcal{V}_0}\int dT_+ dT_-\int_{\mathcal{V}_0} \frac{d^3x_{+,\text{lc}}d^3x_{-,\text{lc}}}{8p'_+k_+p_+}\\
&\;\;\times\bar{u}^{(\text{out})}_{p',s'}(x_T)\hat{e}^*_{k,l}u^{(\text{in})}_{p,s}(x_T)\bar{u}^{(\text{in})}_{p,s}(x_{T'})\hat{e}_{k,l}u^{(\text{out})}_{p',s'}(x_{T'})\\
&\;\;\times e^{i\big[-S_{p'}^{(\text{out})}(x_T)-\bm{\nabla}_{\perp}S_{p'}^{(\text{out})}(x_T)\cdot\frac{\bm{x}_{-,\perp}}{2}-\partial_{\phi}S_{p'}^{(\text{out})}(x_T)\frac{\phi_-}{2}+(kx)+S_p^{(\text{in})}(x_T)+\bm{\nabla}_{\perp}S_p^{(\text{in})}(x_T)\cdot\frac{\bm{x}_{-,\perp}}{2}+\partial_{\phi}S_p^{(\text{in})}(x_T)\frac{\phi_-}{2}\big]}\\
&\;\;\times e^{i\big[S_{p'}^{(\text{out})}(x_{T'})-\bm{\nabla}_{\perp}S_{p'}^{(\text{out})}(x_{T'})\cdot\frac{\bm{x}_{-,\perp}}{2}-\partial_{\phi}S_{p'}^{(\text{out})}(x_{T'})\frac{\phi_-}{2}-(kx')-S_p^{(\text{in})}(x_{T'})+\bm{\nabla}_{\perp}S_p^{(\text{in})}(x_{T'})\cdot\frac{\bm{x}_{-,\perp}}{2}+\partial_{\phi}S_p^{(\text{in})}(x_{T'})\frac{\phi_-}{2}\big]},
\end{split}
\end{equation}
where $x_T=(T,\bm{x}_{+,\text{lc}})$, $x_{T'}=(T',\bm{x}_{+,\text{lc}})$, and where all the derivatives are assumed to be with respect to the plus variables (note that we have ignored complications due to the shape of the light-cone volume $\mathcal{V}_0$ as we will ultimately perform the limit $\mathcal{V}_0\to\infty$). Now, the integrals over the variables $\bm{x}_{-,\text{lc}}$ can be taken and, by using the relation between the derivatives of the action and the kinetic momentum of the electron in the field, enforce the corresponding conservation laws in the limit $\mathcal{V}_0\to\infty$:
\begin{equation}
\begin{split}
\frac{dP^{(e^-\to e^-\gamma)}}{d^3k_{\text{lc}}}&\approx\lim_{\mathcal{V}_0\to\infty}\frac{e^2}{2}\sum_{s,s',l}\int \frac{d^3p'_{\text{lc}}}{(2\pi)^6\mathcal{V}_0}\int dT_+ dT_-\int_{\mathcal{V}_0} \frac{d^3x_{+,\text{lc}}}{8p'_+k_+p_+}\\
&\quad\times\bar{u}^{(\text{out})}_{p',s'}(x_T)\hat{e}^*_{k,l}u^{(\text{in})}_{p,s}(x_T)\bar{u}^{(\text{in})}_{p,s}(x_{T'})\hat{e}_{k,l}u^{(\text{out})}_{p',s'}(x_{T'}) e^{i\int_{T'}^Td\tilde{T}[P^{\prime(\text{out})}_{e,-}(x_{\tilde{T}})+k_--P^{(\text{in})}_{e,-}(x_{\tilde{T}})]}\\
&\quad\times (2\pi)^2\delta^{(2)}[\bm{P}^{\prime(\text{out})}_{e,\perp}(x_+)+\bm{k}_{\perp}-\bm{P}^{(\text{in})}_{e,\perp}(x_+)](2\pi)\delta[P^{\prime(\text{out})}_{e,+}(x_+)+k_+-P^{(\text{in})}_{e,+}(x_+)],
\end{split}
\end{equation}
where we have used the identities $S_{p'}^{(\text{out})}(x_{T'})-S_{p'}^{(\text{out})}(x_T)=-\int_{T'}^Td\tilde{T}\partial_{\tilde{T}}S_{p'}^{(\text{out})}(x_{\tilde{T}})$ and $S_p^{(\text{in})}(x_{T'})-S_p^{(\text{in})}(x_T)=-\int_{T'}^Td\tilde{T}\partial_{\tilde{T}}S_p^{(\text{in})}(x_{\tilde{T}})$, and the corresponding relation between the derivative of the action with respect to the variable $T$ with the minus component of the kinetic momentum.

At this point we observe that the mathematical meaning of the van Vleck determinant (see Par. \ref{vV}) and the fact that up to the leading order in $\eta$ we could have set the real exponential functions corresponding to the van Vleck determinants in Eq. (\ref{dP_NCS}) equal to unity allow us to change the integrals over $d^3p'_{\text{lc}}$ into the integrals over the corresponding local momenta of the outgoing electron in the field [see Eq. (\ref{vV_det_3})] as well as the integrals over $d^3x_{\text{lc}}$ into the integrals over the corresponding initial coordinates of the incoming electron [see Eq. (\ref{vV_det_2})]. By exploiting the delta functions to take the integrals over the corresponding local momenta of the outgoing electron, we obtain
\begin{equation}
\label{dP_NCS_f}
\begin{split}
\frac{dP^{(e^-\to e^-\gamma)}}{d^3k_{\text{lc}}}&\approx\lim_{\mathcal{V}_0\to\infty}\int_{\mathcal{V}_0} \frac{d^3x_{0,\text{lc}}}{\mathcal{V}_0}\frac{\alpha}{16\pi^2}\frac{1}{2p'_+k_+p_+}\sum_{s,s',l}\int dTdT'\bar{u}^{(\text{out})}_{p',s'}(x_T)\hat{e}^*_{k,l}u^{(\text{in})}_{p,s}(x_T)\\
&\quad\times \bar{u}^{(\text{in})}_{p,s}(x_{T'})\hat{e}_{k,l}u^{(\text{out})}_{p',s'}(x_{T'})\\
&\quad\times \exp\left\{i\int_{T'}^Td\tilde{T}\left[\frac{m^2+\bm{\Pi}^{\prime(\text{out})\,2}_{e,\perp}(\tilde{T})}{2p'_+}+\frac{\bm{k}^2_{\perp}}{2k_+}-\frac{m^2+\bm{\Pi}^{(\text{in})\,2}_{e,\perp}(\tilde{T})}{2p_+}\right]\right\},
\end{split}
\end{equation}
where $\alpha=e^2/4\pi\approx 1/137$ is the fine-structure constant, where $\bm{\Pi}^{\prime(\text{out})}_{e,\perp}(T)=\bm{\Pi}^{(\text{in})}_{e,\perp}(T)-\bm{k}_{\perp}$ and $p'_+=p_+-k_+$, and where the coordinates $\bm{x}_{\text{lc}}$ in the spinors have to be expressed in terms of the initial coordinates $\bm{x}_{0,\text{lc}}$ following the corresponding classical electron trajectory (see the Appendix \ref{App_B}). The fact that three components of the (on-shell) four-momentum of the final electron in the field can be written in terms of the corresponding initial ones already indicates that in order to compute the probability in Eq. (\ref{dP_NCS_f}) only the classical trajectory of the incoming electron is necessary, with the probability being obtained by averaging over the initial electron trajectories identified by the initial position of the electron. 

Now, we show that the above expression of the differential probability of nonlinear Compton scattering is identical to the corresponding Baier's formula \cite{Baier_b_1998}. Actually, in order to obtain Baier's formula the remaining task is to manipulate the spinor matrix elements. In fact, concerning the phase, it is easily shown that, by using the conservation laws, it can be cast in the form
\begin{equation}
\int_{T'}^Td\tilde{T}\left[\frac{m^2+\bm{\Pi}^{\prime(\text{out})\,2}_{e,\perp}(\tilde{T})}{2p'_+}+\frac{\bm{k}^2_{\perp}}{2k_+}-\frac{m^2+\bm{\Pi}^{(\text{in})\,2}_{e,\perp}(\tilde{T})}{2p_+}\right]=\frac{1}{p'_+}\int_{T'}^Td\tilde{T}(k\Pi_e^{(\text{in})}(\tilde{T})),
\end{equation}
as in Baier's formula \cite{Baier_b_1998}. The manipulation of the spinor matrix elements is also straightforward. With the help of the definitions in Eqs. (\ref{u_in})-(\ref{u_out}) and by assuming that $e^{\mu}_{k,l}=(0,\bm{e}_{k,l})$, we obtain
\begin{equation}
\begin{split}
\bar{u}^{(\text{out})}_{p',s'}(x_T)\hat{e}^*_{k,l}u^{(\text{in})}_{p,s}(x_T)&\approx-\sqrt{\frac{p'_++m}{p_++m}}\rho^{(\text{out})\dag}_{p',s'}(T)(\bm{\sigma}\cdot\bm{e}^*_{k,l})[\bm{\sigma}\cdot\bm{\Pi}_e^{(\text{in})}(T)]\rho^{(\text{in})}_{p,s}(T)\\
&\quad-\sqrt{\frac{p_++m}{p'_++m}}\rho^{(\text{out})\dag}_{p',s'}(T)[\bm{\sigma}\cdot\bm{\Pi}_e^{\prime(\text{out})}(T)](\bm{\sigma}\cdot\bm{e}^*_{k,l})\rho^{(\text{in})}_{p,s}(T).
\end{split}
\end{equation}
Now, by evaluating Eq. (\ref{Eq_rho}) along the characteristics parametrized via the light-cone time $T$, the two-dimensional spinors $\rho^{(\text{in})}_{p,s}(T)$ and $\rho^{(\text{out})}_{p',s'}(T)$, one sees that at the leading order in $\eta$, one can ignore the evolution of these spinors and use their initial expressions, which we indicate as $\rho_{0;p,s}$ and $\rho'_{0;p',s'}$, respectively. In this way, after using the well-known properties of the Pauli matrices, one obtains
\begin{equation}
\begin{split}
&\bar{u}^{(\text{out})}_{p',s'}(x_T)\hat{e}^*_{k,l}u^{(\text{in})}_{p,s}(x_T)\approx-\left(\sqrt{\frac{p'_++m}{p_++m}}+\sqrt{\frac{p_++m}{p'_++m}}\right)\bm{\Pi}_e^{(\text{in})}(T)\cdot\bm{e}^*_{k,l}\rho^{\prime\dag}_{0;p',s'}\rho_{0;p,s}\\
&\quad-i\rho^{\prime\dag}_{0;p',s'}\bm{\sigma}\rho_{0;p,s}\cdot\left[\sqrt{\frac{p'_++m}{p_++m}}\bm{e}^*_{k,l}\times\bm{\Pi}_e^{(\text{in})}(T)-\sqrt{\frac{p_++m}{p'_++m}}\bm{e}^*_{k,l}\times\bm{\Pi}_e^{\prime(\text{out})}(T)\right].
\end{split}
\end{equation}
By applying this expression for both spinor matrix elements in Eq. (\ref{dP_NCS_f}) and by indicating as $\bm{s}$ [$\bm{s}'$] the initial [final] spin of the incoming [outgoing] electron, with $\rho_{0;p,s}\rho^{\dag}_{0;p,s}=(1+\bm{\sigma}\cdot\bm{s})/2$ [$\rho'_{0;p',s'}\rho^{\prime\dag}_{0;p',s'}=(1+\bm{\sigma}\cdot\bm{s}')/2$], we have that the differential emission probability $dP^{(e^-\to e^-\gamma)}/d^3k_{\text{lc}}$ can be written as
\begin{equation}
\label{dP_NCS_Baier_1}
\frac{dP^{(e^-\to e^-\gamma)}}{d^3k_{\text{lc}}}=\frac{1}{2}\sum_{s,s',l}\frac{dP_{s,s',l}^{(e^-\to e^-\gamma)}}{d^3k_{\text{lc}}},
\end{equation}
where
\begin{equation}
\label{dP_NCS_Baier_2}
\begin{split}
&\frac{dP_{s,s',l}^{(e^-\to e^-\gamma)}}{d^3k_{\text{lc}}}\approx\lim_{\mathcal{V}_0\to\infty}\int_{\mathcal{V}_0}  \frac{d^3x_{0,\text{lc}}}{\mathcal{V}_0}\frac{\alpha}{16\pi^2}\frac{1}{p'_+k_+p_+}\int dTdT' \exp\left[i\frac{1}{p'_+}\int_{T'}^Td\tilde{T}(k\Pi_e^{(\text{in})}(\tilde{T}))\right]\\
&\quad\times\text{tr}\left\{\frac{1+\bm{\sigma}\cdot\bm{s}}{2}[R_C(T')-i\bm{\sigma}\cdot\bm{Q}_C(T')]\frac{1+\bm{\sigma}\cdot\bm{s}'}{2}[R^*_C(T)+i\bm{\sigma}\cdot\bm{Q}^*_C(T)]\right\},
\end{split}
\end{equation}
with
\begin{align}
R_C(T)&=\left(\sqrt{\frac{p'_++m}{p_++m}}+\sqrt{\frac{p_++m}{p'_++m}}\right)\bm{\Pi}_e^{(\text{in})}(T)\cdot\bm{e}_{k,l},\\
\label{Q_C}
\bm{Q}_C(T)&=\bm{e}_{k,l}\times\left[\sqrt{\frac{p'_++m}{p_++m}}\bm{\Pi}_e^{(\text{in})}(T)-\sqrt{\frac{p_++m}{p'_++m}}\bm{\Pi}_e^{\prime(\text{out})}(T)\right].
\end{align}
Equations (\ref{dP_NCS_Baier_1})-(\ref{dP_NCS_Baier_2}) indeed coincide with Baier's formula in Ref. \cite{Baier_b_1998} apart from the averaging over the initial coordinates that is not automatically obtained via Baier's method but has to be implemented by hand \cite{Baier_b_1998, Akhiezer_b_1996}. 

Finally, we note that the average $\lim_{\mathcal{V}_0\to\infty}\int_{\mathcal{V}_0} d^3x_{0,\text{lc}}/\mathcal{V}_0$ over the initial positions within the light-cone volume $\mathcal{V}_0$ can be in practice taken as an average $N^{-1}\sum_{n=1}^N$ over a large number $N$ trajectories all with the same incoming electron momentum (and spin quantum number).

\section{Nonlinear Breit-Wheeler pair production}

Analogous considerations as those in the previous section can be presented in the case of nonlinear Breit-Wheeler pair production. In order to keep the notation similar to that in the previous section, we indicate as $p^{\mu}$ ($p^{\prime\,\mu}$), and $s$ ($s'$) the four-momentum and the spin quantum number of the outgoing positron (electron), respectively, and as $k^{\mu}$ and $l$ the four-momentum and the polarization index of the incoming photon, respectively. The $S$-matrix transition element $S^{(\gamma\to e^-e^+)}$ is given by
\begin{equation}
S^{(\gamma\to e^-e^+)}=-ie\int dT\int_{\mathcal{V}_0} d^3x_{\text{lc}}\,\bar{U}^{(\text{out})}_{p',s'}(x)\frac{\hat{e}_{k,l}e^{-i(kx)}}{\sqrt{2k_+\mathcal{V}_0}}V^{(\text{out})}_{p,s}(x),
\end{equation}
where the positive-energy states in Eqs. (\ref{U_in})-(\ref{u_out}) and the negative-energy states in Eqs. (\ref{V_in})-(\ref{v_out}) are employed. Again, the upper index $(0)$ has been omitted and $\hbar$ has been set equal to unity for the sake of notational simplicity.

The differential pair-production probability with respect to the positron light-cone three-momentum $\bm{p}_{\text{lc}}$ and averaged (summed) over the initial (final) discrete quantum numbers, is given by [see also Eqs. (\ref{V_in})-(\ref{v_out}) for the negative-energy states]
\begin{equation}
\label{dP_NBWPP}
\begin{split}
&\frac{dP^{(\gamma\to e^-e^+)}}{d^3p_{\text{lc}}}=\lim_{\mathcal{V}_0\to\infty}\frac{e^2}{2}\mathcal{V}_0^2\sum_{s,s',l}\int \frac{d^3p'_{\text{lc}}}{(2\pi)^6}\int dTdT'\int_{\mathcal{V}_0} d^3x_{\text{lc}}d^3x'_{\text{lc}}\bar{U}^{(\text{out})}_{p',s'}(x)\frac{\hat{e}_{k,l}e^{-i(kx)}}{\sqrt{2k_+\mathcal{V}_0}}V^{(\text{out})}_{p,s}(x)\\
&\quad\times\bar{V}^{(\text{out})}_{p,s}(x')\frac{\hat{e}^*_{k,l}e^{i(kx')}}{\sqrt{2k_+\mathcal{V}_0}}U^{(\text{out})}_{p',s'}(x')\\
&\;=\lim_{\mathcal{V}_0\to\infty}\frac{e^2}{2}\sum_{s,s',l}\int \frac{d^3p'_{\text{lc}}}{(2\pi)^6\mathcal{V}_0}\int dTdT'\int_{\mathcal{V}_0} \frac{d^3x_{\text{lc}}d^3x'_{\text{lc}}}{8p'_+k_+p_+}\bar{u}^{(\text{out})}_{p',s'}(x)\hat{e}^*_{k,l}v^{(\text{out})}_{p,s}(x)\bar{v}^{(\text{out})}_{p,s}(x')\hat{e}_{k,l}u^{(\text{out})}_{p',s'}(x')\\
&\;\times e^{-\int_{\infty}^Td\tilde{T}(\partial P^{\prime(\text{out})}_e)/2P^{\prime(\text{out})}_{e,+}-\int_{\infty}^Td\tilde{T}(\partial P^{(\text{out})}_p)/2P^{(\text{out})}_{p,+}-\int_{\infty}^{T'}d\tilde{T}(\partial P^{\prime(\text{out})}_e)/2P^{\prime(\text{out})}_{e,+}-\int_{\infty}^{T'}d\tilde{T}(\partial P^{(\text{out})}_p)/2P^{(\text{out})}_{p,+}}\\
&\quad\times e^{i[-S_{p'}^{(\text{out})}(x)-(kx)+S_{-p}^{(\text{out})}(x)+S_{p'}^{(\text{out})}(x')+(kx')-S_{-p}^{(\text{out})}(x')]}.
\end{split}
\end{equation}
By applying exactly the same reasoning as in the previous section, one passes to the average and relative coordinates, expands the actions with respect to $\bm{x}_{-,\text{lc}}$ up to the first order while setting $\bm{x}_{-,\text{lc}}=\bm{0}$ in all other quantities, enforces energy-momentum conservation relations, neglects the real exponential functions corresponding to the van Vleck determinants, and arrives at
\begin{equation}
\label{dP_NBWPP_f}
\begin{split}
\frac{dP^{(\gamma\to e^-e^+)}}{d^3p_{\text{lc}}}&\approx\lim_{\mathcal{V}_0\to\infty}\int\frac{d^3x_{0,\text{lc}}}{\mathcal{V}_0}\frac{\alpha}{16\pi^2}\frac{1}{2p'_+k_+p_+}\sum_{s,s',l} \int dTdT'\bar{u}^{(\text{out})}_{p',s'}(x_T)\hat{e}_{k,l}v^{(\text{out})}_{p,s}(x_T)\\
&\quad\times \bar{v}^{(\text{out})}_{p,s}(x_{T'})\hat{e}^*_{k,l}u^{(\text{out})}_{p',s'}(x_{T'})\\
&\quad\times \exp\left\{i\int_{T'}^Td\tilde{T}\left[\frac{m^2+\bm{\Pi}^{\prime(\text{out})\,2}_{e,\perp}(\tilde{T})}{2p'_+}-\frac{\bm{k}^2_{\perp}}{2k_+}+\frac{m^2+\bm{\Pi}^{(\text{out})\,2}_{p,\perp}(\tilde{T})}{2p_+}\right]\right\},
\end{split}
\end{equation}
where $\bm{\Pi}^{\prime(\text{out})}_{e,\perp}(T)=\bm{k}_{\perp}-\bm{\Pi}^{(\text{out})}_{p,\perp}(T)$ and $p'_+=k_+-p_+$, and where all coordinates $\bm{x}_{\text{lc}}$ in the spinor have to be expressed in terms of the final coordinates $\bm{x}_{0,\text{lc}}$ following the corresponding positron classical trajectory in the external field. The interpretation of $\bm{x}_{0,\text{lc}}$ as the final coordinates of the positron follows from the interpretation of the van Vleck determinants originally included in the expression of the probability because the three integrals over the asymptotic electron momenta have to be transformed into integrals over the local electron momenta and therefore the coordinates $\bm{x}_{0,\text{lc}}$ pertain to the positron.

Now, by noticing that
\begin{equation}
\int_{T'}^Td\tilde{T}\left[\frac{m^2+\bm{\Pi}^{\prime(\text{out})\,2}_{e,\perp}(\tilde{T})}{2p'_+}-\frac{\bm{k}^2_{\perp}}{2k_+}+\frac{m^2+\bm{\Pi}^{(\text{out})\,2}_{p,\perp}(\tilde{T})}{2p_+}\right]=\frac{1}{p'_+}\int_{T'}^Td\tilde{T}(k\Pi_p^{(\text{out})}(\tilde{T})),
\end{equation}
and by manipulating the spinor matrix elements analogously as in the previous section, we finally obtain that the differential pair-production probability $dP^{(\gamma\to e^-e^+)}/d^3p_{\text{lc}}$ can be written as
\begin{equation}
\label{dP_NBWPP_Baier_1}
\frac{dP^{(\gamma\to e^-e^+)}}{d^3p_{\text{lc}}}=\frac{1}{2}\sum_{s,s',l}\frac{dP_{s,s',l}^{(\gamma\to e^-e^+)}}{d^3p_{\text{lc}}},
\end{equation}
where
\begin{equation}
\label{dP_NBWPP_Baier_2}
\begin{split}
&\frac{dP_{s,s',l}^{(\gamma\to e^-e^+)}}{d^3p_{\text{lc}}}\approx \lim_{\mathcal{V}_0\to\infty}\int_{\mathcal{V}_0} \frac{d^3x_{0,\text{lc}}}{\mathcal{V}_0}\frac{\alpha}{16\pi^2}\frac{1}{p'_+k_+p_+}\int dTdT' \exp\left[i\frac{1}{p'_+}\int_{T'}^Td\tilde{T}(k\Pi_p^{(\text{out})}(\tilde{T}))\right]\\
&\;\;\times\text{tr}\left\{\frac{1+\bm{\sigma}\cdot\bm{s}}{2}[R^*_{BW}(T')+i\bm{\sigma}\cdot\bm{Q}^*_{BW}(T')]\frac{1+\bm{\sigma}\cdot\bm{s}'}{2}[R_{BW}(T)-i\bm{\sigma}\cdot\bm{Q}_{BW}(T)]\right\},
\end{split}
\end{equation}
with $\bm{s}$ ($\bm{s}'$) being the asymptotic spin of the outgoing positron (electron) and with
\begin{align}
R_{BW}(T)&=\frac{\bm{e}_{k,l}\cdot[\bm{\Pi}_p^{(\text{out})}(T)\times\bm{\Pi}_e^{\prime(\text{out})}(T)]}{\sqrt{(p_++m)(p'_++m)}},\\
\label{Q_BW}
\begin{split}
\bm{Q}_{BW}(T)&=\left[\sqrt{(p_++m)(p'_++m)}-\frac{\bm{\Pi}_e^{\prime(\text{out})}(T)\cdot\bm{\Pi}_p^{(\text{out})}(T)}{\sqrt{(p_++m)(p'_++m)}}\right]\bm{e}_{k,l}\\
&\quad+\frac{[\bm{\Pi}_e^{\prime(\text{out})}(T)-\bm{\Pi}_p^{(\text{out})}(T)]\cdot\bm{e}_{k,l}}{2\sqrt{(p_++m)(p'_++m)}}[\bm{\Pi}_p^{(\text{out})}(T)-\bm{\Pi}_e^{\prime(\text{out})}(T)].
\end{split}
\end{align}
It is easily seen that Eqs. (\ref{dP_NBWPP_Baier_1})-(\ref{dP_NBWPP_Baier_2}) also coincide with the corresponding Baier's formula in Ref. \cite{Baier_b_1998} again apart from the averaging, this time, over the final coordinates of the positron. Also in this case, the average can be taken over a large number of positron trajectories all with the same asymptotic final momentum (and spin quantum number).

\section{Conclusions and outlook}
We have computed WKB wave functions for electrons in the presence of electromagnetic fields of arbitrary spacetime structure having in mind the case of tightly focused laser beams. The present expressions of the wave functions generalize those found in Refs. \cite{Di_Piazza_2014,Di_Piazza_2015} because they only rely on the validity of the WKB approximation. In fact, we have found the three $\hbar$-dependent conditions $\lambda^2_C/\sigma_0^2\ll \chi/\xi$, $\lambda_0\lambda_C/\sigma_0^2\ll \xi$, and $\lambda_C/\sigma_0\ll 1$ for the validity of the obtained states in the case of a tightly focused Gaussian with the above-discussed parameters. The additional approximation on the electron trajectory inside the laser field gave the possibility in Refs. \cite{Di_Piazza_2014,Di_Piazza_2015} to obtain explicit expressions of the wave functions in terms of the external field, which in turn allowed us analytical computations. The present wave functions are instead more suitable for numerical approaches and can be obtained starting from the classical equations of motion for the electron trajectory (Lorentz equation) and its spin (equation for the two-dimensional spinor leading to the BMT equation for the average spin/magnetic moment). 

In addition, we have shown that in the case of a plane-wave background field the found WKB wave functions exactly reduce to the Volkov states, which have been written in a new form, where also the spinor structure has a manifest quasiclassical form. 

Finally, by computing the probabilities of nonlinear Compton scattering and nonlinear Breit-Wheeler pair production, we have been able to reproduce Baier's formulas for generic electron trajectories. 

Among others, we plan to use the present wave functions to further investigate the effects of the transverse formation length on the photon emission spectrum in order to obtain more quantitative results than the analytical estimations presented in Ref. \cite{Di_Piazza_2021} concerning nonlinear Compton scattering in a flying focus beam \cite{Sainte-Marie_2017,Froula_2018,Palastro_2018,Howard_2019,Palastro_2020}.

\begin{acknowledgments}
I thank the Yukawa Institute for Theoretical Physics at Kyoto University as discussions in particular with H. Taya during YITP-T-20-05 ``The Schwinger Effect and Strong-Field Physics Workshop'' were useful to complete this work.

This publication was also supported by the Collaborative Research Centre 1225 funded by Deutsche Forschungsgemeinschaft (DFG, German Research Foundation) - Project-ID 273811115 - SFB 1225.
\end{acknowledgments}

\appendix

\section{Determination of the action in a plane wave via the method of characteristics}
\label{App_A}
In this appendix we report, as an application of the method of characteristics, the determination of the classical action of an electron in a plane wave and we use the same notation as in the main text. Since the plane wave is assumed to propagate along the negative $z$ direction, it can be described by the four-vector potential $A^{\mu}(T)=(0,\bm{A}_{\perp}(T),0)$, which follows from the choice $A_-(T)=0$ within the Lorenz gauge [$(\partial A)=dA_+(T)/dT=0$] and from the initial condition $A^{\mu}(T_0)=0$.

In this case it is convenient to use directly the coordinate $T$ to parametrize the electron trajectory $\bm{x}_{\text{lc}}=\bm{x}_{\text{lc}}(T;T_0)$ for $T\ge T_0$. As in the main text, the on-shell electron four-momentum $\Pi_e^{\mu}(T;T_0)$ is set equal to $p^{\mu}$ at $T=T_0$, whereas its initial light-cone coordinates are $\bm{x}_{\text{lc}}(T_0;T_0)=\bm{x}_{0,\text{lc}}$. By using the three conservation laws of the plus and the perpendicular components of the canonical four-momentum together with the on-shell condition $\Pi_e^2(T;T_0)=2\Pi_{e,+}(T;T_0)\Pi_{e,-}(T;T_0)-\bm{\Pi}^2_{e,\perp}(T;T_0)=m^2$, one can directly write the four-momentum $\Pi_e^{\mu}(T;T_0)$ at an arbitrary $T$ in a covariant form as
\begin{equation}
\label{P_e_PW}
\Pi_e^{\mu}(T;T_0)=p^{\mu}-eA^{\mu}(T)+e\frac{(pA(T))}{p_+}\tilde{n}^{\mu}-e^2\frac{A^2(T)}{2p_+}\tilde{n}
^{\mu}
\end{equation}
in terms of the initial four-momentum $p^{\mu}$. Correspondingly, the $T$-dependent electron coordinates $\bm{x}_{\text{lc}}(T;T_0)$ can be derived from the equation $p_+dx^{\mu}/dT=\Pi_e^{\mu}$ and the result is
\begin{align}
\label{x_perp_T}
\bm{x}_{\perp}(T;T_0)&=\bm{x}_{0,\perp}+\frac{1}{p_+}\int_{T_0}^TdT'[\bm{p}_{\perp}-e\bm{A}_{\perp}(T')],\\
\label{phi_T}
\begin{split}
\phi(T;T_0)&=\phi_0+\frac{1}{p_+}\int_{T_0}^TdT'\left[p_-+e\frac{(pA(T'))}{p_+}-e^2\frac{A^2(T')}{2p_+}\right]\\
&=\phi_0+\int_{T_0}^TdT'\frac{m^2+[\bm{p}_{\perp}-e\bm{A}_{\perp}(T')]^2}{2p^2_+}.
\end{split}
\end{align}
At this point the action $\Sigma_p(T;T_0)$ along the electron trajectory at hand can be found from the equation [see Eq. (\ref{Eq_mot_Sigma})]
\begin{equation}
\frac{d\Sigma_p}{dT}=-\frac{m^2}{p_+}+e\frac{\bm{\Pi}_{e,\perp}\cdot \bm{A}_{\perp}}{p_+}
\end{equation}
with the initial condition $\Sigma_p(T_0;T_0)=-(p_+\phi_0+p_-T_0-\bm{p}_{\perp}\cdot\bm{x}_{0,\perp})$. In this way, we obtain the action $\Sigma_p(T;x_0)$ along the specific trajectory with initial conditions $\bm{x}_{0,\text{lc}}$:
\begin{equation}
\label{Sigma_PW}
\begin{split}
\Sigma_p(T;x_0)&=\Sigma_p(T_0;T_0)-\frac{m^2}{p_+}(T-T_0)+\frac{e}{p_+}\int_{T_0}^TdT'\,\bm{\Pi}_{e,\perp}(T')\cdot \bm{A}_{\perp}(T')\\
&=-(p_+\phi_0+p_-T_0-\bm{p}_{\perp}\cdot\bm{x}_{0,\perp})-\frac{m^2}{p_+}(T-T_0)\\
&\quad+\frac{e}{p_+}\int_{T_0}^TdT'[\bm{p}_{\perp}-e\bm{A}_{\perp}(T')]\cdot \bm{A}_{\perp}(T').
\end{split}
\end{equation}
Finally, in order to obtain the action $S_p(x;T_0)$ we have to express the initial coordinates $\bm{x}_{0,\text{lc}}$ of the electron in terms of the generic coordinates $\bm{x}_{\text{lc}}$ by inverting the functions $\bm{x}_{\text{lc}}=\bm{x}_{\text{lc}}(T;T_0,\bm{x}_{0,\text{lc}},\bm{p}_{\text{lc}})$ via Eqs. (\ref{x_perp_T})-(\ref{phi_T}). The resulting action [see Eq. (\ref{Sigma_PW})]
\begin{equation}
\label{S_PW}
\begin{split}
S_p(x;T_0)&=-\left\{p_+\phi-\int_{T_0}^TdT'\frac{m^2+[\bm{p}_{\perp}-e\bm{A}_{\perp}(T')]^2}{2p_+}+\frac{m^2+\bm{p}_{\perp}^2}{2p_+}T_0-\bm{p}_{\perp}\cdot\bm{x}_{\perp}\right.\\
&\left.\quad+\frac{1}{p_+}\int_{T_0}^TdT'\bm{p}_{\perp}\cdot[\bm{p}_{\perp}-e\bm{A}_{\perp}(T')]\right\}-\frac{m^2}{p_+}(T-T_0)\\
&\quad+\frac{e}{p_+}\int_{T_0}^TdT'[\bm{p}_{\perp}-e\bm{A}_{\perp}(T')]\cdot \bm{A}_{\perp}(T')\\
&=-(p_+\phi+p_-T-\bm{p}_{\perp}\cdot\bm{x}_{\perp})+\int_{T_0}^TdT'\left[e\frac{\bm{p}_{\perp}\cdot\bm{A}_{\perp}(T')}{p_+}-e^2\frac{\bm{A}_{\perp}^2(T')}{2p_+}\right]\\
&=-(p_+\phi+p_-T-\bm{p}_{\perp}\cdot\bm{x}_{\perp})-\int_{T_0}^TdT'\left[e\frac{(pA(T'))}{p_+}-e^2\frac{A^2(T')}{2p_+}\right]
\end{split}
\end{equation}
coincides with the result in, e.g., Ref. \cite{Landau_b_2_1975}. Finally, it can be easily verified that the general relations $\bm{x}_{0,\perp}-(\bm{p}_{\perp}/p_+)T_0=\bm{\nabla}_{\bm{p}_{\perp}}S_p(x;T_0)$ and $-\phi_0+(p_-/p_+)T_0=\partial_{p_+}S_p(x;T_0)$, with $p_-=(m^2+\bm{p}_{\perp}^2)/2p_+$ hold in this case [see the discussion above Eq. (\ref{vV_det_2})].

\section{Ultrarelativistic electron trajectory in an arbitrarily focused laser beam}
\label{App_B}
In this appendix we report some general considerations on the Lorentz equation of an ultrarelativistic electron in the presence of an arbitrary background electromagnetic field described by the four-vector potential $A^{\mu}(x)$ introduced in the main text and having in mind the case of a tightly focused laser beam. In the first part we draw conclusions solely based on the structure of the equations and in the last part we report an approximated analytical expression of the trajectory of the electron, pointing out the requirements on the field for the approximations to be valid.

We choose the light-cone coordinates introduced in the main text and we describe the electron light-cone position $\bm{x}_{\text{lc}}(T;T_0)$ and kinetic momentum $\bm{\Pi}_{e,\text{lc}}(T;T_0)$ as functions of the light-cone time $T$. The remaining light-cone component $\Pi_{e,-}(T;T_0)$ of the four-momentum $\Pi_e^{\mu}(T;T_0)$ is obtained from the on-shell condition as $\Pi_{e,-}(T;T_0)=[m^2+\bm{\Pi}^2_{e,\perp}(T;T_0)]/2\Pi_{e,+}(T;T_0)$. The trajectory is determined by solving the Lorentz equation, which can be written as
\begin{equation}
\label{L_Eq}
\Pi_{e,+}\frac{d\Pi_e^{\mu}}{dT}=eF^{\mu\nu}\Pi_{e,\nu}.
\end{equation}
The initial conditions are fixed at $T=T_0$ as $\bm{x}_{\text{lc}}(T_0;T_0)=\bm{x}_{0,\text{lc}}$ and $\bm{\Pi}_{e,\text{lc}}(T_0;T_0)=\bm{p}_{\text{lc}}$ [$p_-=(m^2+\bm{p}^2_{\perp})/2p_+$], where the four-vector potential is assumed to vanish for all $\bm{x}_{\text{lc}}$.

We assume that the electron is ultrarelativistic and almost counterpropagating with respect to the laser field, i.e., that the hierarchy $\Pi_{e,+}(T;T_0)\gg\max(m,|\bm{\Pi}_{e,\perp}(T;T_0)|)\gg \Pi_{e,-}(T;T_0)$ among the light-cone components of the electron four-momentum is verified at all light-cone times $T\ge T_0$. For the sake of definiteness, by referring to the standard situation of an electron in a plane-wave field, we can estimate $|\bm{\Pi}_{e,\perp}(T;T_0)|$ as $|\bm{\Pi}_{e,\perp}(T;T_0)|\lesssim m\xi$ and $\Pi_{e,+}(T;T_0)$ as $\Pi_{e,+}(T;T_0)\sim \varepsilon$, and we arrive to the more transparent condition $\eta=\max(m,m\xi)/\varepsilon\ll 1$.

By expressing the Lorentz equation (\ref{L_Eq}) in light-cone coordinates, one obtains
\begin{align}
\label{Pi_p_c}
\frac{d\bm{\Pi}_{e,\perp}}{dT}&=-\frac{d\bm{\mathcal{A}}_{\perp}}{dT}+\frac{\bm{\Pi}_{e,\perp}}{\Pi_{e,+}}\cdot\bm{\nabla}_{\perp}\bm{\mathcal{A}}_{\perp}+\frac{(\bm{\Pi}_e\times\bm{\mathcal{B}}_z)_{\perp}}{\Pi_{e,+}}-\frac{m^2+\bm{\Pi}^2_{e,\perp}}{2\Pi^2_{e,+}}\bm{\nabla}_{\perp}\mathcal{A}_+,\\\label{Pi_+_c}
\frac{d\Pi_{e,+}}{dT}&=-\frac{d\mathcal{A}_+}{dT}-\frac{\bm{\Pi}_{e,\perp}}{\Pi_{e,+}}\cdot\frac{\partial \bm{\mathcal{A}}_{\perp}}{\partial\phi}+\frac{m^2+\bm{\Pi}^2_{e,\perp}}{2\Pi^2_{e,+}}\frac{\partial\mathcal{A}_+}{\partial\phi},
\end{align}
where $\mathcal{A}^{\mu}(x)=eA^{\mu}(x)$, where $\bm{\mathcal{B}}_z(x)=[(\bm{\nabla}\times \bm{\mathcal{A}}(x))\cdot\bm{z}]\bm{z}$, and where, unlike in Ref. \cite{Di_Piazza_2014}, we preferred to express the trajectory in terms of the components $\mathcal{A}_+(x)$ and $\bm{\mathcal{A}}_{\perp}(x)$ of the four-vector potential [recall that $\mathcal{A}_-(x)=0$]. In addition, the equations for the coordinates $\bm{x}_{\perp}(T;T_0)$ and $\phi(T;T_0)$ are
\begin{align}
\label{x_p_c}
\frac{d\bm{x}_{\perp}}{dT}&=\frac{\bm{\Pi}_{e,\perp}}{\Pi_{e,+}},\\
\label{phi_c}
\frac{d\phi}{dT}&=\frac{m^2+\bm{\Pi}^2_{e,\perp}}{2\Pi^2_{e,+}}.
\end{align}
Since we implicitly assume that the light-cone components of the external field are such that $|\mathcal{A}_+(x)|,|\bm{\mathcal{A}}_{\perp}(x)|\lesssim m\xi$, the equations (\ref{Pi_p_c})-(\ref{Pi_+_c}) for the four-momentum components indicate that the variation of the plus component due to the external field is typically much smaller than $\Pi_{e,+}(T;T_0)$. On the contrary, the external field can substantially change the transverse momenta, which is already known in the analytically solvable plane-wave case. Finally, the equations (\ref{x_p_c})-(\ref{phi_c}) for the variations in $T$ of the coordinates instead show that these are always suppressed for ultrarelativistic electrons, with the variation in $|\bm{x}_{\perp}(T;T_0)|$ [$\phi(T;T_0)$] scaling as $1/\Pi_{e,+}(T;T_0)$ [$1/\Pi^2_{e,+}(T;T_0)$]. These features are exploited to estimate the formation region on these coordinates in strong-field QED processes like nonlinear Compton scattering and nonlinear Breit-Wheeler pair production.

We would like to conclude by making a step further and determine an approximated analytical solution of Eqs. (\ref{Pi_p_c})-(\ref{phi_c}). Unlike in Ref. \cite{Di_Piazza_2014}, we allow for the initial transverse momentum of the electron not to vanish: $\bm{p}_{\perp}\ne \bm{0}$. We only rely on an expansion in powers of $1/p_+$ and we compute for each quantity only the leading-order correction (which is slightly different from what we did in Ref. \cite{Di_Piazza_2014}). We comment on the conditions allowing for such an expansion afterwards. 

Concerning the independent light-cone components of the four-momentum, we obtain
\begin{align}
\begin{split}
\bm{\Pi}_{e,\perp}(T;T_0)&=\bm{p}_{\perp}-\bm{\mathcal{A}}_{\perp}+\frac{1}{p_+}\int_{T_0}^TdT'\{\bm{\Pi}_{e,\perp}\cdot\bm{\nabla}_{\perp}\bm{\mathcal{A}}_{\perp}+(\bm{\Pi}_e\times\bm{\mathcal{B}}_z)_{\perp}\}+O\left(\frac{1}{p^2_+}\right)\\
&=\bm{p}_{\perp}-\bm{\mathcal{A}}_{\perp}\\
&\quad+\frac{1}{p_+}\int_{T_0}^TdT'\{(\bm{p}_{\perp}-\bm{\mathcal{A}}_{\perp})\cdot\bm{\nabla}_{\perp}\bm{\mathcal{A}}_{\perp}+[(\bm{p}_{\perp}-\bm{\mathcal{A}}_{\perp})\times\bm{\mathcal{B}}_z]_{\perp}\}+O\left(\frac{1}{p^2_+}\right),
\end{split}\\
\Pi_{e,+}(T;T_0)&=p_+-\mathcal{A}_++O\left(\frac{1}{p_+}\right),
\end{align}
where all the fields are still computed along the electron trajectory. Analogously, starting from Eqs. (\ref{x_p_c})-(\ref{phi_c}), we can write the approximated expressions of the coordinates as 
\begin{align}
\bm{x}_{\perp}(T;T_0)&=\bm{x}_{0,\perp}+\frac{1}{p_+}\int_{T_0}^TdT'(\bm{p}_{\perp}-\bm{\mathcal{A}}_{\perp})+O\left(\frac{1}{p^2_+}\right),\\
\phi(T;T_0)&=\phi_0+O\left(\frac{1}{p^2_+}\right).
\end{align}
We can use these expressions to expand the fields, which are still computed along the electron trajectory. By introducing the quantities $\bm{\mathcal{A}}^{(0)}_{\text{lc}}(T)=\bm{\mathcal{A}}_{\text{lc}}(T,\bm{x}_{0,\text{lc}})$ and $\bm{\mathcal{B}}^{(0)}_z(T)=\bm{\mathcal{B}}_z(T,\bm{x}_{0,\text{lc}})$, we can write
\begin{align}
\label{x_p_sol}
\bm{x}_{\perp}(T;T_0)&=\bm{x}_{0,\perp}+\frac{1}{p_+}\int_{T_0}^TdT'[\bm{p}_{\perp}-\bm{\mathcal{A}}^{(0)}_{\perp}(T')]+O\left(\frac{1}{p^2_+}\right),\\
\label{phi_sol}
\phi(T;T_0)&=\phi_0+O\left(\frac{1}{p^2_+}\right)
\end{align}
and
\begin{align}
\label{Pi_p_sol}
\begin{split}
\bm{\Pi}_{e,\perp}(T;T_0)&=\bm{p}_{\perp}-\bm{\mathcal{A}}^{(0)}_{\perp}(T)-\frac{1}{p_+}\int_{T_0}^TdT'[\bm{p}_{\perp}-\bm{\mathcal{A}}^{(0)}_{\perp}(T')]\cdot\bm{\nabla}_{\perp}\bm{\mathcal{A}}^{(0)}_{\perp}(T)\\
&\quad+\frac{1}{p_+}\int_{T_0}^TdT'\{[\bm{p}_{\perp}-\bm{\mathcal{A}}^{(0)}_{\perp}(T')]\cdot\bm{\nabla}_{\perp}\bm{\mathcal{A}}^{(0)}_{\perp}(T')\\
&\qquad+[[\bm{p}_{\perp}-\bm{\mathcal{A}}^{(0)}_{\perp}(T')]\times\bm{\mathcal{B}}^{(0)}_z(T')]_{\perp}\}+O\left(\frac{1}{p^2_+}\right),\\
\Pi_{e,+}(T;T_0)&=p_+-\mathcal{A}^{(0)}_+(T)+O\left(\frac{1}{p_+}\right).
\end{split}
\end{align}
These expressions of the independent light-cone coordinates and momenta of the electron are an explicit approximated solution of the equation of motion in the ultrarelativistic regime as they are expressed in terms of the external field. 

Some additional remarks are in order, concerning the validity of the used approximation and it is convenient to consider the definite example of an electron initially almost counterpropagating with a tightly focused Gaussian optical beam, characterized by a central angular frequency $\omega_0$ (central wavelength $\lambda_0=2\pi/\omega_0$), transverse spot radius $\sigma_0$ (Rayleigh length $l_R=\pi \sigma_0^2/\lambda_0$), pulse duration $\tau$, and electromagnetic field amplitude $F_0$. By saying that the electron is initially ``almost'' counterpropagating with respect to the laser field, we mean that the initial transverse momentum is less than or of the order of the transverse momentum that the laser field can impart to the electron, i.e., $|\bm{p}_{\perp}|\lesssim m\xi$. Even though in a Gaussian beam the field decreases along the longitudinal direction only as the inverse of the distance from the focal area, for the sake of definiteness, we estimate as $\tau_i=\max(2l_R,\tau)$ the maximum time that the electron spends inside the strong field (interaction time). Thus, it is reasonable to require that the above approximated expressions for the trajectory and the four-momentum components are valid for $0\le T-T_0\lesssim \tau_i$. First, notice that the constant initial transverse momentum $\bm{p}_{\perp}$ induces a drift term in the transverse position which increases linearly with $T$. By imposing that the electron does not exit sideways the laser pulse, we obtain the condition $|\bm{p}_{\perp}|\tau_i/p_+\lesssim \sigma_0$. This condition is fulfilled because it is equivalent to the condition $\eta\lesssim \lambda_0/\sigma_0$, which is fulfilled because $\eta\ll 1$ and because for a tightly focused laser beam it is $\lambda_0\sim \sigma_0$. Correspondingly, the terms proportional to $1/p_+$ in Eq. (\ref{x_p_sol}) are less than or of the order of $\eta(\sigma_0/\lambda_0)\sigma_0$ and then, assuming that $|\bm{x}_{0,\perp}|\lesssim \sigma_0$, we see that they do not alter the hierarchy determined by the large momentum scale $p_+$. The condition $\eta\lesssim \lambda_0/\sigma_0$ also ensures that the external field can be expanded around $\bm{x}_{0,\perp}$. Finally, we observe that accumulation effects arising in Eq. (\ref{Pi_p_sol}) due to the terms quadratic in the fields and integrated over $T'$ might affect the discussed hierarchy. However, under the above conditions, one can estimate that these terms are in order of magnitude about $\eta(\sigma_0/\lambda_0)$ smaller than the leading-order terms in agreement with the hierarchy.

%


\begin{thebibliography}{131}%
\makeatletter
\providecommand \@ifxundefined [1]{%
 \@ifx{#1\undefined}
}%
\providecommand \@ifnum [1]{%
 \ifnum #1\expandafter \@firstoftwo
 \else \expandafter \@secondoftwo
 \fi
}%
\providecommand \@ifx [1]{%
 \ifx #1\expandafter \@firstoftwo
 \else \expandafter \@secondoftwo
 \fi
}%
\providecommand \natexlab [1]{#1}%
\providecommand \enquote  [1]{``#1''}%
\providecommand \bibnamefont  [1]{#1}%
\providecommand \bibfnamefont [1]{#1}%
\providecommand \citenamefont [1]{#1}%
\providecommand \href@noop [0]{\@secondoftwo}%
\providecommand \href [0]{\begingroup \@sanitize@url \@href}%
\providecommand \@href[1]{\@@startlink{#1}\@@href}%
\providecommand \@@href[1]{\endgroup#1\@@endlink}%
\providecommand \@sanitize@url [0]{\catcode `\\12\catcode `\$12\catcode
  `\&12\catcode `\#12\catcode `\^12\catcode `\_12\catcode `\%12\relax}%
\providecommand \@@startlink[1]{}%
\providecommand \@@endlink[0]{}%
\providecommand \url  [0]{\begingroup\@sanitize@url \@url }%
\providecommand \@url [1]{\endgroup\@href {#1}{\urlprefix }}%
\providecommand \urlprefix  [0]{URL }%
\providecommand \Eprint [0]{\href }%
\providecommand \doibase [0]{https://doi.org/}%
\providecommand \selectlanguage [0]{\@gobble}%
\providecommand \bibinfo  [0]{\@secondoftwo}%
\providecommand \bibfield  [0]{\@secondoftwo}%
\providecommand \translation [1]{[#1]}%
\providecommand \BibitemOpen [0]{}%
\providecommand \bibitemStop [0]{}%
\providecommand \bibitemNoStop [0]{.\EOS\space}%
\providecommand \EOS [0]{\spacefactor3000\relax}%
\providecommand \BibitemShut  [1]{\csname bibitem#1\endcsname}%
\let\auto@bib@innerbib\@empty
\bibitem [{\citenamefont {Reiss}(1962)}]{Reiss_1962}%
  \BibitemOpen
  \bibfield  {author} {\bibinfo {author} {\bibfnamefont {H.~R.}\ \bibnamefont
  {Reiss}},\ }\href@noop {} {\bibfield  {journal} {\bibinfo  {journal} {J.
  Math. Phys. (N.Y.)}\ }\textbf {\bibinfo {volume} {3}},\ \bibinfo {pages} {59}
  (\bibinfo {year} {1962})}\BibitemShut {NoStop}%
\bibitem [{\citenamefont {Nikishov}\ and\ \citenamefont
  {Ritus}(1964)}]{Nikishov_1964}%
  \BibitemOpen
  \bibfield  {author} {\bibinfo {author} {\bibfnamefont {A.~I.}\ \bibnamefont
  {Nikishov}}\ and\ \bibinfo {author} {\bibfnamefont {V.~I.}\ \bibnamefont
  {Ritus}},\ }\href@noop {} {\bibfield  {journal} {\bibinfo  {journal} {Sov.
  Phys.-JETP}\ }\textbf {\bibinfo {volume} {19}},\ \bibinfo {pages} {529}
  (\bibinfo {year} {1964})}\BibitemShut {NoStop}%
\bibitem [{\citenamefont {Gol'dman}(1964)}]{Goldman_1964}%
  \BibitemOpen
  \bibfield  {author} {\bibinfo {author} {\bibfnamefont {I.~I.}\ \bibnamefont
  {Gol'dman}},\ }\href@noop {} {\bibfield  {journal} {\bibinfo  {journal}
  {Phys. Lett.}\ }\textbf {\bibinfo {volume} {8}},\ \bibinfo {pages} {103}
  (\bibinfo {year} {1964})}\BibitemShut {NoStop}%
\bibitem [{\citenamefont {Brown}\ and\ \citenamefont
  {Kibble}(1964)}]{Brown_1964}%
  \BibitemOpen
  \bibfield  {author} {\bibinfo {author} {\bibfnamefont {L.~S.}\ \bibnamefont
  {Brown}}\ and\ \bibinfo {author} {\bibfnamefont {T.~W.~B.}\ \bibnamefont
  {Kibble}},\ }\href@noop {} {\bibfield  {journal} {\bibinfo  {journal} {Phys.
  Rev.}\ }\textbf {\bibinfo {volume} {133}},\ \bibinfo {pages} {A705} (\bibinfo
  {year} {1964})}\BibitemShut {NoStop}%
\bibitem [{\citenamefont {Furry}(1951)}]{Furry_1951}%
  \BibitemOpen
  \bibfield  {author} {\bibinfo {author} {\bibfnamefont {W.~H.}\ \bibnamefont
  {Furry}},\ }\href@noop {} {\bibfield  {journal} {\bibinfo  {journal} {Phys.
  Rev.}\ }\textbf {\bibinfo {volume} {81}},\ \bibinfo {pages} {115} (\bibinfo
  {year} {1951})}\BibitemShut {NoStop}%
\bibitem [{\citenamefont {Berestetskii}\ \emph {et~al.}(1982)\citenamefont
  {Berestetskii}, \citenamefont {Lifshitz},\ and\ \citenamefont
  {Pitaevskii}}]{Landau_b_4_1982}%
  \BibitemOpen
  \bibfield  {author} {\bibinfo {author} {\bibfnamefont {V.~B.}\ \bibnamefont
  {Berestetskii}}, \bibinfo {author} {\bibfnamefont {E.~M.}\ \bibnamefont
  {Lifshitz}},\ and\ \bibinfo {author} {\bibfnamefont {L.~P.}\ \bibnamefont
  {Pitaevskii}},\ }\href@noop {} {\emph {\bibinfo {title} {Quantum
  Electrodynamics}}}\ (\bibinfo  {publisher} {Elsevier Butterworth-Heinemann,
  Oxford},\ \bibinfo {year} {1982})\BibitemShut {NoStop}%
\bibitem [{\citenamefont {Volkov}(1935)}]{Volkov_1935}%
  \BibitemOpen
  \bibfield  {author} {\bibinfo {author} {\bibfnamefont {D.~M.}\ \bibnamefont
  {Volkov}},\ }\href@noop {} {\bibfield  {journal} {\bibinfo  {journal} {Z.
  Phys.}\ }\textbf {\bibinfo {volume} {94}},\ \bibinfo {pages} {250} (\bibinfo
  {year} {1935})}\BibitemShut {NoStop}%
\bibitem [{\citenamefont {Mitter}(1975)}]{Mitter_1975}%
  \BibitemOpen
  \bibfield  {author} {\bibinfo {author} {\bibfnamefont {H.}~\bibnamefont
  {Mitter}},\ }\href@noop {} {\bibfield  {journal} {\bibinfo  {journal} {Acta
  Phys. Austriaca}\ }\textbf {\bibinfo {volume} {XIV}},\ \bibinfo {pages} {397}
  (\bibinfo {year} {1975})}\BibitemShut {NoStop}%
\bibitem [{\citenamefont {Ritus}(1985)}]{Ritus_1985}%
  \BibitemOpen
  \bibfield  {author} {\bibinfo {author} {\bibfnamefont {V.~I.}\ \bibnamefont
  {Ritus}},\ }\href@noop {} {\bibfield  {journal} {\bibinfo  {journal} {J. Sov.
  Laser Res.}\ }\textbf {\bibinfo {volume} {6}},\ \bibinfo {pages} {497}
  (\bibinfo {year} {1985})}\BibitemShut {NoStop}%
\bibitem [{\citenamefont {Ehlotzky}\ \emph {et~al.}(2009)\citenamefont
  {Ehlotzky}, \citenamefont {Krajewska},\ and\ \citenamefont
  {Kami\'{n}ski}}]{Ehlotzky_2009}%
  \BibitemOpen
  \bibfield  {author} {\bibinfo {author} {\bibfnamefont {F.}~\bibnamefont
  {Ehlotzky}}, \bibinfo {author} {\bibfnamefont {K.}~\bibnamefont
  {Krajewska}},\ and\ \bibinfo {author} {\bibfnamefont {J.~Z.}\ \bibnamefont
  {Kami\'{n}ski}},\ }\href@noop {} {\bibfield  {journal} {\bibinfo  {journal}
  {Rep. Prog. Phys.}\ }\textbf {\bibinfo {volume} {72}},\ \bibinfo {pages}
  {046401} (\bibinfo {year} {2009})}\BibitemShut {NoStop}%
\bibitem [{\citenamefont {Reiss}(2009)}]{Reiss_2009}%
  \BibitemOpen
  \bibfield  {author} {\bibinfo {author} {\bibfnamefont {H.~R.}\ \bibnamefont
  {Reiss}},\ }\href@noop {} {\bibfield  {journal} {\bibinfo  {journal} {Eur.
  Phys. J. D}\ }\textbf {\bibinfo {volume} {55}},\ \bibinfo {pages} {365}
  (\bibinfo {year} {2009})}\BibitemShut {NoStop}%
\bibitem [{\citenamefont {Di~Piazza}\ \emph {et~al.}(2012)\citenamefont
  {Di~Piazza}, \citenamefont {M\"{u}ller}, \citenamefont {Hatsagortsyan},\ and\
  \citenamefont {Keitel}}]{Di_Piazza_2012}%
  \BibitemOpen
  \bibfield  {author} {\bibinfo {author} {\bibfnamefont {A.}~\bibnamefont
  {Di~Piazza}}, \bibinfo {author} {\bibfnamefont {C.}~\bibnamefont
  {M\"{u}ller}}, \bibinfo {author} {\bibfnamefont {K.~Z.}\ \bibnamefont
  {Hatsagortsyan}},\ and\ \bibinfo {author} {\bibfnamefont {C.~H.}\
  \bibnamefont {Keitel}},\ }\href@noop {} {\bibfield  {journal} {\bibinfo
  {journal} {Rev. Mod. Phys.}\ }\textbf {\bibinfo {volume} {84}},\ \bibinfo
  {pages} {1177} (\bibinfo {year} {2012})}\BibitemShut {NoStop}%
\bibitem [{\citenamefont {Dunne}(2014)}]{Dunne_2014}%
  \BibitemOpen
  \bibfield  {author} {\bibinfo {author} {\bibfnamefont {G.~V.}\ \bibnamefont
  {Dunne}},\ }\href@noop {} {\bibfield  {journal} {\bibinfo  {journal} {Eur.
  Phys. J. Special Topics}\ }\textbf {\bibinfo {volume} {223}},\ \bibinfo
  {pages} {1055} (\bibinfo {year} {2014})}\BibitemShut {NoStop}%
\bibitem [{\citenamefont {Greiner}\ \emph {et~al.}(1985)\citenamefont
  {Greiner}, \citenamefont {M\"{u}ller},\ and\ \citenamefont
  {Rafelski}}]{Greiner_b_1985}%
  \BibitemOpen
  \bibfield  {author} {\bibinfo {author} {\bibfnamefont {W.}~\bibnamefont
  {Greiner}}, \bibinfo {author} {\bibfnamefont {B.}~\bibnamefont
  {M\"{u}ller}},\ and\ \bibinfo {author} {\bibfnamefont {J.}~\bibnamefont
  {Rafelski}},\ }\href@noop {} {\emph {\bibinfo {title} {Quantum
  Electrodynamics of Strong Fields}}}\ (\bibinfo  {publisher} {Springer-Verlag,
  Berlin},\ \bibinfo {year} {1985})\BibitemShut {NoStop}%
\bibitem [{\citenamefont {Fradkin}\ \emph {et~al.}(1991)\citenamefont
  {Fradkin}, \citenamefont {Gitman},\ and\ \citenamefont
  {Shvartsman}}]{Fradkin_b_1991}%
  \BibitemOpen
  \bibfield  {author} {\bibinfo {author} {\bibfnamefont {E.~S.}\ \bibnamefont
  {Fradkin}}, \bibinfo {author} {\bibfnamefont {D.~M.}\ \bibnamefont
  {Gitman}},\ and\ \bibinfo {author} {\bibfnamefont {{\relax Sh}.~M.}\
  \bibnamefont {Shvartsman}},\ }\href@noop {} {\emph {\bibinfo {title} {Quantum
  Electrodynamics with Unstable Vacuum}}}\ (\bibinfo  {publisher} {Springer,
  Berlin},\ \bibinfo {year} {1991})\BibitemShut {NoStop}%
\bibitem [{\citenamefont {Baier}\ \emph {et~al.}(1998)\citenamefont {Baier},
  \citenamefont {Katkov},\ and\ \citenamefont {Strakhovenko}}]{Baier_b_1998}%
  \BibitemOpen
  \bibfield  {author} {\bibinfo {author} {\bibfnamefont {V.~N.}\ \bibnamefont
  {Baier}}, \bibinfo {author} {\bibfnamefont {V.~M.}\ \bibnamefont {Katkov}},\
  and\ \bibinfo {author} {\bibfnamefont {V.~M.}\ \bibnamefont {Strakhovenko}},\
  }\href@noop {} {\emph {\bibinfo {title} {Electromagnetic Processes at High
  Energies in Oriented Single Crystals}}}\ (\bibinfo  {publisher} {World
  Scientific, Singapore},\ \bibinfo {year} {1998})\BibitemShut {NoStop}%
\bibitem [{\citenamefont {Bula}\ \emph {et~al.}(1996)\citenamefont {Bula},
  \citenamefont {McDonald}, \citenamefont {Prebys}, \citenamefont {Bamber},
  \citenamefont {Boege}, \citenamefont {Kotseroglou}, \citenamefont
  {Melissinos}, \citenamefont {Meyerhofer}, \citenamefont {Ragg}, \citenamefont
  {Burke}, \citenamefont {Field}, \citenamefont {Horton-Smith}, \citenamefont
  {Odian}, \citenamefont {Spencer}, \citenamefont {Walz}, \citenamefont
  {Berridge}, \citenamefont {Bugg}, \citenamefont {Shmakov},\ and\
  \citenamefont {Weidemann}}]{Bula_1996}%
  \BibitemOpen
  \bibfield  {author} {\bibinfo {author} {\bibfnamefont {C.}~\bibnamefont
  {Bula}}, \bibinfo {author} {\bibfnamefont {K.~T.}\ \bibnamefont {McDonald}},
  \bibinfo {author} {\bibfnamefont {E.~J.}\ \bibnamefont {Prebys}}, \bibinfo
  {author} {\bibfnamefont {C.}~\bibnamefont {Bamber}}, \bibinfo {author}
  {\bibfnamefont {S.}~\bibnamefont {Boege}}, \bibinfo {author} {\bibfnamefont
  {T.}~\bibnamefont {Kotseroglou}}, \bibinfo {author} {\bibfnamefont {A.~C.}\
  \bibnamefont {Melissinos}}, \bibinfo {author} {\bibfnamefont {D.~D.}\
  \bibnamefont {Meyerhofer}}, \bibinfo {author} {\bibfnamefont
  {W.}~\bibnamefont {Ragg}}, \bibinfo {author} {\bibfnamefont {D.~L.}\
  \bibnamefont {Burke}}, \bibinfo {author} {\bibfnamefont {R.~C.}\ \bibnamefont
  {Field}}, \bibinfo {author} {\bibfnamefont {G.}~\bibnamefont {Horton-Smith}},
  \bibinfo {author} {\bibfnamefont {A.~C.}\ \bibnamefont {Odian}}, \bibinfo
  {author} {\bibfnamefont {J.~E.}\ \bibnamefont {Spencer}}, \bibinfo {author}
  {\bibfnamefont {D.}~\bibnamefont {Walz}}, \bibinfo {author} {\bibfnamefont
  {S.~C.}\ \bibnamefont {Berridge}}, \bibinfo {author} {\bibfnamefont {W.~M.}\
  \bibnamefont {Bugg}}, \bibinfo {author} {\bibfnamefont {K.}~\bibnamefont
  {Shmakov}},\ and\ \bibinfo {author} {\bibfnamefont {A.~W.}\ \bibnamefont
  {Weidemann}},\ }\href@noop {} {\bibfield  {journal} {\bibinfo  {journal}
  {Phys. Rev. Lett.}\ }\textbf {\bibinfo {volume} {76}},\ \bibinfo {pages}
  {3116} (\bibinfo {year} {1996})}\BibitemShut {NoStop}%
\bibitem [{\citenamefont {Burke}\ \emph {et~al.}(1997)\citenamefont {Burke},
  \citenamefont {Field}, \citenamefont {Horton-Smith}, \citenamefont {Spencer},
  \citenamefont {Walz}, \citenamefont {Berridge}, \citenamefont {Bugg},
  \citenamefont {Shmakov}, \citenamefont {Weidemann}, \citenamefont {Bula},
  \citenamefont {McDonald}, \citenamefont {Prebys}, \citenamefont {Bamber},
  \citenamefont {Boege}, \citenamefont {Koffas}, \citenamefont {Kotseroglou},
  \citenamefont {Melissinos}, \citenamefont {Meyerhofer}, \citenamefont
  {Reis},\ and\ \citenamefont {Ragg}}]{Burke_1997}%
  \BibitemOpen
  \bibfield  {author} {\bibinfo {author} {\bibfnamefont {D.~L.}\ \bibnamefont
  {Burke}}, \bibinfo {author} {\bibfnamefont {R.~C.}\ \bibnamefont {Field}},
  \bibinfo {author} {\bibfnamefont {G.}~\bibnamefont {Horton-Smith}}, \bibinfo
  {author} {\bibfnamefont {J.~E.}\ \bibnamefont {Spencer}}, \bibinfo {author}
  {\bibfnamefont {D.}~\bibnamefont {Walz}}, \bibinfo {author} {\bibfnamefont
  {S.~C.}\ \bibnamefont {Berridge}}, \bibinfo {author} {\bibfnamefont {W.~M.}\
  \bibnamefont {Bugg}}, \bibinfo {author} {\bibfnamefont {K.}~\bibnamefont
  {Shmakov}}, \bibinfo {author} {\bibfnamefont {A.~W.}\ \bibnamefont
  {Weidemann}}, \bibinfo {author} {\bibfnamefont {C.}~\bibnamefont {Bula}},
  \bibinfo {author} {\bibfnamefont {K.~T.}\ \bibnamefont {McDonald}}, \bibinfo
  {author} {\bibfnamefont {E.~J.}\ \bibnamefont {Prebys}}, \bibinfo {author}
  {\bibfnamefont {C.}~\bibnamefont {Bamber}}, \bibinfo {author} {\bibfnamefont
  {S.~J.}\ \bibnamefont {Boege}}, \bibinfo {author} {\bibfnamefont
  {T.}~\bibnamefont {Koffas}}, \bibinfo {author} {\bibfnamefont
  {T.}~\bibnamefont {Kotseroglou}}, \bibinfo {author} {\bibfnamefont {A.~C.}\
  \bibnamefont {Melissinos}}, \bibinfo {author} {\bibfnamefont {D.~D.}\
  \bibnamefont {Meyerhofer}}, \bibinfo {author} {\bibfnamefont {D.~A.}\
  \bibnamefont {Reis}},\ and\ \bibinfo {author} {\bibfnamefont
  {W.}~\bibnamefont {Ragg}},\ }\href@noop {} {\bibfield  {journal} {\bibinfo
  {journal} {Phys. Rev. Lett.}\ }\textbf {\bibinfo {volume} {79}},\ \bibinfo
  {pages} {1626} (\bibinfo {year} {1997})}\BibitemShut {NoStop}%
\bibitem [{\citenamefont {Yoon}\ \emph {et~al.}(2019)\citenamefont {Yoon},
  \citenamefont {Jeon}, \citenamefont {Shin}, \citenamefont {Lee},
  \citenamefont {Lee}, \citenamefont {Choi}, \citenamefont {Kim}, \citenamefont
  {Sung},\ and\ \citenamefont {Nam}}]{Yoon_2019}%
  \BibitemOpen
  \bibfield  {author} {\bibinfo {author} {\bibfnamefont {J.~W.}\ \bibnamefont
  {Yoon}}, \bibinfo {author} {\bibfnamefont {C.}~\bibnamefont {Jeon}}, \bibinfo
  {author} {\bibfnamefont {J.}~\bibnamefont {Shin}}, \bibinfo {author}
  {\bibfnamefont {S.~K.}\ \bibnamefont {Lee}}, \bibinfo {author} {\bibfnamefont
  {H.~W.}\ \bibnamefont {Lee}}, \bibinfo {author} {\bibfnamefont {I.~W.}\
  \bibnamefont {Choi}}, \bibinfo {author} {\bibfnamefont {H.~T.}\ \bibnamefont
  {Kim}}, \bibinfo {author} {\bibfnamefont {J.~H.}\ \bibnamefont {Sung}},\ and\
  \bibinfo {author} {\bibfnamefont {C.~H.}\ \bibnamefont {Nam}},\ }\href@noop
  {} {\bibfield  {journal} {\bibinfo  {journal} {Opt. Express}\ }\textbf
  {\bibinfo {volume} {27}},\ \bibinfo {pages} {20412} (\bibinfo {year}
  {2019})}\BibitemShut {NoStop}%
\bibitem [{\citenamefont {Papadopoulos}\ \emph {et~al.}(2016)\citenamefont
  {Papadopoulos}, \citenamefont {Zou}, \citenamefont {Le~Blanc}, \citenamefont
  {Ch\'{e}riaux}, \citenamefont {Georges}, \citenamefont {Druon}, \citenamefont
  {Mennerat}, \citenamefont {Ramirez}, \citenamefont {Martin}, \citenamefont
  {Fr\'{e}neaux}, \citenamefont {Beluze}, \citenamefont {Lebas}, \citenamefont
  {Monot}, \citenamefont {Mathieu},\ and\ \citenamefont
  {Audebert}}]{APOLLON_10P}%
  \BibitemOpen
  \bibfield  {author} {\bibinfo {author} {\bibfnamefont {D.~N.}\ \bibnamefont
  {Papadopoulos}}, \bibinfo {author} {\bibfnamefont {J.~P.}\ \bibnamefont
  {Zou}}, \bibinfo {author} {\bibfnamefont {C.}~\bibnamefont {Le~Blanc}},
  \bibinfo {author} {\bibfnamefont {G.}~\bibnamefont {Ch\'{e}riaux}}, \bibinfo
  {author} {\bibfnamefont {P.}~\bibnamefont {Georges}}, \bibinfo {author}
  {\bibfnamefont {F.}~\bibnamefont {Druon}}, \bibinfo {author} {\bibfnamefont
  {G.}~\bibnamefont {Mennerat}}, \bibinfo {author} {\bibfnamefont
  {P.}~\bibnamefont {Ramirez}}, \bibinfo {author} {\bibfnamefont
  {L.}~\bibnamefont {Martin}}, \bibinfo {author} {\bibfnamefont
  {A.}~\bibnamefont {Fr\'{e}neaux}}, \bibinfo {author} {\bibfnamefont
  {A.}~\bibnamefont {Beluze}}, \bibinfo {author} {\bibfnamefont
  {N.}~\bibnamefont {Lebas}}, \bibinfo {author} {\bibfnamefont
  {P.}~\bibnamefont {Monot}}, \bibinfo {author} {\bibfnamefont
  {F.}~\bibnamefont {Mathieu}},\ and\ \bibinfo {author} {\bibfnamefont
  {P.}~\bibnamefont {Audebert}},\ }\href@noop {} {\bibfield  {journal}
  {\bibinfo  {journal} {High Power Laser Sci. Eng.}\ }\textbf {\bibinfo
  {volume} {4}},\ \bibinfo {pages} {e34} (\bibinfo {year} {2016})}\BibitemShut
  {NoStop}%
\bibitem [{ELI()}]{ELI}%
  \BibitemOpen
  \href@noop {} {\bibinfo {title} {{Extreme Light Infrastructure (ELI)}}},\
  \bibinfo {howpublished} {\url{https://eli-laser.eu/}}\BibitemShut {NoStop}%
\bibitem [{CoR()}]{CoReLS}%
  \BibitemOpen
  \href@noop {} {\bibinfo {title} {{Center for Relativistic Laser Science
  (CoReLS)}}},\ \bibinfo {howpublished}
  {\url{https://www.ibs.re.kr/eng/sub02_03_05.do}}\BibitemShut {NoStop}%
\bibitem [{\citenamefont {Bromage}\ \emph {et~al.}(2019)\citenamefont
  {Bromage}, \citenamefont {Bahk}, \citenamefont {Begishev}, \citenamefont
  {Dorrer}, \citenamefont {Guardalben}, \citenamefont {Hoffman}, \citenamefont
  {Oliver}, \citenamefont {Roides}, \citenamefont {Schiesser}, \citenamefont
  {Shoup~III}, \citenamefont {Spilatro}, \citenamefont {Webb}, \citenamefont
  {Weiner},\ and\ \citenamefont {Zuegel}}]{Bromage_2019}%
  \BibitemOpen
  \bibfield  {author} {\bibinfo {author} {\bibfnamefont {J.}~\bibnamefont
  {Bromage}}, \bibinfo {author} {\bibfnamefont {S.-W.}\ \bibnamefont {Bahk}},
  \bibinfo {author} {\bibfnamefont {I.~A.}\ \bibnamefont {Begishev}}, \bibinfo
  {author} {\bibfnamefont {C.}~\bibnamefont {Dorrer}}, \bibinfo {author}
  {\bibfnamefont {M.~J.}\ \bibnamefont {Guardalben}}, \bibinfo {author}
  {\bibfnamefont {B.~N.}\ \bibnamefont {Hoffman}}, \bibinfo {author}
  {\bibfnamefont {J.~B.}\ \bibnamefont {Oliver}}, \bibinfo {author}
  {\bibfnamefont {R.~G.}\ \bibnamefont {Roides}}, \bibinfo {author}
  {\bibfnamefont {E.~M.}\ \bibnamefont {Schiesser}}, \bibinfo {author}
  {\bibfnamefont {M.~J.}\ \bibnamefont {Shoup~III}}, \bibinfo {author}
  {\bibfnamefont {M.}~\bibnamefont {Spilatro}}, \bibinfo {author}
  {\bibfnamefont {B.}~\bibnamefont {Webb}}, \bibinfo {author} {\bibfnamefont
  {D.}~\bibnamefont {Weiner}},\ and\ \bibinfo {author} {\bibfnamefont {J.~D.}\
  \bibnamefont {Zuegel}},\ }\href@noop {} {\bibfield  {journal} {\bibinfo
  {journal} {High Power Laser Sci. Eng.}\ }\textbf {\bibinfo {volume} {7}},\
  \bibinfo {pages} {e4} (\bibinfo {year} {2019})}\BibitemShut {NoStop}%
\bibitem [{XCE()}]{XCELS}%
  \BibitemOpen
  \href@noop {} {\bibinfo {title} {{Exawatt Center for Extreme Light Studies
  (XCELS)}}},\ \bibinfo {howpublished}
  {\url{http://www.xcels.iapras.ru/}}\BibitemShut {NoStop}%
\bibitem [{\citenamefont {Leemans}\ \emph {et~al.}(2014)\citenamefont
  {Leemans}, \citenamefont {Gonsalves}, \citenamefont {Mao}, \citenamefont
  {Nakamura}, \citenamefont {Benedetti}, \citenamefont {Schroeder},
  \citenamefont {T\'oth}, \citenamefont {Daniels}, \citenamefont
  {Mittelberger}, \citenamefont {Bulanov}, \citenamefont {Vay}, \citenamefont
  {Geddes},\ and\ \citenamefont {Esarey}}]{Leemans_2014}%
  \BibitemOpen
  \bibfield  {author} {\bibinfo {author} {\bibfnamefont {W.~P.}\ \bibnamefont
  {Leemans}}, \bibinfo {author} {\bibfnamefont {A.~J.}\ \bibnamefont
  {Gonsalves}}, \bibinfo {author} {\bibfnamefont {H.-S.}\ \bibnamefont {Mao}},
  \bibinfo {author} {\bibfnamefont {K.}~\bibnamefont {Nakamura}}, \bibinfo
  {author} {\bibfnamefont {C.}~\bibnamefont {Benedetti}}, \bibinfo {author}
  {\bibfnamefont {C.~B.}\ \bibnamefont {Schroeder}}, \bibinfo {author}
  {\bibfnamefont {C.}~\bibnamefont {T\'oth}}, \bibinfo {author} {\bibfnamefont
  {J.}~\bibnamefont {Daniels}}, \bibinfo {author} {\bibfnamefont {D.~E.}\
  \bibnamefont {Mittelberger}}, \bibinfo {author} {\bibfnamefont {S.~S.}\
  \bibnamefont {Bulanov}}, \bibinfo {author} {\bibfnamefont {J.-L.}\
  \bibnamefont {Vay}}, \bibinfo {author} {\bibfnamefont {C.~G.~R.}\
  \bibnamefont {Geddes}},\ and\ \bibinfo {author} {\bibfnamefont
  {E.}~\bibnamefont {Esarey}},\ }\href@noop {} {\bibfield  {journal} {\bibinfo
  {journal} {Phys. Rev. Lett.}\ }\textbf {\bibinfo {volume} {113}},\ \bibinfo
  {pages} {245002} (\bibinfo {year} {2014})}\BibitemShut {NoStop}%
\bibitem [{\citenamefont {Cole}\ \emph {et~al.}(2018)\citenamefont {Cole},
  \citenamefont {Behm}, \citenamefont {Gerstmayr}, \citenamefont {Blackburn},
  \citenamefont {Wood}, \citenamefont {Baird}, \citenamefont {Duff},
  \citenamefont {Harvey}, \citenamefont {Ilderton}, \citenamefont {Joglekar},
  \citenamefont {Krushelnick}, \citenamefont {Kuschel}, \citenamefont
  {Marklund}, \citenamefont {McKenna}, \citenamefont {Murphy}, \citenamefont
  {Poder}, \citenamefont {Ridgers}, \citenamefont {Samarin}, \citenamefont
  {Sarri}, \citenamefont {Symes}, \citenamefont {Thomas}, \citenamefont
  {Warwick}, \citenamefont {Zepf}, \citenamefont {Najmudin},\ and\
  \citenamefont {Mangles}}]{Cole_2018}%
  \BibitemOpen
  \bibfield  {author} {\bibinfo {author} {\bibfnamefont {J.~M.}\ \bibnamefont
  {Cole}}, \bibinfo {author} {\bibfnamefont {K.~T.}\ \bibnamefont {Behm}},
  \bibinfo {author} {\bibfnamefont {E.}~\bibnamefont {Gerstmayr}}, \bibinfo
  {author} {\bibfnamefont {T.~G.}\ \bibnamefont {Blackburn}}, \bibinfo {author}
  {\bibfnamefont {J.~C.}\ \bibnamefont {Wood}}, \bibinfo {author}
  {\bibfnamefont {C.~D.}\ \bibnamefont {Baird}}, \bibinfo {author}
  {\bibfnamefont {M.~J.}\ \bibnamefont {Duff}}, \bibinfo {author}
  {\bibfnamefont {C.}~\bibnamefont {Harvey}}, \bibinfo {author} {\bibfnamefont
  {A.}~\bibnamefont {Ilderton}}, \bibinfo {author} {\bibfnamefont {A.~S.}\
  \bibnamefont {Joglekar}}, \bibinfo {author} {\bibfnamefont {K.}~\bibnamefont
  {Krushelnick}}, \bibinfo {author} {\bibfnamefont {S.}~\bibnamefont
  {Kuschel}}, \bibinfo {author} {\bibfnamefont {M.}~\bibnamefont {Marklund}},
  \bibinfo {author} {\bibfnamefont {P.}~\bibnamefont {McKenna}}, \bibinfo
  {author} {\bibfnamefont {C.~D.}\ \bibnamefont {Murphy}}, \bibinfo {author}
  {\bibfnamefont {K.}~\bibnamefont {Poder}}, \bibinfo {author} {\bibfnamefont
  {C.~P.}\ \bibnamefont {Ridgers}}, \bibinfo {author} {\bibfnamefont {G.~M.}\
  \bibnamefont {Samarin}}, \bibinfo {author} {\bibfnamefont {G.}~\bibnamefont
  {Sarri}}, \bibinfo {author} {\bibfnamefont {D.~R.}\ \bibnamefont {Symes}},
  \bibinfo {author} {\bibfnamefont {A.~G.~R.}\ \bibnamefont {Thomas}}, \bibinfo
  {author} {\bibfnamefont {J.}~\bibnamefont {Warwick}}, \bibinfo {author}
  {\bibfnamefont {M.}~\bibnamefont {Zepf}}, \bibinfo {author} {\bibfnamefont
  {Z.}~\bibnamefont {Najmudin}},\ and\ \bibinfo {author} {\bibfnamefont
  {S.~P.~D.}\ \bibnamefont {Mangles}},\ }\href@noop {} {\bibfield  {journal}
  {\bibinfo  {journal} {Phys. Rev. X}\ }\textbf {\bibinfo {volume} {8}},\
  \bibinfo {pages} {011020} (\bibinfo {year} {2018})}\BibitemShut {NoStop}%
\bibitem [{\citenamefont {Poder}\ \emph {et~al.}(2018)\citenamefont {Poder},
  \citenamefont {Tamburini}, \citenamefont {Sarri}, \citenamefont {Di~Piazza},
  \citenamefont {Kuschel}, \citenamefont {Baird}, \citenamefont {Behm},
  \citenamefont {Bohlen}, \citenamefont {Cole}, \citenamefont {Corvan},
  \citenamefont {Duff}, \citenamefont {Gerstmayr}, \citenamefont {Keitel},
  \citenamefont {Krushelnick}, \citenamefont {Mangles}, \citenamefont
  {McKenna}, \citenamefont {Murphy}, \citenamefont {Najmudin}, \citenamefont
  {Ridgers}, \citenamefont {Samarin}, \citenamefont {Symes}, \citenamefont
  {Thomas}, \citenamefont {Warwick},\ and\ \citenamefont {Zepf}}]{Poder_2018}%
  \BibitemOpen
  \bibfield  {author} {\bibinfo {author} {\bibfnamefont {K.}~\bibnamefont
  {Poder}}, \bibinfo {author} {\bibfnamefont {M.}~\bibnamefont {Tamburini}},
  \bibinfo {author} {\bibfnamefont {G.}~\bibnamefont {Sarri}}, \bibinfo
  {author} {\bibfnamefont {A.}~\bibnamefont {Di~Piazza}}, \bibinfo {author}
  {\bibfnamefont {S.}~\bibnamefont {Kuschel}}, \bibinfo {author} {\bibfnamefont
  {C.~D.}\ \bibnamefont {Baird}}, \bibinfo {author} {\bibfnamefont
  {K.}~\bibnamefont {Behm}}, \bibinfo {author} {\bibfnamefont {S.}~\bibnamefont
  {Bohlen}}, \bibinfo {author} {\bibfnamefont {J.~M.}\ \bibnamefont {Cole}},
  \bibinfo {author} {\bibfnamefont {D.~J.}\ \bibnamefont {Corvan}}, \bibinfo
  {author} {\bibfnamefont {M.}~\bibnamefont {Duff}}, \bibinfo {author}
  {\bibfnamefont {E.}~\bibnamefont {Gerstmayr}}, \bibinfo {author}
  {\bibfnamefont {C.~H.}\ \bibnamefont {Keitel}}, \bibinfo {author}
  {\bibfnamefont {K.}~\bibnamefont {Krushelnick}}, \bibinfo {author}
  {\bibfnamefont {S.~P.~D.}\ \bibnamefont {Mangles}}, \bibinfo {author}
  {\bibfnamefont {P.}~\bibnamefont {McKenna}}, \bibinfo {author} {\bibfnamefont
  {C.~D.}\ \bibnamefont {Murphy}}, \bibinfo {author} {\bibfnamefont
  {Z.}~\bibnamefont {Najmudin}}, \bibinfo {author} {\bibfnamefont {C.~P.}\
  \bibnamefont {Ridgers}}, \bibinfo {author} {\bibfnamefont {G.~M.}\
  \bibnamefont {Samarin}}, \bibinfo {author} {\bibfnamefont {D.~R.}\
  \bibnamefont {Symes}}, \bibinfo {author} {\bibfnamefont {A.~G.~R.}\
  \bibnamefont {Thomas}}, \bibinfo {author} {\bibfnamefont {J.}~\bibnamefont
  {Warwick}},\ and\ \bibinfo {author} {\bibfnamefont {M.}~\bibnamefont
  {Zepf}},\ }\href@noop {} {\bibfield  {journal} {\bibinfo  {journal} {Phys.
  Rev. X}\ }\textbf {\bibinfo {volume} {8}},\ \bibinfo {pages} {031004}
  (\bibinfo {year} {2018})}\BibitemShut {NoStop}%
\bibitem [{\citenamefont {Wistisen}\ \emph {et~al.}(2018)\citenamefont
  {Wistisen}, \citenamefont {Di~Piazza}, \citenamefont {Knudsen},\ and\
  \citenamefont {Uggerh{\o}j}}]{Wistisen_2018}%
  \BibitemOpen
  \bibfield  {author} {\bibinfo {author} {\bibfnamefont {T.~N.}\ \bibnamefont
  {Wistisen}}, \bibinfo {author} {\bibfnamefont {A.}~\bibnamefont {Di~Piazza}},
  \bibinfo {author} {\bibfnamefont {H.~V.}\ \bibnamefont {Knudsen}},\ and\
  \bibinfo {author} {\bibfnamefont {U.~I.}\ \bibnamefont {Uggerh{\o}j}},\
  }\href@noop {} {\bibfield  {journal} {\bibinfo  {journal} {Nat. Commun.}\
  }\textbf {\bibinfo {volume} {9}},\ \bibinfo {pages} {795} (\bibinfo {year}
  {2018})}\BibitemShut {NoStop}%
\bibitem [{\citenamefont {Wistisen}\ \emph {et~al.}(2019)\citenamefont
  {Wistisen}, \citenamefont {Di~Piazza}, \citenamefont {Nielsen}, \citenamefont
  {S\o{}rensen},\ and\ \citenamefont {Uggerh\o{}j}}]{Wistisen_2019}%
  \BibitemOpen
  \bibfield  {author} {\bibinfo {author} {\bibfnamefont {T.~N.}\ \bibnamefont
  {Wistisen}}, \bibinfo {author} {\bibfnamefont {A.}~\bibnamefont {Di~Piazza}},
  \bibinfo {author} {\bibfnamefont {C.~F.}\ \bibnamefont {Nielsen}}, \bibinfo
  {author} {\bibfnamefont {A.~H.}\ \bibnamefont {S\o{}rensen}},\ and\ \bibinfo
  {author} {\bibfnamefont {U.~I.}\ \bibnamefont {Uggerh\o{}j}},\ }\href@noop {}
  {\bibfield  {journal} {\bibinfo  {journal} {Phys. Rev. Research}\ }\textbf
  {\bibinfo {volume} {1}},\ \bibinfo {pages} {033014} (\bibinfo {year}
  {2019})}\BibitemShut {NoStop}%
\bibitem [{\citenamefont {Abramowicz}\ \emph {et~al.}()\citenamefont
  {Abramowicz}, \citenamefont {Altarelli}, \citenamefont {A{\ss}mann},
  \citenamefont {Behnke}, \citenamefont {Benhammou}, \citenamefont {Borysov},
  \citenamefont {Borysova}, \citenamefont {Brinkmann}, \citenamefont {Burkart},
  \citenamefont {B\"{u}{\ss}er}, \citenamefont {Davidi}, \citenamefont
  {Decking}, \citenamefont {Elkina}, \citenamefont {Harsh}, \citenamefont
  {Hartin}, \citenamefont {Hartl}, \citenamefont {Heinemann}, \citenamefont
  {Heinzl}, \citenamefont {TalHod}, \citenamefont {Hoffmann}, \citenamefont
  {Ilderton}, \citenamefont {King}, \citenamefont {Levy}, \citenamefont {List},
  \citenamefont {Maier}, \citenamefont {Negodin}, \citenamefont {Perez},
  \citenamefont {Pomerantz}, \citenamefont {Ringwald}, \citenamefont
  {R\"{o}del}, \citenamefont {Saimpert}, \citenamefont {Salgado}, \citenamefont
  {Sarri}, \citenamefont {Savoray}, \citenamefont {Teter}, \citenamefont
  {Wing},\ and\ \citenamefont {Zepf}}]{Abramowicz_2019}%
  \BibitemOpen
  \bibfield  {author} {\bibinfo {author} {\bibfnamefont {H.}~\bibnamefont
  {Abramowicz}}, \bibinfo {author} {\bibfnamefont {M.}~\bibnamefont
  {Altarelli}}, \bibinfo {author} {\bibfnamefont {R.}~\bibnamefont
  {A{\ss}mann}}, \bibinfo {author} {\bibfnamefont {T.}~\bibnamefont {Behnke}},
  \bibinfo {author} {\bibfnamefont {Y.}~\bibnamefont {Benhammou}}, \bibinfo
  {author} {\bibfnamefont {O.}~\bibnamefont {Borysov}}, \bibinfo {author}
  {\bibfnamefont {M.}~\bibnamefont {Borysova}}, \bibinfo {author}
  {\bibfnamefont {R.}~\bibnamefont {Brinkmann}}, \bibinfo {author}
  {\bibfnamefont {F.}~\bibnamefont {Burkart}}, \bibinfo {author} {\bibfnamefont
  {K.}~\bibnamefont {B\"{u}{\ss}er}}, \bibinfo {author} {\bibfnamefont
  {O.}~\bibnamefont {Davidi}}, \bibinfo {author} {\bibfnamefont
  {W.}~\bibnamefont {Decking}}, \bibinfo {author} {\bibfnamefont
  {N.}~\bibnamefont {Elkina}}, \bibinfo {author} {\bibfnamefont
  {H.}~\bibnamefont {Harsh}}, \bibinfo {author} {\bibfnamefont
  {A.}~\bibnamefont {Hartin}}, \bibinfo {author} {\bibfnamefont
  {I.}~\bibnamefont {Hartl}}, \bibinfo {author} {\bibfnamefont
  {B.}~\bibnamefont {Heinemann}}, \bibinfo {author} {\bibfnamefont
  {T.}~\bibnamefont {Heinzl}}, \bibinfo {author} {\bibfnamefont
  {N.}~\bibnamefont {TalHod}}, \bibinfo {author} {\bibfnamefont
  {M.}~\bibnamefont {Hoffmann}}, \bibinfo {author} {\bibfnamefont
  {A.}~\bibnamefont {Ilderton}}, \bibinfo {author} {\bibfnamefont
  {B.}~\bibnamefont {King}}, \bibinfo {author} {\bibfnamefont {A.}~\bibnamefont
  {Levy}}, \bibinfo {author} {\bibfnamefont {J.}~\bibnamefont {List}}, \bibinfo
  {author} {\bibfnamefont {A.~R.}\ \bibnamefont {Maier}}, \bibinfo {author}
  {\bibfnamefont {E.}~\bibnamefont {Negodin}}, \bibinfo {author} {\bibfnamefont
  {G.}~\bibnamefont {Perez}}, \bibinfo {author} {\bibfnamefont
  {I.}~\bibnamefont {Pomerantz}}, \bibinfo {author} {\bibfnamefont
  {A.}~\bibnamefont {Ringwald}}, \bibinfo {author} {\bibfnamefont
  {C.}~\bibnamefont {R\"{o}del}}, \bibinfo {author} {\bibfnamefont
  {M.}~\bibnamefont {Saimpert}}, \bibinfo {author} {\bibfnamefont
  {F.}~\bibnamefont {Salgado}}, \bibinfo {author} {\bibfnamefont
  {G.}~\bibnamefont {Sarri}}, \bibinfo {author} {\bibfnamefont
  {I.}~\bibnamefont {Savoray}}, \bibinfo {author} {\bibfnamefont
  {T.}~\bibnamefont {Teter}}, \bibinfo {author} {\bibfnamefont
  {M.}~\bibnamefont {Wing}},\ and\ \bibinfo {author} {\bibfnamefont
  {M.}~\bibnamefont {Zepf}},\ }\href@noop {} {\bibinfo  {journal}
  {arXiv:1909.00860}\ }\BibitemShut {NoStop}%
\bibitem [{\citenamefont {Meuren}\ \emph {et~al.}()\citenamefont {Meuren},
  \citenamefont {Bucksbaum}, \citenamefont {Fisch}, \citenamefont {Fi\'{u}za},
  \citenamefont {Glenzer}, \citenamefont {Hogan}, \citenamefont {Qu},
  \citenamefont {Reis}, \citenamefont {White},\ and\ \citenamefont
  {Yakimenko}}]{Meuren_2020}%
  \BibitemOpen
\bibfield  {journal} {  }\bibfield  {author} {\bibinfo {author} {\bibfnamefont
  {S.}~\bibnamefont {Meuren}}, \bibinfo {author} {\bibfnamefont {P.~H.}\
  \bibnamefont {Bucksbaum}}, \bibinfo {author} {\bibfnamefont {N.~J.}\
  \bibnamefont {Fisch}}, \bibinfo {author} {\bibfnamefont {F.}~\bibnamefont
  {Fi\'{u}za}}, \bibinfo {author} {\bibfnamefont {S.}~\bibnamefont {Glenzer}},
  \bibinfo {author} {\bibfnamefont {M.~J.}\ \bibnamefont {Hogan}}, \bibinfo
  {author} {\bibfnamefont {K.}~\bibnamefont {Qu}}, \bibinfo {author}
  {\bibfnamefont {D.~A.}\ \bibnamefont {Reis}}, \bibinfo {author}
  {\bibfnamefont {G.}~\bibnamefont {White}},\ and\ \bibinfo {author}
  {\bibfnamefont {V.}~\bibnamefont {Yakimenko}},\ }\href@noop {} {\bibinfo
  {journal} {arXiv:2002.10051}\ }\BibitemShut {NoStop}%
\bibitem [{\citenamefont {Narozhny}\ and\ \citenamefont
  {Fofanov}(2000)}]{Narozhny_2000}%
  \BibitemOpen
\bibfield  {journal} {  }\bibfield  {author} {\bibinfo {author} {\bibfnamefont
  {N.~B.}\ \bibnamefont {Narozhny}}\ and\ \bibinfo {author} {\bibfnamefont
  {M.~S.}\ \bibnamefont {Fofanov}},\ }\href@noop {} {\bibfield  {journal}
  {\bibinfo  {journal} {J. Exp. Theor. Phys.}\ }\textbf {\bibinfo {volume}
  {90}},\ \bibinfo {pages} {415} (\bibinfo {year} {2000})}\BibitemShut
  {NoStop}%
\bibitem [{\citenamefont {Ivanov}\ \emph {et~al.}(2004)\citenamefont {Ivanov},
  \citenamefont {Kotkin},\ and\ \citenamefont {Serbo}}]{Ivanov_2004}%
  \BibitemOpen
  \bibfield  {author} {\bibinfo {author} {\bibfnamefont {D.~{\relax Yu}.}\
  \bibnamefont {Ivanov}}, \bibinfo {author} {\bibfnamefont {G.~L.}\
  \bibnamefont {Kotkin}},\ and\ \bibinfo {author} {\bibfnamefont {V.~G.}\
  \bibnamefont {Serbo}},\ }\href@noop {} {\bibfield  {journal} {\bibinfo
  {journal} {Eur. Phys. J. C}\ }\textbf {\bibinfo {volume} {36}},\ \bibinfo
  {pages} {127} (\bibinfo {year} {2004})}\BibitemShut {NoStop}%
\bibitem [{\citenamefont {Boca}\ and\ \citenamefont
  {Florescu}(2009)}]{Boca_2009}%
  \BibitemOpen
  \bibfield  {author} {\bibinfo {author} {\bibfnamefont {M.}~\bibnamefont
  {Boca}}\ and\ \bibinfo {author} {\bibfnamefont {V.}~\bibnamefont
  {Florescu}},\ }\href@noop {} {\bibfield  {journal} {\bibinfo  {journal}
  {Phys. Rev. A}\ }\textbf {\bibinfo {volume} {80}},\ \bibinfo {pages} {053403}
  (\bibinfo {year} {2009})}\BibitemShut {NoStop}%
\bibitem [{\citenamefont {Harvey}\ \emph {et~al.}(2009)\citenamefont {Harvey},
  \citenamefont {Heinzl},\ and\ \citenamefont {Ilderton}}]{Harvey_2009}%
  \BibitemOpen
  \bibfield  {author} {\bibinfo {author} {\bibfnamefont {C.}~\bibnamefont
  {Harvey}}, \bibinfo {author} {\bibfnamefont {T.}~\bibnamefont {Heinzl}},\
  and\ \bibinfo {author} {\bibfnamefont {A.}~\bibnamefont {Ilderton}},\
  }\href@noop {} {\bibfield  {journal} {\bibinfo  {journal} {Phys. Rev. A}\
  }\textbf {\bibinfo {volume} {79}},\ \bibinfo {pages} {063407} (\bibinfo
  {year} {2009})}\BibitemShut {NoStop}%
\bibitem [{\citenamefont {Mackenroth}\ \emph {et~al.}(2010)\citenamefont
  {Mackenroth}, \citenamefont {Di~Piazza},\ and\ \citenamefont
  {Keitel}}]{Mackenroth_2010}%
  \BibitemOpen
  \bibfield  {author} {\bibinfo {author} {\bibfnamefont {F.}~\bibnamefont
  {Mackenroth}}, \bibinfo {author} {\bibfnamefont {A.}~\bibnamefont
  {Di~Piazza}},\ and\ \bibinfo {author} {\bibfnamefont {C.~H.}\ \bibnamefont
  {Keitel}},\ }\href@noop {} {\bibfield  {journal} {\bibinfo  {journal} {Phys.
  Rev. Lett.}\ }\textbf {\bibinfo {volume} {105}},\ \bibinfo {pages} {063903}
  (\bibinfo {year} {2010})}\BibitemShut {NoStop}%
\bibitem [{\citenamefont {Boca}\ and\ \citenamefont
  {Florescu}(2011)}]{Boca_2011}%
  \BibitemOpen
  \bibfield  {author} {\bibinfo {author} {\bibfnamefont {M.}~\bibnamefont
  {Boca}}\ and\ \bibinfo {author} {\bibfnamefont {V.}~\bibnamefont
  {Florescu}},\ }\href@noop {} {\bibfield  {journal} {\bibinfo  {journal} {Eur.
  Phys. J. D}\ }\textbf {\bibinfo {volume} {61}},\ \bibinfo {pages} {449}
  (\bibinfo {year} {2011})}\BibitemShut {NoStop}%
\bibitem [{\citenamefont {Mackenroth}\ and\ \citenamefont
  {Di~Piazza}(2011)}]{Mackenroth_2011}%
  \BibitemOpen
  \bibfield  {author} {\bibinfo {author} {\bibfnamefont {F.}~\bibnamefont
  {Mackenroth}}\ and\ \bibinfo {author} {\bibfnamefont {A.}~\bibnamefont
  {Di~Piazza}},\ }\href@noop {} {\bibfield  {journal} {\bibinfo  {journal}
  {Phys. Rev. A}\ }\textbf {\bibinfo {volume} {83}},\ \bibinfo {pages} {032106}
  (\bibinfo {year} {2011})}\BibitemShut {NoStop}%
\bibitem [{\citenamefont {Seipt}\ and\ \citenamefont
  {K\"ampfer}(2011{\natexlab{a}})}]{Seipt_2011}%
  \BibitemOpen
  \bibfield  {author} {\bibinfo {author} {\bibfnamefont {D.}~\bibnamefont
  {Seipt}}\ and\ \bibinfo {author} {\bibfnamefont {B.}~\bibnamefont
  {K\"ampfer}},\ }\href@noop {} {\bibfield  {journal} {\bibinfo  {journal}
  {Phys. Rev. A}\ }\textbf {\bibinfo {volume} {83}},\ \bibinfo {pages} {022101}
  (\bibinfo {year} {2011}{\natexlab{a}})}\BibitemShut {NoStop}%
\bibitem [{\citenamefont {Seipt}\ and\ \citenamefont
  {K\"ampfer}(2011{\natexlab{b}})}]{Seipt_2011b}%
  \BibitemOpen
  \bibfield  {author} {\bibinfo {author} {\bibfnamefont {D.}~\bibnamefont
  {Seipt}}\ and\ \bibinfo {author} {\bibfnamefont {B.}~\bibnamefont
  {K\"ampfer}},\ }\href@noop {} {\bibfield  {journal} {\bibinfo  {journal}
  {Phys. Rev. ST Accel. Beams}\ }\textbf {\bibinfo {volume} {14}},\ \bibinfo
  {pages} {040704} (\bibinfo {year} {2011}{\natexlab{b}})}\BibitemShut
  {NoStop}%
\bibitem [{\citenamefont {Dinu}\ \emph {et~al.}(2012)\citenamefont {Dinu},
  \citenamefont {Heinzl},\ and\ \citenamefont {Ilderton}}]{Dinu_2012}%
  \BibitemOpen
  \bibfield  {author} {\bibinfo {author} {\bibfnamefont {V.}~\bibnamefont
  {Dinu}}, \bibinfo {author} {\bibfnamefont {T.}~\bibnamefont {Heinzl}},\ and\
  \bibinfo {author} {\bibfnamefont {A.}~\bibnamefont {Ilderton}},\ }\href@noop
  {} {\bibfield  {journal} {\bibinfo  {journal} {Phys. Rev. D}\ }\textbf
  {\bibinfo {volume} {86}},\ \bibinfo {pages} {085037} (\bibinfo {year}
  {2012})}\BibitemShut {NoStop}%
\bibitem [{\citenamefont {Krajewska}\ and\ \citenamefont
  {Kami\ifmmode~\acute{n}\else \'{n}\fi{}ski}(2012)}]{Krajewska_2012}%
  \BibitemOpen
  \bibfield  {author} {\bibinfo {author} {\bibfnamefont {K.}~\bibnamefont
  {Krajewska}}\ and\ \bibinfo {author} {\bibfnamefont {J.~Z.}\ \bibnamefont
  {Kami\ifmmode~\acute{n}\else \'{n}\fi{}ski}},\ }\href@noop {} {\bibfield
  {journal} {\bibinfo  {journal} {Phys. Rev. A}\ }\textbf {\bibinfo {volume}
  {85}},\ \bibinfo {pages} {062102} (\bibinfo {year} {2012})}\BibitemShut
  {NoStop}%
\bibitem [{\citenamefont {Dinu}(2013)}]{Dinu_2013}%
  \BibitemOpen
  \bibfield  {author} {\bibinfo {author} {\bibfnamefont {V.}~\bibnamefont
  {Dinu}},\ }\href@noop {} {\bibfield  {journal} {\bibinfo  {journal} {Phys.
  Rev. A}\ }\textbf {\bibinfo {volume} {87}},\ \bibinfo {pages} {052101}
  (\bibinfo {year} {2013})}\BibitemShut {NoStop}%
\bibitem [{\citenamefont {Seipt}\ and\ \citenamefont
  {K\"ampfer}(2013)}]{Seipt_2013}%
  \BibitemOpen
  \bibfield  {author} {\bibinfo {author} {\bibfnamefont {D.}~\bibnamefont
  {Seipt}}\ and\ \bibinfo {author} {\bibfnamefont {B.}~\bibnamefont
  {K\"ampfer}},\ }\href@noop {} {\bibfield  {journal} {\bibinfo  {journal}
  {Phys. Rev. A}\ }\textbf {\bibinfo {volume} {88}},\ \bibinfo {pages} {012127}
  (\bibinfo {year} {2013})}\BibitemShut {NoStop}%
\bibitem [{\citenamefont {Krajewska}\ \emph {et~al.}(2014)\citenamefont
  {Krajewska}, \citenamefont {Twardy},\ and\ \citenamefont
  {Kami\ifmmode~\acute{n}\else \'{n}\fi{}ski}}]{Krajewska_2014}%
  \BibitemOpen
  \bibfield  {author} {\bibinfo {author} {\bibfnamefont {K.}~\bibnamefont
  {Krajewska}}, \bibinfo {author} {\bibfnamefont {M.}~\bibnamefont {Twardy}},\
  and\ \bibinfo {author} {\bibfnamefont {J.~Z.}\ \bibnamefont
  {Kami\ifmmode~\acute{n}\else \'{n}\fi{}ski}},\ }\href@noop {} {\bibfield
  {journal} {\bibinfo  {journal} {Phys. Rev. A}\ }\textbf {\bibinfo {volume}
  {89}},\ \bibinfo {pages} {032125} (\bibinfo {year} {2014})}\BibitemShut
  {NoStop}%
\bibitem [{\citenamefont {Wistisen}(2014)}]{Wistisen_2014}%
  \BibitemOpen
  \bibfield  {author} {\bibinfo {author} {\bibfnamefont {T.~N.}\ \bibnamefont
  {Wistisen}},\ }\href@noop {} {\bibfield  {journal} {\bibinfo  {journal}
  {Phys. Rev. D}\ }\textbf {\bibinfo {volume} {90}},\ \bibinfo {pages} {125008}
  (\bibinfo {year} {2014})}\BibitemShut {NoStop}%
\bibitem [{\citenamefont {Harvey}\ \emph {et~al.}(2015)\citenamefont {Harvey},
  \citenamefont {Ilderton},\ and\ \citenamefont {King}}]{Harvey_2015}%
  \BibitemOpen
  \bibfield  {author} {\bibinfo {author} {\bibfnamefont {C.~N.}\ \bibnamefont
  {Harvey}}, \bibinfo {author} {\bibfnamefont {A.}~\bibnamefont {Ilderton}},\
  and\ \bibinfo {author} {\bibfnamefont {B.}~\bibnamefont {King}},\ }\href@noop
  {} {\bibfield  {journal} {\bibinfo  {journal} {Phys. Rev. A}\ }\textbf
  {\bibinfo {volume} {91}},\ \bibinfo {pages} {013822} (\bibinfo {year}
  {2015})}\BibitemShut {NoStop}%
\bibitem [{\citenamefont {Seipt}\ \emph
  {et~al.}(2016{\natexlab{a}})\citenamefont {Seipt}, \citenamefont {Kharin},
  \citenamefont {Rykovanov}, \citenamefont {Surzhykov},\ and\ \citenamefont
  {Fritzsche}}]{Seipt_2016}%
  \BibitemOpen
  \bibfield  {author} {\bibinfo {author} {\bibfnamefont {D.}~\bibnamefont
  {Seipt}}, \bibinfo {author} {\bibfnamefont {V.}~\bibnamefont {Kharin}},
  \bibinfo {author} {\bibfnamefont {S.}~\bibnamefont {Rykovanov}}, \bibinfo
  {author} {\bibfnamefont {A.}~\bibnamefont {Surzhykov}},\ and\ \bibinfo
  {author} {\bibfnamefont {S.}~\bibnamefont {Fritzsche}},\ }\href@noop {}
  {\bibfield  {journal} {\bibinfo  {journal} {J. Plasma Phys.}\ }\textbf
  {\bibinfo {volume} {82}},\ \bibinfo {pages} {655820203} (\bibinfo {year}
  {2016}{\natexlab{a}})}\BibitemShut {NoStop}%
\bibitem [{\citenamefont {Seipt}\ \emph
  {et~al.}(2016{\natexlab{b}})\citenamefont {Seipt}, \citenamefont {Surzhykov},
  \citenamefont {Fritzsche},\ and\ \citenamefont {K\"{a}mpfer}}]{Seipt_2016b}%
  \BibitemOpen
  \bibfield  {author} {\bibinfo {author} {\bibfnamefont {D.}~\bibnamefont
  {Seipt}}, \bibinfo {author} {\bibfnamefont {A.}~\bibnamefont {Surzhykov}},
  \bibinfo {author} {\bibfnamefont {S.}~\bibnamefont {Fritzsche}},\ and\
  \bibinfo {author} {\bibfnamefont {B.}~\bibnamefont {K\"{a}mpfer}},\
  }\href@noop {} {\bibfield  {journal} {\bibinfo  {journal} {New J. Phys.}\
  }\textbf {\bibinfo {volume} {18}},\ \bibinfo {pages} {023044} (\bibinfo
  {year} {2016}{\natexlab{b}})}\BibitemShut {NoStop}%
\bibitem [{\citenamefont {Angioi}\ \emph {et~al.}(2016)\citenamefont {Angioi},
  \citenamefont {Mackenroth},\ and\ \citenamefont {Di~Piazza}}]{Angioi_2016}%
  \BibitemOpen
  \bibfield  {author} {\bibinfo {author} {\bibfnamefont {A.}~\bibnamefont
  {Angioi}}, \bibinfo {author} {\bibfnamefont {F.}~\bibnamefont {Mackenroth}},\
  and\ \bibinfo {author} {\bibfnamefont {A.}~\bibnamefont {Di~Piazza}},\
  }\href@noop {} {\bibfield  {journal} {\bibinfo  {journal} {Phys. Rev. A}\
  }\textbf {\bibinfo {volume} {93}},\ \bibinfo {pages} {052102} (\bibinfo
  {year} {2016})}\BibitemShut {NoStop}%
\bibitem [{\citenamefont {Harvey}\ \emph
  {et~al.}(2016{\natexlab{a}})\citenamefont {Harvey}, \citenamefont {Gonoskov},
  \citenamefont {Marklund},\ and\ \citenamefont {Wallin}}]{Harvey_2016b}%
  \BibitemOpen
  \bibfield  {author} {\bibinfo {author} {\bibfnamefont {C.~N.}\ \bibnamefont
  {Harvey}}, \bibinfo {author} {\bibfnamefont {A.}~\bibnamefont {Gonoskov}},
  \bibinfo {author} {\bibfnamefont {M.}~\bibnamefont {Marklund}},\ and\
  \bibinfo {author} {\bibfnamefont {E.}~\bibnamefont {Wallin}},\ }\href@noop {}
  {\bibfield  {journal} {\bibinfo  {journal} {Phys. Rev. A}\ }\textbf {\bibinfo
  {volume} {93}},\ \bibinfo {pages} {022112} (\bibinfo {year}
  {2016}{\natexlab{a}})}\BibitemShut {NoStop}%
\bibitem [{\citenamefont {Angioi}\ and\ \citenamefont
  {Di~Piazza}(2018)}]{Angioi_2018}%
  \BibitemOpen
  \bibfield  {author} {\bibinfo {author} {\bibfnamefont {A.}~\bibnamefont
  {Angioi}}\ and\ \bibinfo {author} {\bibfnamefont {A.}~\bibnamefont
  {Di~Piazza}},\ }\href@noop {} {\bibfield  {journal} {\bibinfo  {journal}
  {Phys. Rev. Lett.}\ }\textbf {\bibinfo {volume} {121}},\ \bibinfo {pages}
  {010402} (\bibinfo {year} {2018})}\BibitemShut {NoStop}%
\bibitem [{\citenamefont {Di~Piazza}\ \emph {et~al.}(2018)\citenamefont
  {Di~Piazza}, \citenamefont {Tamburini}, \citenamefont {Meuren},\ and\
  \citenamefont {Keitel}}]{Di_Piazza_2018}%
  \BibitemOpen
  \bibfield  {author} {\bibinfo {author} {\bibfnamefont {A.}~\bibnamefont
  {Di~Piazza}}, \bibinfo {author} {\bibfnamefont {M.}~\bibnamefont
  {Tamburini}}, \bibinfo {author} {\bibfnamefont {S.}~\bibnamefont {Meuren}},\
  and\ \bibinfo {author} {\bibfnamefont {C.~H.}\ \bibnamefont {Keitel}},\
  }\href@noop {} {\bibfield  {journal} {\bibinfo  {journal} {Phys. Rev. A}\
  }\textbf {\bibinfo {volume} {98}},\ \bibinfo {pages} {012134} (\bibinfo
  {year} {2018})}\BibitemShut {NoStop}%
\bibitem [{\citenamefont {Aleksandrov}\ \emph {et~al.}(2019)\citenamefont
  {Aleksandrov}, \citenamefont {Plunien},\ and\ \citenamefont
  {Shabaev}}]{Alexandrov_2019}%
  \BibitemOpen
  \bibfield  {author} {\bibinfo {author} {\bibfnamefont {I.~A.}\ \bibnamefont
  {Aleksandrov}}, \bibinfo {author} {\bibfnamefont {G.}~\bibnamefont
  {Plunien}},\ and\ \bibinfo {author} {\bibfnamefont {V.~M.}\ \bibnamefont
  {Shabaev}},\ }\href@noop {} {\bibfield  {journal} {\bibinfo  {journal} {Phys.
  Rev. D}\ }\textbf {\bibinfo {volume} {99}},\ \bibinfo {pages} {016020}
  (\bibinfo {year} {2019})}\BibitemShut {NoStop}%
\bibitem [{\citenamefont {Di~Piazza}\ \emph {et~al.}(2019)\citenamefont
  {Di~Piazza}, \citenamefont {Tamburini}, \citenamefont {Meuren},\ and\
  \citenamefont {Keitel}}]{Di_Piazza_2019}%
  \BibitemOpen
  \bibfield  {author} {\bibinfo {author} {\bibfnamefont {A.}~\bibnamefont
  {Di~Piazza}}, \bibinfo {author} {\bibfnamefont {M.}~\bibnamefont
  {Tamburini}}, \bibinfo {author} {\bibfnamefont {S.}~\bibnamefont {Meuren}},\
  and\ \bibinfo {author} {\bibfnamefont {C.~H.}\ \bibnamefont {Keitel}},\
  }\href@noop {} {\bibfield  {journal} {\bibinfo  {journal} {Phys. Rev. A}\
  }\textbf {\bibinfo {volume} {99}},\ \bibinfo {pages} {022125} (\bibinfo
  {year} {2019})}\BibitemShut {NoStop}%
\bibitem [{\citenamefont {Ilderton}\ \emph {et~al.}(2019)\citenamefont
  {Ilderton}, \citenamefont {King},\ and\ \citenamefont
  {Seipt}}]{Ilderton_2019_b}%
  \BibitemOpen
  \bibfield  {author} {\bibinfo {author} {\bibfnamefont {A.}~\bibnamefont
  {Ilderton}}, \bibinfo {author} {\bibfnamefont {B.}~\bibnamefont {King}},\
  and\ \bibinfo {author} {\bibfnamefont {D.}~\bibnamefont {Seipt}},\
  }\href@noop {} {\bibfield  {journal} {\bibinfo  {journal} {Phys. Rev. A}\
  }\textbf {\bibinfo {volume} {99}},\ \bibinfo {pages} {042121} (\bibinfo
  {year} {2019})}\BibitemShut {NoStop}%
\bibitem [{\citenamefont {Seipt}\ and\ \citenamefont
  {King}(2020)}]{Seipt_2020}%
  \BibitemOpen
  \bibfield  {author} {\bibinfo {author} {\bibfnamefont {D.}~\bibnamefont
  {Seipt}}\ and\ \bibinfo {author} {\bibfnamefont {B.}~\bibnamefont {King}},\
  }\href@noop {} {\bibfield  {journal} {\bibinfo  {journal} {Phys. Rev. A}\
  }\textbf {\bibinfo {volume} {102}},\ \bibinfo {pages} {052805} (\bibinfo
  {year} {2020})}\BibitemShut {NoStop}%
\bibitem [{\citenamefont {King}\ and\ \citenamefont
  {Tang}(2020)}]{King_2020_b}%
  \BibitemOpen
  \bibfield  {author} {\bibinfo {author} {\bibfnamefont {B.}~\bibnamefont
  {King}}\ and\ \bibinfo {author} {\bibfnamefont {S.}~\bibnamefont {Tang}},\
  }\href@noop {} {\bibfield  {journal} {\bibinfo  {journal} {Phys. Rev. A}\
  }\textbf {\bibinfo {volume} {102}},\ \bibinfo {pages} {022809} (\bibinfo
  {year} {2020})}\BibitemShut {NoStop}%
\bibitem [{\citenamefont {Roshchupkin}(2001)}]{Roshchupkin_2001}%
  \BibitemOpen
  \bibfield  {author} {\bibinfo {author} {\bibfnamefont {S.~P.}\ \bibnamefont
  {Roshchupkin}},\ }\href@noop {} {\bibfield  {journal} {\bibinfo  {journal}
  {Phys. At. Nucl.}\ }\textbf {\bibinfo {volume} {64}},\ \bibinfo {pages} {243}
  (\bibinfo {year} {2001})}\BibitemShut {NoStop}%
\bibitem [{\citenamefont {Heinzl}\ \emph {et~al.}(2010)\citenamefont {Heinzl},
  \citenamefont {Ilderton},\ and\ \citenamefont {Marklund}}]{Heinzl_2010b}%
  \BibitemOpen
  \bibfield  {author} {\bibinfo {author} {\bibfnamefont {T.}~\bibnamefont
  {Heinzl}}, \bibinfo {author} {\bibfnamefont {A.}~\bibnamefont {Ilderton}},\
  and\ \bibinfo {author} {\bibfnamefont {M.}~\bibnamefont {Marklund}},\
  }\href@noop {} {\bibfield  {journal} {\bibinfo  {journal} {Phys. Lett. B}\
  }\textbf {\bibinfo {volume} {692}},\ \bibinfo {pages} {250} (\bibinfo {year}
  {2010})}\BibitemShut {NoStop}%
\bibitem [{\citenamefont {M\"{u}ller}\ and\ \citenamefont
  {M\"{u}ller}(2011)}]{Mueller_2011b}%
  \BibitemOpen
  \bibfield  {author} {\bibinfo {author} {\bibfnamefont {T.-O.}\ \bibnamefont
  {M\"{u}ller}}\ and\ \bibinfo {author} {\bibfnamefont {C.}~\bibnamefont
  {M\"{u}ller}},\ }\href@noop {} {\bibfield  {journal} {\bibinfo  {journal}
  {Phys. Lett. B}\ }\textbf {\bibinfo {volume} {696}},\ \bibinfo {pages} {201}
  (\bibinfo {year} {2011})}\BibitemShut {NoStop}%
\bibitem [{\citenamefont {Titov}\ \emph {et~al.}(2012)\citenamefont {Titov},
  \citenamefont {Takabe}, \citenamefont {K\"ampfer},\ and\ \citenamefont
  {Hosaka}}]{Titov_2012}%
  \BibitemOpen
  \bibfield  {author} {\bibinfo {author} {\bibfnamefont {A.~I.}\ \bibnamefont
  {Titov}}, \bibinfo {author} {\bibfnamefont {H.}~\bibnamefont {Takabe}},
  \bibinfo {author} {\bibfnamefont {B.}~\bibnamefont {K\"ampfer}},\ and\
  \bibinfo {author} {\bibfnamefont {A.}~\bibnamefont {Hosaka}},\ }\href@noop {}
  {\bibfield  {journal} {\bibinfo  {journal} {Phys. Rev. Lett.}\ }\textbf
  {\bibinfo {volume} {108}},\ \bibinfo {pages} {240406} (\bibinfo {year}
  {2012})}\BibitemShut {NoStop}%
\bibitem [{\citenamefont {Nousch}\ \emph {et~al.}(2012)\citenamefont {Nousch},
  \citenamefont {Seipt}, \citenamefont {K\"{a}mpfer},\ and\ \citenamefont
  {Titov}}]{Nousch_2012}%
  \BibitemOpen
  \bibfield  {author} {\bibinfo {author} {\bibfnamefont {T.}~\bibnamefont
  {Nousch}}, \bibinfo {author} {\bibfnamefont {D.}~\bibnamefont {Seipt}},
  \bibinfo {author} {\bibfnamefont {B.}~\bibnamefont {K\"{a}mpfer}},\ and\
  \bibinfo {author} {\bibfnamefont {A.}~\bibnamefont {Titov}},\ }\href@noop {}
  {\bibfield  {journal} {\bibinfo  {journal} {Phys. Lett. B}\ }\textbf
  {\bibinfo {volume} {715}},\ \bibinfo {pages} {246} (\bibinfo {year}
  {2012})}\BibitemShut {NoStop}%
\bibitem [{\citenamefont {Krajewska}\ \emph {et~al.}(2013)\citenamefont
  {Krajewska}, \citenamefont {M\"uller},\ and\ \citenamefont
  {Kami\ifmmode~\acute{n}\else \'{n}\fi{}ski}}]{Krajewska_2013b}%
  \BibitemOpen
  \bibfield  {author} {\bibinfo {author} {\bibfnamefont {K.}~\bibnamefont
  {Krajewska}}, \bibinfo {author} {\bibfnamefont {C.}~\bibnamefont
  {M\"uller}},\ and\ \bibinfo {author} {\bibfnamefont {J.~Z.}\ \bibnamefont
  {Kami\ifmmode~\acute{n}\else \'{n}\fi{}ski}},\ }\href@noop {} {\bibfield
  {journal} {\bibinfo  {journal} {Phys. Rev. A}\ }\textbf {\bibinfo {volume}
  {87}},\ \bibinfo {pages} {062107} (\bibinfo {year} {2013})}\BibitemShut
  {NoStop}%
\bibitem [{\citenamefont {Jansen}\ and\ \citenamefont
  {M\"uller}(2013)}]{Jansen_2013}%
  \BibitemOpen
  \bibfield  {author} {\bibinfo {author} {\bibfnamefont {M.~J.~A.}\
  \bibnamefont {Jansen}}\ and\ \bibinfo {author} {\bibfnamefont
  {C.}~\bibnamefont {M\"uller}},\ }\href@noop {} {\bibfield  {journal}
  {\bibinfo  {journal} {Phys. Rev. A}\ }\textbf {\bibinfo {volume} {88}},\
  \bibinfo {pages} {052125} (\bibinfo {year} {2013})}\BibitemShut {NoStop}%
\bibitem [{\citenamefont {Augustin}\ and\ \citenamefont
  {M\"{u}ller}(2014)}]{Augustin_2014}%
  \BibitemOpen
  \bibfield  {author} {\bibinfo {author} {\bibfnamefont {S.}~\bibnamefont
  {Augustin}}\ and\ \bibinfo {author} {\bibfnamefont {C.}~\bibnamefont
  {M\"{u}ller}},\ }\href@noop {} {\bibfield  {journal} {\bibinfo  {journal}
  {Phys. Lett. B}\ }\textbf {\bibinfo {volume} {737}},\ \bibinfo {pages} {114}
  (\bibinfo {year} {2014})}\BibitemShut {NoStop}%
\bibitem [{\citenamefont {Meuren}\ \emph {et~al.}(2015)\citenamefont {Meuren},
  \citenamefont {Hatsagortsyan}, \citenamefont {Keitel},\ and\ \citenamefont
  {Di~Piazza}}]{Meuren_2015}%
  \BibitemOpen
  \bibfield  {author} {\bibinfo {author} {\bibfnamefont {S.}~\bibnamefont
  {Meuren}}, \bibinfo {author} {\bibfnamefont {K.~Z.}\ \bibnamefont
  {Hatsagortsyan}}, \bibinfo {author} {\bibfnamefont {C.~H.}\ \bibnamefont
  {Keitel}},\ and\ \bibinfo {author} {\bibfnamefont {A.}~\bibnamefont
  {Di~Piazza}},\ }\href@noop {} {\bibfield  {journal} {\bibinfo  {journal}
  {Phys. Rev. D}\ }\textbf {\bibinfo {volume} {91}},\ \bibinfo {pages} {013009}
  (\bibinfo {year} {2015})}\BibitemShut {NoStop}%
\bibitem [{\citenamefont {Meuren}\ \emph {et~al.}(2016)\citenamefont {Meuren},
  \citenamefont {Keitel},\ and\ \citenamefont {Di~Piazza}}]{Meuren_2016}%
  \BibitemOpen
  \bibfield  {author} {\bibinfo {author} {\bibfnamefont {S.}~\bibnamefont
  {Meuren}}, \bibinfo {author} {\bibfnamefont {C.~H.}\ \bibnamefont {Keitel}},\
  and\ \bibinfo {author} {\bibfnamefont {A.}~\bibnamefont {Di~Piazza}},\
  }\href@noop {} {\bibfield  {journal} {\bibinfo  {journal} {Phys. Rev. D}\
  }\textbf {\bibinfo {volume} {93}},\ \bibinfo {pages} {085028} (\bibinfo
  {year} {2016})}\BibitemShut {NoStop}%
\bibitem [{\citenamefont {King}(2020)}]{King_2020}%
  \BibitemOpen
  \bibfield  {author} {\bibinfo {author} {\bibfnamefont {B.}~\bibnamefont
  {King}},\ }\href@noop {} {\bibfield  {journal} {\bibinfo  {journal} {Phys.
  Rev. A}\ }\textbf {\bibinfo {volume} {101}},\ \bibinfo {pages} {042508}
  (\bibinfo {year} {2020})}\BibitemShut {NoStop}%
\bibitem [{\citenamefont {Roshchupkin}\ \emph {et~al.}(2012)\citenamefont
  {Roshchupkin}, \citenamefont {Lebed'}, \citenamefont {Padusenko},\ and\
  \citenamefont {Voroshilo}}]{Roshchupkin_2012}%
  \BibitemOpen
  \bibfield  {author} {\bibinfo {author} {\bibfnamefont {S.~P.}\ \bibnamefont
  {Roshchupkin}}, \bibinfo {author} {\bibfnamefont {A.~A.}\ \bibnamefont
  {Lebed'}}, \bibinfo {author} {\bibfnamefont {E.~A.}\ \bibnamefont
  {Padusenko}},\ and\ \bibinfo {author} {\bibfnamefont {A.~I.}\ \bibnamefont
  {Voroshilo}},\ }\href@noop {} {\bibfield  {journal} {\bibinfo  {journal}
  {Laser Phys.}\ }\textbf {\bibinfo {volume} {22}},\ \bibinfo {pages} {1113}
  (\bibinfo {year} {2012})}\BibitemShut {NoStop}%
\bibitem [{\citenamefont {L\"otstedt}\ and\ \citenamefont
  {Jentschura}(2009)}]{Loetstedt_2009}%
  \BibitemOpen
  \bibfield  {author} {\bibinfo {author} {\bibfnamefont {E.}~\bibnamefont
  {L\"otstedt}}\ and\ \bibinfo {author} {\bibfnamefont {U.~D.}\ \bibnamefont
  {Jentschura}},\ }\href@noop {} {\bibfield  {journal} {\bibinfo  {journal}
  {Phys. Rev. Lett.}\ }\textbf {\bibinfo {volume} {103}},\ \bibinfo {pages}
  {110404} (\bibinfo {year} {2009})}\BibitemShut {NoStop}%
\bibitem [{\citenamefont {Seipt}\ and\ \citenamefont
  {K\"ampfer}(2012)}]{Seipt_2012}%
  \BibitemOpen
  \bibfield  {author} {\bibinfo {author} {\bibfnamefont {D.}~\bibnamefont
  {Seipt}}\ and\ \bibinfo {author} {\bibfnamefont {B.}~\bibnamefont
  {K\"ampfer}},\ }\href@noop {} {\bibfield  {journal} {\bibinfo  {journal}
  {Phys. Rev. D}\ }\textbf {\bibinfo {volume} {85}},\ \bibinfo {pages}
  {101701(R)} (\bibinfo {year} {2012})}\BibitemShut {NoStop}%
\bibitem [{\citenamefont {Mackenroth}\ and\ \citenamefont
  {Di~Piazza}(2013)}]{Mackenroth_2013}%
  \BibitemOpen
  \bibfield  {author} {\bibinfo {author} {\bibfnamefont {F.}~\bibnamefont
  {Mackenroth}}\ and\ \bibinfo {author} {\bibfnamefont {A.}~\bibnamefont
  {Di~Piazza}},\ }\href@noop {} {\bibfield  {journal} {\bibinfo  {journal}
  {Phys. Rev. Lett.}\ }\textbf {\bibinfo {volume} {110}},\ \bibinfo {pages}
  {070402} (\bibinfo {year} {2013})}\BibitemShut {NoStop}%
\bibitem [{\citenamefont {King}(2015)}]{King_2015}%
  \BibitemOpen
  \bibfield  {author} {\bibinfo {author} {\bibfnamefont {B.}~\bibnamefont
  {King}},\ }\href@noop {} {\bibfield  {journal} {\bibinfo  {journal} {Phys.
  Rev. A}\ }\textbf {\bibinfo {volume} {91}},\ \bibinfo {pages} {033415}
  (\bibinfo {year} {2015})}\BibitemShut {NoStop}%
\bibitem [{\citenamefont {Dinu}\ and\ \citenamefont
  {Torgrimsson}(2019)}]{Dinu_2019}%
  \BibitemOpen
  \bibfield  {author} {\bibinfo {author} {\bibfnamefont {V.}~\bibnamefont
  {Dinu}}\ and\ \bibinfo {author} {\bibfnamefont {G.}~\bibnamefont
  {Torgrimsson}},\ }\href@noop {} {\bibfield  {journal} {\bibinfo  {journal}
  {Phys. Rev. D}\ }\textbf {\bibinfo {volume} {99}},\ \bibinfo {pages} {096018}
  (\bibinfo {year} {2019})}\BibitemShut {NoStop}%
\bibitem [{\citenamefont {Hu}\ \emph {et~al.}(2010)\citenamefont {Hu},
  \citenamefont {M\"uller},\ and\ \citenamefont {Keitel}}]{Hu_2010}%
  \BibitemOpen
  \bibfield  {author} {\bibinfo {author} {\bibfnamefont {H.}~\bibnamefont
  {Hu}}, \bibinfo {author} {\bibfnamefont {C.}~\bibnamefont {M\"uller}},\ and\
  \bibinfo {author} {\bibfnamefont {C.~H.}\ \bibnamefont {Keitel}},\
  }\href@noop {} {\bibfield  {journal} {\bibinfo  {journal} {Phys. Rev. Lett.}\
  }\textbf {\bibinfo {volume} {105}},\ \bibinfo {pages} {080401} (\bibinfo
  {year} {2010})}\BibitemShut {NoStop}%
\bibitem [{\citenamefont {Ilderton}(2011)}]{Ilderton_2011}%
  \BibitemOpen
  \bibfield  {author} {\bibinfo {author} {\bibfnamefont {A.}~\bibnamefont
  {Ilderton}},\ }\href@noop {} {\bibfield  {journal} {\bibinfo  {journal}
  {Phys. Rev. Lett.}\ }\textbf {\bibinfo {volume} {106}},\ \bibinfo {pages}
  {020404} (\bibinfo {year} {2011})}\BibitemShut {NoStop}%
\bibitem [{\citenamefont {King}\ \emph {et~al.}(2013)\citenamefont {King},
  \citenamefont {Elkina},\ and\ \citenamefont {Ruhl}}]{King_2013}%
  \BibitemOpen
  \bibfield  {author} {\bibinfo {author} {\bibfnamefont {B.}~\bibnamefont
  {King}}, \bibinfo {author} {\bibfnamefont {N.}~\bibnamefont {Elkina}},\ and\
  \bibinfo {author} {\bibfnamefont {H.}~\bibnamefont {Ruhl}},\ }\href@noop {}
  {\bibfield  {journal} {\bibinfo  {journal} {Phys. Rev. A}\ }\textbf {\bibinfo
  {volume} {87}},\ \bibinfo {pages} {042117} (\bibinfo {year}
  {2013})}\BibitemShut {NoStop}%
\bibitem [{\citenamefont {Dinu}\ and\ \citenamefont
  {Torgrimsson}(2018)}]{Dinu_2018}%
  \BibitemOpen
  \bibfield  {author} {\bibinfo {author} {\bibfnamefont {V.}~\bibnamefont
  {Dinu}}\ and\ \bibinfo {author} {\bibfnamefont {G.}~\bibnamefont
  {Torgrimsson}},\ }\href@noop {} {\bibfield  {journal} {\bibinfo  {journal}
  {Phys. Rev. D}\ }\textbf {\bibinfo {volume} {97}},\ \bibinfo {pages} {036021}
  (\bibinfo {year} {2018})}\BibitemShut {NoStop}%
\bibitem [{\citenamefont {Mackenroth}\ and\ \citenamefont
  {Di~Piazza}(2018)}]{Mackenroth_2018}%
  \BibitemOpen
  \bibfield  {author} {\bibinfo {author} {\bibfnamefont {F.}~\bibnamefont
  {Mackenroth}}\ and\ \bibinfo {author} {\bibfnamefont {A.}~\bibnamefont
  {Di~Piazza}},\ }\href@noop {} {\bibfield  {journal} {\bibinfo  {journal}
  {Phys. Rev. D}\ }\textbf {\bibinfo {volume} {98}},\ \bibinfo {pages} {116002}
  (\bibinfo {year} {2018})}\BibitemShut {NoStop}%
\bibitem [{\citenamefont {Dinu}\ and\ \citenamefont
  {Torgrimsson}(2020)}]{Dinu_2020}%
  \BibitemOpen
  \bibfield  {author} {\bibinfo {author} {\bibfnamefont {V.}~\bibnamefont
  {Dinu}}\ and\ \bibinfo {author} {\bibfnamefont {G.}~\bibnamefont
  {Torgrimsson}},\ }\href@noop {} {\bibfield  {journal} {\bibinfo  {journal}
  {Phys. Rev. D}\ }\textbf {\bibinfo {volume} {101}},\ \bibinfo {pages}
  {056017} (\bibinfo {year} {2020})}\BibitemShut {NoStop}%
\bibitem [{\citenamefont {Torgrimsson}(2020)}]{Torgrimsson_2020}%
  \BibitemOpen
  \bibfield  {author} {\bibinfo {author} {\bibfnamefont {G.}~\bibnamefont
  {Torgrimsson}},\ }\href@noop {} {\bibfield  {journal} {\bibinfo  {journal}
  {Phys. Rev. D}\ }\textbf {\bibinfo {volume} {102}},\ \bibinfo {pages}
  {096008} (\bibinfo {year} {2020})}\BibitemShut {NoStop}%
\bibitem [{\citenamefont {Baier}\ \emph {et~al.}(1989)\citenamefont {Baier},
  \citenamefont {Katkov},\ and\ \citenamefont {Strakhovenko}}]{Baier_1989}%
  \BibitemOpen
  \bibfield  {author} {\bibinfo {author} {\bibfnamefont {V.~N.}\ \bibnamefont
  {Baier}}, \bibinfo {author} {\bibfnamefont {V.~M.}\ \bibnamefont {Katkov}},\
  and\ \bibinfo {author} {\bibfnamefont {V.~M.}\ \bibnamefont {Strakhovenko}},\
  }\href@noop {} {\bibfield  {journal} {\bibinfo  {journal} {Nucl. Phys.}\
  }\textbf {\bibinfo {volume} {B328}},\ \bibinfo {pages} {387} (\bibinfo {year}
  {1989})}\BibitemShut {NoStop}%
\bibitem [{\citenamefont {Khokonov}\ and\ \citenamefont
  {Nitta}(2002)}]{Khokonov_2002}%
  \BibitemOpen
  \bibfield  {author} {\bibinfo {author} {\bibfnamefont {M.~K.}\ \bibnamefont
  {Khokonov}}\ and\ \bibinfo {author} {\bibfnamefont {H.}~\bibnamefont
  {Nitta}},\ }\href@noop {} {\bibfield  {journal} {\bibinfo  {journal} {Phys.
  Rev. Lett.}\ }\textbf {\bibinfo {volume} {89}},\ \bibinfo {pages} {094801}
  (\bibinfo {year} {2002})}\BibitemShut {NoStop}%
\bibitem [{\citenamefont {Di~Piazza}\ \emph {et~al.}(2007)\citenamefont
  {Di~Piazza}, \citenamefont {Milstein},\ and\ \citenamefont
  {Keitel}}]{Di_Piazza_2007}%
  \BibitemOpen
  \bibfield  {author} {\bibinfo {author} {\bibfnamefont {A.}~\bibnamefont
  {Di~Piazza}}, \bibinfo {author} {\bibfnamefont {A.~I.}\ \bibnamefont
  {Milstein}},\ and\ \bibinfo {author} {\bibfnamefont {C.~H.}\ \bibnamefont
  {Keitel}},\ }\href@noop {} {\bibfield  {journal} {\bibinfo  {journal} {Phys.
  Rev. A}\ }\textbf {\bibinfo {volume} {76}},\ \bibinfo {pages} {032103}
  (\bibinfo {year} {2007})}\BibitemShut {NoStop}%
\bibitem [{\citenamefont {Wistisen}(2015)}]{Wistisen_2015}%
  \BibitemOpen
  \bibfield  {author} {\bibinfo {author} {\bibfnamefont {T.~N.}\ \bibnamefont
  {Wistisen}},\ }\href@noop {} {\bibfield  {journal} {\bibinfo  {journal}
  {Phys. Rev. D}\ }\textbf {\bibinfo {volume} {92}},\ \bibinfo {pages} {045045}
  (\bibinfo {year} {2015})}\BibitemShut {NoStop}%
\bibitem [{\citenamefont {Dinu}\ \emph {et~al.}(2016)\citenamefont {Dinu},
  \citenamefont {Harvey}, \citenamefont {Ilderton}, \citenamefont {Marklund},\
  and\ \citenamefont {Torgrimsson}}]{Dinu_2016}%
  \BibitemOpen
  \bibfield  {author} {\bibinfo {author} {\bibfnamefont {V.}~\bibnamefont
  {Dinu}}, \bibinfo {author} {\bibfnamefont {C.}~\bibnamefont {Harvey}},
  \bibinfo {author} {\bibfnamefont {A.}~\bibnamefont {Ilderton}}, \bibinfo
  {author} {\bibfnamefont {M.}~\bibnamefont {Marklund}},\ and\ \bibinfo
  {author} {\bibfnamefont {G.}~\bibnamefont {Torgrimsson}},\ }\href@noop {}
  {\bibfield  {journal} {\bibinfo  {journal} {Phys. Rev. Lett.}\ }\textbf
  {\bibinfo {volume} {116}},\ \bibinfo {pages} {044801} (\bibinfo {year}
  {2016})}\BibitemShut {NoStop}%
\bibitem [{\citenamefont {Blackburn}\ \emph {et~al.}(2018)\citenamefont
  {Blackburn}, \citenamefont {Seipt}, \citenamefont {Bulanov},\ and\
  \citenamefont {Marklund}}]{Blackburn_2018}%
  \BibitemOpen
  \bibfield  {author} {\bibinfo {author} {\bibfnamefont {T.~G.}\ \bibnamefont
  {Blackburn}}, \bibinfo {author} {\bibfnamefont {D.}~\bibnamefont {Seipt}},
  \bibinfo {author} {\bibfnamefont {S.~S.}\ \bibnamefont {Bulanov}},\ and\
  \bibinfo {author} {\bibfnamefont {M.}~\bibnamefont {Marklund}},\ }\href@noop
  {} {\bibfield  {journal} {\bibinfo  {journal} {Phys. Plasmas}\ }\textbf
  {\bibinfo {volume} {25}},\ \bibinfo {pages} {083108} (\bibinfo {year}
  {2018})}\BibitemShut {NoStop}%
\bibitem [{\citenamefont {Podszus}\ and\ \citenamefont
  {Di~Piazza}(2019)}]{Podszus_2019}%
  \BibitemOpen
  \bibfield  {author} {\bibinfo {author} {\bibfnamefont {T.}~\bibnamefont
  {Podszus}}\ and\ \bibinfo {author} {\bibfnamefont {A.}~\bibnamefont
  {Di~Piazza}},\ }\href@noop {} {\bibfield  {journal} {\bibinfo  {journal}
  {Phys. Rev. D}\ }\textbf {\bibinfo {volume} {99}},\ \bibinfo {pages} {076004}
  (\bibinfo {year} {2019})}\BibitemShut {NoStop}%
\bibitem [{\citenamefont {Ilderton}(2019)}]{Ilderton_2019}%
  \BibitemOpen
  \bibfield  {author} {\bibinfo {author} {\bibfnamefont {A.}~\bibnamefont
  {Ilderton}},\ }\href@noop {} {\bibfield  {journal} {\bibinfo  {journal}
  {Phys. Rev. D}\ }\textbf {\bibinfo {volume} {99}},\ \bibinfo {pages} {085002}
  (\bibinfo {year} {2019})}\BibitemShut {NoStop}%
\bibitem [{\citenamefont {Lv}\ \emph {et~al.}(2021)\citenamefont {Lv},
  \citenamefont {Raicher}, \citenamefont {Keitel},\ and\ \citenamefont
  {Hatsagortsyan}}]{Raicher_2021}%
  \BibitemOpen
  \bibfield  {author} {\bibinfo {author} {\bibfnamefont {Q.~Z.}\ \bibnamefont
  {Lv}}, \bibinfo {author} {\bibfnamefont {E.}~\bibnamefont {Raicher}},
  \bibinfo {author} {\bibfnamefont {C.~H.}\ \bibnamefont {Keitel}},\ and\
  \bibinfo {author} {\bibfnamefont {K.~Z.}\ \bibnamefont {Hatsagortsyan}},\
  }\href@noop {} {\bibfield  {journal} {\bibinfo  {journal} {Phys. Rev.
  Research}\ }\textbf {\bibinfo {volume} {3}},\ \bibinfo {pages} {013214}
  (\bibinfo {year} {2021})}\BibitemShut {NoStop}%
\bibitem [{\citenamefont {Baier}\ and\ \citenamefont
  {Katkov}(1967)}]{Baier_1967}%
  \BibitemOpen
  \bibfield  {author} {\bibinfo {author} {\bibfnamefont {V.~N.}\ \bibnamefont
  {Baier}}\ and\ \bibinfo {author} {\bibfnamefont {V.~M.}\ \bibnamefont
  {Katkov}},\ }\href@noop {} {\bibfield  {journal} {\bibinfo  {journal} {Phys.
  Lett.}\ }\textbf {\bibinfo {volume} {A25}},\ \bibinfo {pages} {492} (\bibinfo
  {year} {1967})}\BibitemShut {NoStop}%
\bibitem [{\citenamefont {Baier}\ and\ \citenamefont
  {Katkov}(1968)}]{Baier_1968}%
  \BibitemOpen
  \bibfield  {author} {\bibinfo {author} {\bibfnamefont {V.~N.}\ \bibnamefont
  {Baier}}\ and\ \bibinfo {author} {\bibfnamefont {V.~M.}\ \bibnamefont
  {Katkov}},\ }\href@noop {} {\bibfield  {journal} {\bibinfo  {journal} {Sov.
  Phys.-JETP}\ }\textbf {\bibinfo {volume} {26}},\ \bibinfo {pages} {854}
  (\bibinfo {year} {1968})}\BibitemShut {NoStop}%
\bibitem [{\citenamefont {Baier}\ and\ \citenamefont
  {Katkov}(1969)}]{Baier_1969}%
  \BibitemOpen
  \bibfield  {author} {\bibinfo {author} {\bibfnamefont {V.~N.}\ \bibnamefont
  {Baier}}\ and\ \bibinfo {author} {\bibfnamefont {V.~M.}\ \bibnamefont
  {Katkov}},\ }\href@noop {} {\bibfield  {journal} {\bibinfo  {journal} {Sov.
  Phys.-JETP}\ }\textbf {\bibinfo {volume} {28}},\ \bibinfo {pages} {807}
  (\bibinfo {year} {1969})}\BibitemShut {NoStop}%
\bibitem [{\citenamefont {Akhiezer}\ and\ \citenamefont
  {Shul'ga}(1996)}]{Akhiezer_b_1996}%
  \BibitemOpen
  \bibfield  {author} {\bibinfo {author} {\bibfnamefont {A.~I.}\ \bibnamefont
  {Akhiezer}}\ and\ \bibinfo {author} {\bibfnamefont {N.~F.}\ \bibnamefont
  {Shul'ga}},\ }\href@noop {} {\emph {\bibinfo {title} {High-Energy
  Electrodynamics in Matter}}}\ (\bibinfo  {publisher} {Gordon and Breach
  Publishers, Amsterdam},\ \bibinfo {year} {1996})\BibitemShut {NoStop}%
\bibitem [{\citenamefont {Li}\ \emph {et~al.}(2015)\citenamefont {Li},
  \citenamefont {Hatsagortsyan}, \citenamefont {Galow},\ and\ \citenamefont
  {Keitel}}]{Li_2015}%
  \BibitemOpen
  \bibfield  {author} {\bibinfo {author} {\bibfnamefont {J.-X.}\ \bibnamefont
  {Li}}, \bibinfo {author} {\bibfnamefont {K.~Z.}\ \bibnamefont
  {Hatsagortsyan}}, \bibinfo {author} {\bibfnamefont {B.~J.}\ \bibnamefont
  {Galow}},\ and\ \bibinfo {author} {\bibfnamefont {C.~H.}\ \bibnamefont
  {Keitel}},\ }\href@noop {} {\bibfield  {journal} {\bibinfo  {journal} {Phys.
  Rev. Lett.}\ }\textbf {\bibinfo {volume} {115}},\ \bibinfo {pages} {204801}
  (\bibinfo {year} {2015})}\BibitemShut {NoStop}%
\bibitem [{\citenamefont {Harvey}\ \emph
  {et~al.}(2016{\natexlab{b}})\citenamefont {Harvey}, \citenamefont
  {Marklund},\ and\ \citenamefont {Holkundkar}}]{Harvey_2016}%
  \BibitemOpen
  \bibfield  {author} {\bibinfo {author} {\bibfnamefont {C.}~\bibnamefont
  {Harvey}}, \bibinfo {author} {\bibfnamefont {M.}~\bibnamefont {Marklund}},\
  and\ \bibinfo {author} {\bibfnamefont {A.~R.}\ \bibnamefont {Holkundkar}},\
  }\href@noop {} {\bibfield  {journal} {\bibinfo  {journal} {Phys. Rev. Accel.
  Beams}\ }\textbf {\bibinfo {volume} {19}},\ \bibinfo {pages} {094701}
  (\bibinfo {year} {2016}{\natexlab{b}})}\BibitemShut {NoStop}%
\bibitem [{\citenamefont {Heinzl}\ \emph {et~al.}(2016)\citenamefont {Heinzl},
  \citenamefont {Ilderton},\ and\ \citenamefont {King}}]{Heinzl_2016}%
  \BibitemOpen
  \bibfield  {author} {\bibinfo {author} {\bibfnamefont {T.}~\bibnamefont
  {Heinzl}}, \bibinfo {author} {\bibfnamefont {A.}~\bibnamefont {Ilderton}},\
  and\ \bibinfo {author} {\bibfnamefont {B.}~\bibnamefont {King}},\ }\href@noop
  {} {\bibfield  {journal} {\bibinfo  {journal} {Phys. Rev. D}\ }\textbf
  {\bibinfo {volume} {94}},\ \bibinfo {pages} {065039} (\bibinfo {year}
  {2016})}\BibitemShut {NoStop}%
\bibitem [{\citenamefont {King}\ and\ \citenamefont {Hu}(2016)}]{King_2016}%
  \BibitemOpen
  \bibfield  {author} {\bibinfo {author} {\bibfnamefont {B.}~\bibnamefont
  {King}}\ and\ \bibinfo {author} {\bibfnamefont {H.}~\bibnamefont {Hu}},\
  }\href@noop {} {\bibfield  {journal} {\bibinfo  {journal} {Phys. Rev. D}\
  }\textbf {\bibinfo {volume} {94}},\ \bibinfo {pages} {125010} (\bibinfo
  {year} {2016})}\BibitemShut {NoStop}%
\bibitem [{\citenamefont {Heinzl}\ and\ \citenamefont
  {Ilderton}(2017)}]{Heinzl_2017}%
  \BibitemOpen
  \bibfield  {author} {\bibinfo {author} {\bibfnamefont {T.}~\bibnamefont
  {Heinzl}}\ and\ \bibinfo {author} {\bibfnamefont {A.}~\bibnamefont
  {Ilderton}},\ }\href@noop {} {\bibfield  {journal} {\bibinfo  {journal}
  {Phys. Rev. Lett.}\ }\textbf {\bibinfo {volume} {118}},\ \bibinfo {pages}
  {113202} (\bibinfo {year} {2017})}\BibitemShut {NoStop}%
\bibitem [{\citenamefont {Bagrov}\ and\ \citenamefont
  {Gitman}(2014)}]{Bagrov_b_2014}%
  \BibitemOpen
  \bibfield  {author} {\bibinfo {author} {\bibfnamefont {V.~G.}\ \bibnamefont
  {Bagrov}}\ and\ \bibinfo {author} {\bibfnamefont {D.~M.}\ \bibnamefont
  {Gitman}},\ }\href@noop {} {\emph {\bibinfo {title} {The Dirac Equation and
  its Solutions}}}\ (\bibinfo  {publisher} {De Guyter, Berlin},\ \bibinfo
  {year} {2014})\BibitemShut {NoStop}%
\bibitem [{\citenamefont {Orzalesi}(1974)}]{Orzalesi_1974}%
  \BibitemOpen
  \bibfield  {author} {\bibinfo {author} {\bibfnamefont {C.~A.}\ \bibnamefont
  {Orzalesi}},\ }\href@noop {} {\bibfield  {journal} {\bibinfo  {journal} {Ann.
  Phys. (N. Y.)}\ }\textbf {\bibinfo {volume} {88}},\ \bibinfo {pages} {88}
  (\bibinfo {year} {1974})}\BibitemShut {NoStop}%
\bibitem [{\citenamefont {Orzalesi}(1975)}]{Orzalesi_1975}%
  \BibitemOpen
  \bibfield  {author} {\bibinfo {author} {\bibfnamefont {C.~A.}\ \bibnamefont
  {Orzalesi}},\ }\href@noop {} {\bibfield  {journal} {\bibinfo  {journal} {Ann.
  Phys. (N. Y.)}\ }\textbf {\bibinfo {volume} {92}},\ \bibinfo {pages} {44}
  (\bibinfo {year} {1975})}\BibitemShut {NoStop}%
\bibitem [{\citenamefont {Di~Piazza}(2014)}]{Di_Piazza_2014}%
  \BibitemOpen
  \bibfield  {author} {\bibinfo {author} {\bibfnamefont {A.}~\bibnamefont
  {Di~Piazza}},\ }\href@noop {} {\bibfield  {journal} {\bibinfo  {journal}
  {Phys. Rev. Lett.}\ }\textbf {\bibinfo {volume} {113}},\ \bibinfo {pages}
  {040402} (\bibinfo {year} {2014})}\BibitemShut {NoStop}%
\bibitem [{\citenamefont {Di~Piazza}(2015)}]{Di_Piazza_2015}%
  \BibitemOpen
  \bibfield  {author} {\bibinfo {author} {\bibfnamefont {A.}~\bibnamefont
  {Di~Piazza}},\ }\href@noop {} {\bibfield  {journal} {\bibinfo  {journal}
  {Phys. Rev. A}\ }\textbf {\bibinfo {volume} {91}},\ \bibinfo {pages} {042118}
  (\bibinfo {year} {2015})}\BibitemShut {NoStop}%
\bibitem [{\citenamefont {Di~Piazza}(2016)}]{Di_Piazza_2016}%
  \BibitemOpen
  \bibfield  {author} {\bibinfo {author} {\bibfnamefont {A.}~\bibnamefont
  {Di~Piazza}},\ }\href@noop {} {\bibfield  {journal} {\bibinfo  {journal}
  {Phys. Rev. Lett.}\ }\textbf {\bibinfo {volume} {117}},\ \bibinfo {pages}
  {213201} (\bibinfo {year} {2016})}\BibitemShut {NoStop}%
\bibitem [{\citenamefont {Di~Piazza}(2017)}]{Di_Piazza_2017_b}%
  \BibitemOpen
  \bibfield  {author} {\bibinfo {author} {\bibfnamefont {A.}~\bibnamefont
  {Di~Piazza}},\ }\href@noop {} {\bibfield  {journal} {\bibinfo  {journal}
  {Phys. Rev. A}\ }\textbf {\bibinfo {volume} {95}},\ \bibinfo {pages} {032121}
  (\bibinfo {year} {2017})}\BibitemShut {NoStop}%
\bibitem [{\citenamefont {Blankenbecler}\ and\ \citenamefont
  {Drell}(1987)}]{Blankenbecler_1987}%
  \BibitemOpen
  \bibfield  {author} {\bibinfo {author} {\bibfnamefont {R.}~\bibnamefont
  {Blankenbecler}}\ and\ \bibinfo {author} {\bibfnamefont {S.~D.}\ \bibnamefont
  {Drell}},\ }\href@noop {} {\bibfield  {journal} {\bibinfo  {journal}
  {Phys.Rev.D}\ }\textbf {\bibinfo {volume} {36}},\ \bibinfo {pages} {277}
  (\bibinfo {year} {1987})}\BibitemShut {NoStop}%
\bibitem [{\citenamefont {Akhiezer}\ and\ \citenamefont
  {Shul'ga}(1993)}]{Akhiezer_1993}%
  \BibitemOpen
  \bibfield  {author} {\bibinfo {author} {\bibfnamefont {A.~I.}\ \bibnamefont
  {Akhiezer}}\ and\ \bibinfo {author} {\bibfnamefont {N.~F.}\ \bibnamefont
  {Shul'ga}},\ }\href@noop {} {\bibfield  {journal} {\bibinfo  {journal} {Phys.
  Rep.}\ }\textbf {\bibinfo {volume} {234}},\ \bibinfo {pages} {297} (\bibinfo
  {year} {1993})}\BibitemShut {NoStop}%
\bibitem [{Note1()}]{Note1}%
  \BibitemOpen
  \bibinfo {note} {This does not contradict the above statement about the
  expansion of the trajectory with respect to $\eta $ in Refs.~\cite
  {Di_Piazza_2014,Di_Piazza_2015,Di_Piazza_2016,Di_Piazza_2017_b}. As mentioned
  in Ref. \cite {Di_Piazza_2014} there can be situations, where the external
  field features particular symmetries, like in a crystal, where the
  approximated solution of the trajectory presented there is not valid even
  though the electron energy is the largest dynamical energy scale in the
  problem.}\BibitemShut {Stop}%
\bibitem [{\citenamefont {Raicher}\ \emph {et~al.}(2019)\citenamefont
  {Raicher}, \citenamefont {Eliezer}, \citenamefont {Keitel},\ and\
  \citenamefont {Hatsagortsyan}}]{Raicher_2019}%
  \BibitemOpen
  \bibfield  {author} {\bibinfo {author} {\bibfnamefont {E.}~\bibnamefont
  {Raicher}}, \bibinfo {author} {\bibfnamefont {S.}~\bibnamefont {Eliezer}},
  \bibinfo {author} {\bibfnamefont {C.~H.}\ \bibnamefont {Keitel}},\ and\
  \bibinfo {author} {\bibfnamefont {K.~Z.}\ \bibnamefont {Hatsagortsyan}},\
  }\href@noop {} {\bibfield  {journal} {\bibinfo  {journal} {Phys. Rev. A}\
  }\textbf {\bibinfo {volume} {99}},\ \bibinfo {pages} {052513} (\bibinfo
  {year} {2019})}\BibitemShut {NoStop}%
\bibitem [{\citenamefont {Landau}\ and\ \citenamefont
  {Lifshitz}(1977)}]{Landau_b_3_1977}%
  \BibitemOpen
  \bibfield  {author} {\bibinfo {author} {\bibfnamefont {L.~D.}\ \bibnamefont
  {Landau}}\ and\ \bibinfo {author} {\bibfnamefont {E.~M.}\ \bibnamefont
  {Lifshitz}},\ }\href@noop {} {\emph {\bibinfo {title} {Quantum Mechanics
  (Non-Relativistic Theory)}}}\ (\bibinfo  {publisher} {Elsevier, Oxford},\
  \bibinfo {year} {1977})\BibitemShut {NoStop}%
\bibitem [{\citenamefont {Goldstein}\ \emph {et~al.}(2002)\citenamefont
  {Goldstein}, \citenamefont {Poole},\ and\ \citenamefont
  {Safko}}]{Goldstein_b_2002}%
  \BibitemOpen
  \bibfield  {author} {\bibinfo {author} {\bibfnamefont {H.}~\bibnamefont
  {Goldstein}}, \bibinfo {author} {\bibfnamefont {C.~P.~J.}\ \bibnamefont
  {Poole}},\ and\ \bibinfo {author} {\bibfnamefont {J.~L.}\ \bibnamefont
  {Safko}},\ }\href@noop {} {\emph {\bibinfo {title} {Classical Mechanics}}}\
  (\bibinfo  {publisher} {Pearson International Edition, Upper Saddle River,
  NJ},\ \bibinfo {year} {2002})\BibitemShut {NoStop}%
\bibitem [{\citenamefont {Landau}\ and\ \citenamefont
  {Lifshitz}(1975)}]{Landau_b_2_1975}%
  \BibitemOpen
  \bibfield  {author} {\bibinfo {author} {\bibfnamefont {L.~D.}\ \bibnamefont
  {Landau}}\ and\ \bibinfo {author} {\bibfnamefont {E.~M.}\ \bibnamefont
  {Lifshitz}},\ }\href@noop {} {\emph {\bibinfo {title} {The Classical Theory
  of Fields}}}\ (\bibinfo  {publisher} {Elsevier, Oxford},\ \bibinfo {year}
  {1975})\BibitemShut {NoStop}%
\bibitem [{\citenamefont {Evans}(2010)}]{Evans_b_2010}%
  \BibitemOpen
  \bibfield  {author} {\bibinfo {author} {\bibfnamefont {L.~C.}\ \bibnamefont
  {Evans}},\ }\href@noop {} {\emph {\bibinfo {title} {Partial Differential
  Equations}}}\ (\bibinfo  {publisher} {American Mathematical Society,
  Providence},\ \bibinfo {year} {2010})\BibitemShut {NoStop}%
\bibitem [{\citenamefont {van Vleck}(1928)}]{van_Vleck_1928}%
  \BibitemOpen
  \bibfield  {author} {\bibinfo {author} {\bibfnamefont {J.~H.}\ \bibnamefont
  {van Vleck}},\ }\href@noop {} {\bibfield  {journal} {\bibinfo  {journal}
  {Proc. Natl. Acad. Sci. U.S.A.}\ }\textbf {\bibinfo {volume} {14}},\ \bibinfo
  {pages} {178} (\bibinfo {year} {1928})}\BibitemShut {NoStop}%
\bibitem [{\citenamefont {Schiller}(1962{\natexlab{a}})}]{Schiller_1962_a}%
  \BibitemOpen
  \bibfield  {author} {\bibinfo {author} {\bibfnamefont {R.}~\bibnamefont
  {Schiller}},\ }\href@noop {} {\bibfield  {journal} {\bibinfo  {journal}
  {Phys. Rev.}\ }\textbf {\bibinfo {volume} {125}},\ \bibinfo {pages} {1100}
  (\bibinfo {year} {1962}{\natexlab{a}})}\BibitemShut {NoStop}%
\bibitem [{\citenamefont {Schiller}(1962{\natexlab{b}})}]{Schiller_1962_b}%
  \BibitemOpen
  \bibfield  {author} {\bibinfo {author} {\bibfnamefont {R.}~\bibnamefont
  {Schiller}},\ }\href@noop {} {\bibfield  {journal} {\bibinfo  {journal}
  {Phys. Rev.}\ }\textbf {\bibinfo {volume} {128}},\ \bibinfo {pages} {1402}
  (\bibinfo {year} {1962}{\natexlab{b}})}\BibitemShut {NoStop}%
\bibitem [{\citenamefont {Rubinow}\ and\ \citenamefont
  {Keller}(1963)}]{Rubinow_1963}%
  \BibitemOpen
  \bibfield  {author} {\bibinfo {author} {\bibfnamefont {S.~I.}\ \bibnamefont
  {Rubinow}}\ and\ \bibinfo {author} {\bibfnamefont {J.~B.}\ \bibnamefont
  {Keller}},\ }\href@noop {} {\bibfield  {journal} {\bibinfo  {journal} {Phys.
  Rev.}\ }\textbf {\bibinfo {volume} {131}},\ \bibinfo {pages} {2789} (\bibinfo
  {year} {1963})}\BibitemShut {NoStop}%
\bibitem [{\citenamefont {Pauli}(1932)}]{Pauli_1932}%
  \BibitemOpen
  \bibfield  {author} {\bibinfo {author} {\bibfnamefont {W.}~\bibnamefont
  {Pauli}},\ }\href@noop {} {\bibfield  {journal} {\bibinfo  {journal} {Helv.
  Phys. Acta}\ }\textbf {\bibinfo {volume} {5}},\ \bibinfo {pages} {179}
  (\bibinfo {year} {1932})}\BibitemShut {NoStop}%
\bibitem [{\citenamefont {Fock}(1937)}]{Fock_1937}%
  \BibitemOpen
  \bibfield  {author} {\bibinfo {author} {\bibfnamefont {V.~A.}\ \bibnamefont
  {Fock}},\ }\href@noop {} {\bibfield  {journal} {\bibinfo  {journal} {Phys. Z.
  Sowjetunion}\ }\textbf {\bibinfo {volume} {12}},\ \bibinfo {pages} {404}
  (\bibinfo {year} {1937})}\BibitemShut {NoStop}%
\bibitem [{\citenamefont {Taya}\ \emph {et~al.}()\citenamefont {Taya},
  \citenamefont {Fujimori}, \citenamefont {Misumi}, \citenamefont {Nitta},\
  and\ \citenamefont {Sakai}}]{Taya_2021}%
  \BibitemOpen
  \bibfield  {author} {\bibinfo {author} {\bibfnamefont {H.}~\bibnamefont
  {Taya}}, \bibinfo {author} {\bibfnamefont {T.}~\bibnamefont {Fujimori}},
  \bibinfo {author} {\bibfnamefont {T.}~\bibnamefont {Misumi}}, \bibinfo
  {author} {\bibfnamefont {M.}~\bibnamefont {Nitta}},\ and\ \bibinfo {author}
  {\bibfnamefont {N.}~\bibnamefont {Sakai}},\ }\href@noop {} {\bibinfo
  {journal} {arXiv:2010.16080}\ }\BibitemShut {NoStop}%
\bibitem [{\citenamefont {Ter-Mikaelian}(1972)}]{Ter-Mikaelian_b_1972}%
  \BibitemOpen
\bibfield  {journal} {  }\bibfield  {author} {\bibinfo {author} {\bibfnamefont
  {M.~L.}\ \bibnamefont {Ter-Mikaelian}},\ }\href@noop {} {\emph {\bibinfo
  {title} {High-Energy Electromagnetic Processes in Condensed Matter}}}\
  (\bibinfo  {publisher} {Wiley-Interscience, Toronto},\ \bibinfo {year}
  {1972})\BibitemShut {NoStop}%
\bibitem [{\citenamefont {Baier}\ and\ \citenamefont
  {Katkov}(2005)}]{Baier_2005}%
  \BibitemOpen
  \bibfield  {author} {\bibinfo {author} {\bibfnamefont {V.~N.}\ \bibnamefont
  {Baier}}\ and\ \bibinfo {author} {\bibfnamefont {V.~M.}\ \bibnamefont
  {Katkov}},\ }\href@noop {} {\bibfield  {journal} {\bibinfo  {journal} {Phys.
  Rep.}\ }\textbf {\bibinfo {volume} {409}},\ \bibinfo {pages} {261} (\bibinfo
  {year} {2005})}\BibitemShut {NoStop}%
\bibitem [{\citenamefont {Di~Piazza}\ \emph {et~al.}(2017)\citenamefont
  {Di~Piazza}, \citenamefont {Wistisen},\ and\ \citenamefont
  {Uggerh{\o}j}}]{Di_Piazza_2017}%
  \BibitemOpen
  \bibfield  {author} {\bibinfo {author} {\bibfnamefont {A.}~\bibnamefont
  {Di~Piazza}}, \bibinfo {author} {\bibfnamefont {T.~N.}\ \bibnamefont
  {Wistisen}},\ and\ \bibinfo {author} {\bibfnamefont {U.~I.}\ \bibnamefont
  {Uggerh{\o}j}},\ }\href@noop {} {\bibfield  {journal} {\bibinfo  {journal}
  {Phys. Lett. B}\ }\textbf {\bibinfo {volume} {765}},\ \bibinfo {pages} {1}
  (\bibinfo {year} {2017})}\BibitemShut {NoStop}%
\bibitem [{\citenamefont {Di~Piazza}(2021)}]{Di_Piazza_2021}%
  \BibitemOpen
  \bibfield  {author} {\bibinfo {author} {\bibfnamefont {A.}~\bibnamefont
  {Di~Piazza}},\ }\href@noop {} {\bibfield  {journal} {\bibinfo  {journal}
  {Phys. Rev. A}\ }\textbf {\bibinfo {volume} {103}},\ \bibinfo {pages}
  {012215} (\bibinfo {year} {2021})}\BibitemShut {NoStop}%
\bibitem [{\citenamefont {Sainte-Marie}\ \emph {et~al.}(2017)\citenamefont
  {Sainte-Marie}, \citenamefont {Gobert},\ and\ \citenamefont
  {Qu\'{e}r\'{e}}}]{Sainte-Marie_2017}%
  \BibitemOpen
  \bibfield  {author} {\bibinfo {author} {\bibfnamefont {A.}~\bibnamefont
  {Sainte-Marie}}, \bibinfo {author} {\bibfnamefont {O.}~\bibnamefont
  {Gobert}},\ and\ \bibinfo {author} {\bibfnamefont {F.}~\bibnamefont
  {Qu\'{e}r\'{e}}},\ }\href@noop {} {\bibfield  {journal} {\bibinfo  {journal}
  {Optica}\ }\textbf {\bibinfo {volume} {4}},\ \bibinfo {pages} {1298}
  (\bibinfo {year} {2017})}\BibitemShut {NoStop}%
\bibitem [{\citenamefont {Froula}\ \emph {et~al.}(2018)\citenamefont {Froula},
  \citenamefont {Turnbull}, \citenamefont {Davies}, \citenamefont {Kessler},
  \citenamefont {Haberberger}, \citenamefont {Palastro}, \citenamefont {Bahk},
  \citenamefont {Begishev}, \citenamefont {Boni}, \citenamefont {Bucht},
  \citenamefont {Katz},\ and\ \citenamefont {Shaw}}]{Froula_2018}%
  \BibitemOpen
  \bibfield  {author} {\bibinfo {author} {\bibfnamefont {D.~H.}\ \bibnamefont
  {Froula}}, \bibinfo {author} {\bibfnamefont {D.}~\bibnamefont {Turnbull}},
  \bibinfo {author} {\bibfnamefont {A.~S.}\ \bibnamefont {Davies}}, \bibinfo
  {author} {\bibfnamefont {T.~J.}\ \bibnamefont {Kessler}}, \bibinfo {author}
  {\bibfnamefont {D.}~\bibnamefont {Haberberger}}, \bibinfo {author}
  {\bibfnamefont {J.~P.}\ \bibnamefont {Palastro}}, \bibinfo {author}
  {\bibfnamefont {S.-W.}\ \bibnamefont {Bahk}}, \bibinfo {author}
  {\bibfnamefont {I.~A.}\ \bibnamefont {Begishev}}, \bibinfo {author}
  {\bibfnamefont {R.}~\bibnamefont {Boni}}, \bibinfo {author} {\bibfnamefont
  {S.}~\bibnamefont {Bucht}}, \bibinfo {author} {\bibfnamefont
  {J.}~\bibnamefont {Katz}},\ and\ \bibinfo {author} {\bibfnamefont {J.~L.}\
  \bibnamefont {Shaw}},\ }\href@noop {} {\bibfield  {journal} {\bibinfo
  {journal} {Nat. Photonics}\ }\textbf {\bibinfo {volume} {12}},\ \bibinfo
  {pages} {262} (\bibinfo {year} {2018})}\BibitemShut {NoStop}%
\bibitem [{\citenamefont {Palastro}\ \emph {et~al.}(2018)\citenamefont
  {Palastro}, \citenamefont {Turnbull}, \citenamefont {Bahk}, \citenamefont
  {Follett}, \citenamefont {Shaw}, \citenamefont {Haberberger}, \citenamefont
  {Bromage},\ and\ \citenamefont {Froula}}]{Palastro_2018}%
  \BibitemOpen
  \bibfield  {author} {\bibinfo {author} {\bibfnamefont {J.~P.}\ \bibnamefont
  {Palastro}}, \bibinfo {author} {\bibfnamefont {D.}~\bibnamefont {Turnbull}},
  \bibinfo {author} {\bibfnamefont {S.-W.}\ \bibnamefont {Bahk}}, \bibinfo
  {author} {\bibfnamefont {R.~K.}\ \bibnamefont {Follett}}, \bibinfo {author}
  {\bibfnamefont {J.~L.}\ \bibnamefont {Shaw}}, \bibinfo {author}
  {\bibfnamefont {D.}~\bibnamefont {Haberberger}}, \bibinfo {author}
  {\bibfnamefont {J.}~\bibnamefont {Bromage}},\ and\ \bibinfo {author}
  {\bibfnamefont {D.~H.}\ \bibnamefont {Froula}},\ }\href@noop {} {\bibfield
  {journal} {\bibinfo  {journal} {Phys. Rev. A}\ }\textbf {\bibinfo {volume}
  {97}},\ \bibinfo {pages} {033835} (\bibinfo {year} {2018})}\BibitemShut
  {NoStop}%
\bibitem [{\citenamefont {Howard}\ \emph {et~al.}(2019)\citenamefont {Howard},
  \citenamefont {Turnbull}, \citenamefont {Davies}, \citenamefont {Franke},
  \citenamefont {Froula},\ and\ \citenamefont {Palastro}}]{Howard_2019}%
  \BibitemOpen
  \bibfield  {author} {\bibinfo {author} {\bibfnamefont {A.~J.}\ \bibnamefont
  {Howard}}, \bibinfo {author} {\bibfnamefont {D.}~\bibnamefont {Turnbull}},
  \bibinfo {author} {\bibfnamefont {A.~S.}\ \bibnamefont {Davies}}, \bibinfo
  {author} {\bibfnamefont {P.}~\bibnamefont {Franke}}, \bibinfo {author}
  {\bibfnamefont {D.~H.}\ \bibnamefont {Froula}},\ and\ \bibinfo {author}
  {\bibfnamefont {J.~P.}\ \bibnamefont {Palastro}},\ }\href@noop {} {\bibfield
  {journal} {\bibinfo  {journal} {Phys. Rev. Lett.}\ }\textbf {\bibinfo
  {volume} {123}},\ \bibinfo {pages} {124801} (\bibinfo {year}
  {2019})}\BibitemShut {NoStop}%
\bibitem [{\citenamefont {Palastro}\ \emph {et~al.}(2020)\citenamefont
  {Palastro}, \citenamefont {Shaw}, \citenamefont {Franke}, \citenamefont
  {Ramsey}, \citenamefont {Simpson},\ and\ \citenamefont
  {Froula}}]{Palastro_2020}%
  \BibitemOpen
  \bibfield  {author} {\bibinfo {author} {\bibfnamefont {J.~P.}\ \bibnamefont
  {Palastro}}, \bibinfo {author} {\bibfnamefont {J.~L.}\ \bibnamefont {Shaw}},
  \bibinfo {author} {\bibfnamefont {P.}~\bibnamefont {Franke}}, \bibinfo
  {author} {\bibfnamefont {D.}~\bibnamefont {Ramsey}}, \bibinfo {author}
  {\bibfnamefont {T.~T.}\ \bibnamefont {Simpson}},\ and\ \bibinfo {author}
  {\bibfnamefont {D.~H.}\ \bibnamefont {Froula}},\ }\href@noop {} {\bibfield
  {journal} {\bibinfo  {journal} {Phys. Rev. Lett.}\ }\textbf {\bibinfo
  {volume} {124}},\ \bibinfo {pages} {134802} (\bibinfo {year}
  {2020})}\BibitemShut {NoStop}%
\end{thebibliography}

\end{document}